\documentclass[a4paper,fleqn,usenatbib]{mnras}

\usepackage{mathptmx}

\usepackage[T1]{fontenc}

\usepackage{graphicx}	
\usepackage{amsmath}	
\usepackage{amssymb}	

\usepackage[flushleft]{threeparttable}
\usepackage{tabularx}
\usepackage{color}

\title[Powerful quasars, their hosts and quasar winds]{Towards a comprehensive picture of powerful quasars, their host galaxies and quasar winds at $z \sim 0.5$}

\author[D. Wylezalek et al.]{
Dominika Wylezalek,$^{1}$\thanks{E-mail: dwylezalek@jhu.edu}
Nadia L. Zakamska,$^{1}$
Guilin Liu$^{2}$
and Georges Obied$^{1,3}$
\\
$^{1}$Department of Physics \& Astronomy, Johns Hopkins University, Bloomberg Center, 3400 N. Charles St., Baltimore, MD 21218, USA\\
$^{2}$Department of Physics, Virginia Tech, Blacksburg, VA 24061, USA\\
$^{3}$Department of Physics, Harvard University, 17 Oxford Street, Cambridge MA 02138, USA
}

\date{Accepted XXX. Received YYY; in original form ZZZ}

\pubyear{2015}

\begin{document}
\newcolumntype{C}[1]{>{\centering\arraybackslash}p{#1}}

\label{firstpage}
\pagerange{\pageref{firstpage}--\pageref{lastpage}}
\maketitle

\begin{abstract}

Luminous type-2 quasars in which the glow from the central black hole is obscured by dust are ideal targets for studying their host galaxies and the quasars' effect on galaxy evolution. Such feedback appears ubiquitous in luminous obscured quasars where high velocity ionized nebulae have been found. We present rest-frame yellow-band ($\sim 5000~\AA$) observations using the \textit{Hubble Space Telescope} for a sample of 20 luminous quasar host galaxies at $0.2 < z < 0.6$ selected from the Sloan Digital Sky Survey. For the first time, we combine host galaxy observations with geometric measurements of quasar illumination using blue-band HST observations and [OIII] integral field unit observations probing the quasar winds. The HST images reveal bright merger signatures in about half the galaxies; a significantly higher fraction than in comparison inactive ellipticals. We show that the host galaxies are primarily bulge-dominated, with masses close to $M^{*}$, but belong to $<$ 30\% of elliptical galaxies that are highly star-forming at $z \sim 0.5$. Ionized gas signatures are uncorrelated with faint stellar disks (if present), confirming that the ionized gas is not concentrated in a disk. Scattering cones and [OIII] ionized gas velocity field are aligned with the forward scattering cones being co-spatial with the blue-shifted side of the velocity field, suggesting the high velocity gas is indeed photo-ionized by the quasar. Based on the host galaxies' high star-formation rates and bright merger signatures, we suggest that this low-redshift outbreak of luminous quasar activity is triggered by recent minor mergers. Combining these novel observations, we present new quasar unification tests, which are in agreement with expectations of the orientation-based unification model for quasars.
\end{abstract}




\section{Introduction}

The discovery of a tight relationship between the masses of black holes in nearby galaxies and the velocities and masses of their stellar populations \citep[e.g.][]{Magorrian_1998, Gebhardt_2000, Ferrarese_2000} has made it clear that the active (quasar) phase of black hole evolution has profound effects on the formation of the galaxy. One idea -- quasar feedback -- seeks to simultaneously explain several major observational findings and open issues in galaxy formation theory such as the observed relations between black holes and their hosts \citep{Hopkins_2006} and the absence of overly massive galaxies \citep{Croton_2006}. Even a small fraction of energy liberated during accretion of the material on the black hole is sufficient to unbind much of the gas supply of the host galaxy, thus shutting off star formation and launching a ``galactic wind''. However, specific examples of quasar-driven feedback in action have been surprisingly hard to come by, and it is far from clear how exactly the coupling between accretion energy and matter on galactic scales is established. Understanding the mechanism, the physics of quasar feedback and its impact on the quasar host galaxy has been the subject of intense observational effort \citep{Nesvadba_2008, Crenshaw_2003, Arav_2008, Moe_2009}.

Recently, \citet{Liu_2013a, Liu_2013b}  found signatures of such feedback in quasi-spherical, high velocity ionized nebulae that appear to be ubiquitous in luminous ($L_{\rm{[OIII]}} > 10^{42.8}$ erg s$^{-1}$) obscured quasars using Integral Field Unit (IFU) observations of 11 radio-quiet type-2 quasars to determine the kinematics and morphology of ionized gas. They detected extended emission-line nebulae on 20-30 kpc scales in every case via the [OIII]$\lambda5007$\AA\ emission line. The nebulae are almost round, suggestive of outflows with large covering factors, and show well-defined kinematic structures and high velocity dispersions. These observations are indicative of organized galaxy-wide outflows, that cannot be confined by the galaxy potential. Several groups have now started to investigate such high-velocity outflows in luminous quasars to explore various mechanisms of coupling between the black hole radiative output and the gas  \citep[e.g.][]{Zakamska_2014, Brusa_2015}. 

However, few observations of host galaxies of such luminous quasars with signatures of quasar feedback are available and it is unclear, for example, what types of galaxies these luminous quasars are hosted by.  While multiple studies find that host galaxies of luminous type-2 quasars at intermediate redshift are often found to be ellipticals with old stellar populations and little ongoing star formation \citep{Dunlop_2003, Villar_Martin_2012}, there have been a number of contradicting results. \citet{Lacy_2007}, for example, report that all but one type-2 quasar hosts in their sample at $0.3 < z < 0.8$ selected in the mid-IR using the \textit{Spitzer Space Telescope} show dusty star-forming disks and \citet{Letawe_2007} find large amounts of gas, blue colors and mixed morphological types in their sample of 20 luminous quasars.

If many hosts at $z \sim 0.5$ and high quasar luminosities like the objects in \citet{Liu_2013a, Liu_2013b} tend to be ellipticals, it is unclear what is the origin of the extended envelopes of ionized gas and how it has been supplied to the center to activate the quasar. In some of the kinematic IFU data, \citet{Liu_2013b} detected a few low-velocity-dispersion which might be either small companion galaxies that simply happened to be illuminated by the luminous quasar or the less turbulent components within the quasar wind. 

At $z \sim 2$, the peak of quasar activity, mergers and galaxy interactions were more common and major mergers may have played a significant role in the triggering of quasars \citep{Glikman_2015}. At lower redshift, recent work by \citet{Villforth_2015} shows that powerful quasars at $z \sim 0.7$ selected to have clear signs of mergers also show extended ionized gas with fast outflow velocities ($\sim 1500$ km s$^{-1}$). Studying a sample of $\sim 40$ luminous quasars, \citet{Villar_Martin_2012} find that more than half of the quasars' host galaxies show clear signs of morphological disturbances. Major mergers are therefore also a likely scenario for quasar triggering at $z \sim 0.5$. However, some results indicate that they might not be the dominant triggering process at high quasar luminosities \citep{Villforth_2014}. It therefore remains less clear what triggers some of the most luminous quasars at lower redshift \citep{Floyd_2004, Urrutia_2008}. 

The glow from the quasar itself is a limiting factor and has impeded studies of gas distributions around quasars and their host galaxies in the past. Type-2, i.e. obscured, quasars \citep{Zakamska_2003} are ideal targets to circumvent this limitation. In the standard unification model for quasars \citep{Antonucci_1993} this obscuration is produced by an optically thick circumnuclear torus that is viewed edge-on or close to edge-on. Depending on the orientation angle at which the quasar is observed, an unobscured (type-1) or obscured (type-2) quasar is seen. According to this model, type-1 and type-2 quasars are intrinsically the same objects. Recently, an alternative paradigm has been gaining attention in which type-2 quasars evolve into type-1 quasars as their merger-triggered, dust-obscured black holes become more powerful and clear out the environment \cite[e.g.][]{Sanders_1988, Hopkins_2006, Menci_2008, Somerville_2008, Glikman_2015}. Independent of the model, obscuration removes the glare of the nucleus itself allowing the study of objects with extremely high intrinsic luminosities \citep[$M_B<-26.9$ mag,][]{Reyes_2008} and allowing to observe the host galaxy which the quasar would otherwise outshine. 

In this paper, we present a comprehensive study of the host galaxies of two samples of luminous type-2 quasars at $z \sim 0.5$. Using the Advanced Camera for Surveys (ACS) aboard the \textit{Hubble Space Telescope (HST)}, we have obtained high-resolution optical broad band photometry of all 11 quasars in the \citet{Liu_2013a, Liu_2013b} sample to study in detail the morphologies and structure of the quasars' host galaxies. We also investigate the immediate environment of the galaxies and look for merger signs. For the first time, we also analyze the relationsip between the orientation of the quasar wind probed by the IFU observations, the orientation of the host galaxy and the geometry of quasar obscuration probed by scattered light observations \citep{Obied_2015}. We supplement these data with optical and UV HST imaging of a sample of luminous quasars presented in \citet{Zakamska_2006}. Multi-wavelength supporting archival data for all 20 sources help constrain star formation rates and stellar masses. 

The paper is organized as follows: Section 2 gives details on the sample selection, the various observations and data reduction. In Section 3 we describe the analysis of the host galaxies' morphologies, spectral energy distribution (SED) fitting and comparison with IFU data. We discuss our results in Section 4, including a test of the unification-by-orientation model for quasars and draw conclusions in Section 5. Throughout the paper we assume $H_0 = 71$ km s$^{-1}$ Mpc$^{-1}$, $\Omega_m = 0.27, \Omega_{\lambda} = 0.73$. 

\section{Sample Selection, Observations and Data Reduction}
\subsection{HST}

In this paper, we study 20 luminous quasars that were imaged with the \textit{Hubble Space Telescope} at rest-frame UV and rest-frame optical wavelengths. Our total sample consists of two subsamples of 11 (Sample I) and 9 (Sample II) quasars, respectively, that were selected and observed as described in the following.

\subsubsection{Sample I}

This sample consists of all quasars studied by \citet{Liu_2013a, Liu_2013b} that were selected as luminous ($L_{\rm{[OIII]}} > 10^{9.2}$~L$_{\sun}$), radio-quiet quasars from the catalog of \cite{Reyes_2008} of type-2 quasars from the Sloan Digital Sky Survey \citep[SDSS,][]{York_2000}. To maximize spatial information, these objects were selected to be as low redshift as possible while still meeting the luminosity and radio-quiet selection criteria, i.e. they cover redshifts $0.4 < z < 0.6$. Details on sample selection of these 11 objects can be found in \citet{Liu_2013a}. Using integral-field unit (IFU) observations \citet{Liu_2013a} showed that all of these luminous quasars have large, galaxy-wide extended ionized emission-line nebulae whose kinematics are pointing towards fast, wide-angle quasi-spherical outflows. 

We observed this sample using the HST (GO-13307, PI: N. L. Zakamska) and all targets were imaged in two broad bands, rest-frame U and rest-frame yellow (between V and R), carefully chosen to sample the continuum and to avoid strong forbidden emission lines. The observations were performed using the Advanced Camera for Surveys (ACS). Avoiding strong emission lines, we used the ramp filter FR914M for the rest-frame yellow-band observations. The effective wavelength of the ramp filter was slightly adjusted for each object depending on its redshift (see Tab.~\ref{observations}). The wavelength variation across the objects are generally of the order of 10$\AA$ and negligible for the subsequent analysis. For the rest-frame U-band observations we used the filter F475W for all but two objects. For the two lowest redshift objects, we used F435W. The observations are summarized in Tab.~\ref{observations}. The U-band data provide the measurements of the scattered light, the opening angles and the surface brightness profiles and allow us to obtain a rough constraint on the inclination angle of the quasar \citep{Obied_2015}.

In this paper, we focus on the yellow-band images which are used to investigate galaxy morphology and the immediate environments of the host galaxies. All objects were imaged using the ACS with the ramp filter FR914M which is suitably placed between [OIII]$\lambda\lambda$4959,5007 and H$\alpha$. The ACS/WFC imaging consists of four exposures obtained by using the default {\sc acs-wfc-dither-box} pointing pattern designed for optimal half-pixel sampling. To optimize the quality of the data products, we reprocess the data using the AstroDrizzle task in the software package DrizzlePac distributed through PyRAF. As our targets are single compact sources, we set the size of the shrunk pixels (``drops'') in the drizzle algorithm to be half of the native plate scale (0.05\arcsec), following the suggestion in the DrizzlePac Handbook\footnote{http://www.stsci.edu/hst/HST\_overview/drizzlepac.}. The adopted drizzle pattern also facilitates the rejection of cosmic rays and the larger detector artifacts. For the purpose of quality control, we have verified that statistics performed on the drizzled weight images yield a r.m.s.-to-median ratio of $\sim0.1$, satisfying the $<0.2$ requirement for balancing between resolution improvement and background noise increment due to pixel resampling, as per the HST Dither Handbook\footnote{http://www.stsci.edu/hst/HST\_overview/documents/\\dither\_handbook.}. Accordingly, the final pixels of our drizzled images are resampled to 0.025\arcsec.

\subsubsection{Sample II}

We supplement the observations of Sample I with a sample of 9 luminous quasars for which U-band and yellow-band observations are available from the HST archive (GO-9905, PI: M. Strauss). These quasars were originally selected from the parent sample of 294 type-2 quasar candidates from \citet{Zakamska_2003}, out of which 9 were chosen to be observed with the HST based on their high [OIII]$\lambda\lambda$4959,5007 line luminosities ($L_{\rm{[OIII]}} > 10^{9}$~L$_{\sun}$), radio-quiet properties and redshifts to be observed using the available filters on the HST. Each object was imaged with the Wide Field Channel of the ACS in three filters, at rest-frame UV, blue and yellow wavelengths. Details on the sample selection and observations can be found in \citet{Zakamska_2006}. The objects cover a redshift range of $0.2 < z < 0.4$. The native pixel scale for these images is 0.049\arcsec pixel$^{-1}$ and we resampled the drizzled images to 0.025\arcsec\ following the same procedure as described above. In this paper, we only use the rest-frame UV- and yellow-band observations, matching the available data for Sample I (Tab.~\ref{observations}). 
\begin{table*}
\caption{Summary of HST observations}
\begin{center}
\begin{tabular}{lccc C{2.9cm}}
\hline\hline
Object Name  & z & log($L_{\rm{[OIII]}}/L_{\rm{\sun}}$) & Filters (blue, yellow) & rest-frame effective $\lambda$ (\AA): blue, yellow\\
(1) & (2) & (3) & (4) & (5)\\
\hline\vspace{0.05cm}
Sample I: & & & & \\
SDSS J014932.53-004803.7     & 0.566 	  &      9.28 	  &   F475W, FR914M   &  3007,   5092    \\
SDSS J021047.01-100152.9  & 0.540      &      9.89    &    F475W, FR914M    &  3057,   5178  \\
SDSS J031909.61-001916.7    & 0.635      &      9.15    &    F475W, FR914M    &  2880,   4877  \\
SDSS J031950.54-005850.6    & 0.626      &      9.37    &    F475W, FR914M    &  2895,     4904   \\
SDSS J032144.11+001638.2    &  0.643      &      9.51    &    F475W, FR914M    &  2866,     4854   \\
SDSS J075944.64+133945.8    &  0.649      &      9.79    &    F475W, FR914M    &  2855,    4836   \\
SDSS J084130.78+204220.5     &  0.641      &      9.72    &    F475W, FR914M    &  2869,     4859   \\
SDSS J084234.94+362503.1    &  0.561      &      9.97    &    F475W, FR914M    &  3016,     5108   \\
SDSS J085829.59+441734.7   &   0.454      &      9.71    &    F435W, FR914M    &  2979,     5484   \\
SDSS J103927.19+451215.4    & 0.579      &      9.70    &    F475W, FR914M    &  2982,    5050   \\
SDSS J104014.43+474554.8    & 0.486      &      9.93    &    F435W, FR914M    &  2915,   5366  \\
\hline \vspace{0.05cm}
Sample II: & & & & \\
SDSS J012341.47+004435.9   &   0.399   &  9.13  &   F435W, FR914M    &  3086,  5700  \\
SDSS J092014.11+453157.3   &    0.402   &  9.04 &   F435W, FR914M    &  3079,  5688  \\
SDSS J103951.49+643004.2   &    0.402   &  9.41 &   F435W, FR914M    &  3079,  5688  \\
SDSS J110621.96+035747.1   &    0.242   &  9.13 &   FR459M, FR647M   &   3124,   5804  \\
SDSS J124337.34-023200.2    &    0.281   &  9.02 &   FR459M, FR647M   &   3118,   5789   \\
SDSS J130128.76-005804.3    &   0.246   &  9.25 &   FR459M, FR647M   &   3114,   5786  \\
SDSS J132323.33-015941.9    &   0.350   &  9.19 &   F435W, F775W     &  3198, 5699  \\
SDSS J141315.31-014221.0    &  0.380   &  9.25 &   F435W, FR914M    &  3129,  5779  \\
SDSS J235818.87-000919.5    &    0.402   &  9.32 &   F435W, FR914M    &  3079,  5688  \\
 \hline 
\end{tabular}
\end{center}
\label{observations}
\begin{tablenotes}
\item \textit{Notes:}  (1) Object name (2) Redshift from \citet{Zakamska_2003} and \citet{Reyes_2008} (3) Total luminosity of the [OIII]$\lambda 5007 \AA\ $line, logarithmic scale in units of solar luminosity, from \citet{Liu_2013a} (4) HST filters (5) rest-frame effective wavelength probed
        \end{tablenotes}
\end{table*}

\subsubsection{Color Composite images}

\begin{figure*}[]
\begin{center}
\includegraphics[trim = 0cm 2.5cm 0cm 0cm, clip = true, scale = 0.8]{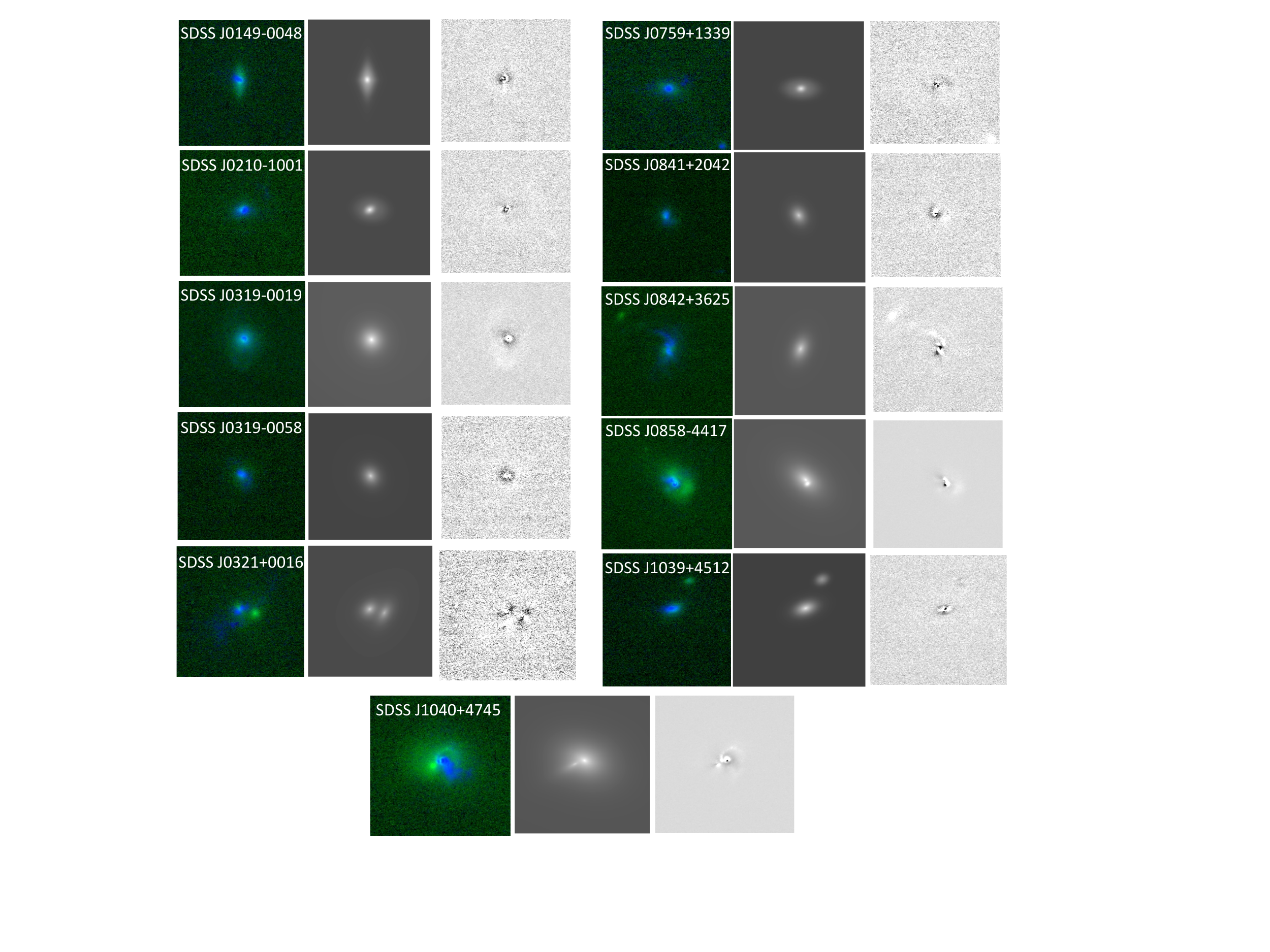}
\caption{\textit{Left panels:} Color-composite images of all sources in Sample I using the algorithm of \citet{Lupton_2004}, showing the HST blue image in blue and the HST yellow image in green. \textit{Middle panels:} GALFIT model to the HST yellow-band images. \textit{Right panels:} Residual yellow-band images subtracting the GALFIT model from the yellow HST image. The postage stamps are 6~arcsec $\times$ 6~arcsec, corresponding to a physical size of 35~kpc~$\times$~35~kpc to 41~kpc~$\times$~41~kpc depending on the redshift of the object.}
\label{stamps}
\end{center}
\end{figure*}

We use {\tt iraf imexam} to measure the centers of two to three stars in the field of view where the U-band and yellow-band images overlap and then use {\tt iraf imalign} to shift the U-band images and align them with the yellow images. The images are then combined to produce a two-color composite image using the asinh stretching and color combining routine from \citet{Lupton_2004}. The resulting images for Sample I are presented in Figure \ref{stamps}. Color-composite images for Sample II are presented in \citet{Zakamska_2006}.

\subsection{Gemini Integral Field Unit Observations}

\citet{Liu_2013a, Liu_2013b} presented IFU observations of the ionized gas around the quasars in Sample I measured by the [OIII]$\lambda$5007\AA\ emission line. Details on the observations and data reduction can be found in \citet{Liu_2013a}. In brief, the observations were taken using GMOS-N IFU in 2010 December covering a field-of-view of 5 $\times$ 7 arcsec (30$\times$42 kpc at $z = 0.5$) for a total exposure time of 3600 s. In order to cover both the H$\beta$ and [OIII] lines, all objects were observed in the $i$-band ($7060-8050$\AA) to cover the rest-frame optical at $\sim 4100-5200$\AA . Data reduction was performed using the Gemini package for IRAF with adaptions to improve on e.g. cosmic ray rejection. The reduced data are resampled and interpolated onto a data cube with  spatial sampling of 0.1~arcsec. The spectra were flux-calibrated using SDSS DR7 \citep{Abazajian_2009} spectra of the same science targets. 

Re-analysis of the data revealed that the world coordinate system (WCS) of the Gemini IFU cubes obtained using the Gemini pipeline were not calibrated correctly. While irrelevant for the analysis of \citet{Liu_2013a, Liu_2013b}, the proper orientation of the cubes is essential for the analysis of this paper, where we investigate the relationships between the IFU and imaging data. We therefore correct the WCS in the Gemini cubes.

We first match the (R.A., Dec.) positions of the quasars as measured from the SDSS images to the (x,y) center pixel values in the Gemini [OIII] intensity maps measured by the IRAF task {\tt imexamine}. We then use the IRAF task {\tt mkcwcs} to update the Gemini WCS. In addition to the sources in Sample I, \citet{Liu_2013a} also observed three radio-loud sources where one has a small companion galaxy and another one has some extended structures. These distinct structures allow for direct comparison between Gemini intensity maps and other data \citep[SDSS images, \textit{HST}/WFPC2 images from][]{Odea_2002}. We find that in order to align the Gemini images correctly to the HST images, a right-handed coordinate system has to be applied, which flips the images and in which position angles are then measured clockwise. Using {\tt mkwcs}, we therefore apply a right-handed coordinate system to all Gemini images of the targets in Sample I and rotate them according to the position angles of the Gemini observations \citep[see Tab. 1 in][]{Liu_2013a}. 

Re-analysing of the IFU data also revealed a PSF anomaly problem involving 3 obscured quasars (SDSS J0149$-$0048, SDSS J0319$-$0019 and SDSS J0842+3625) that were analyzed in our previous Gemini 2-slit IFU campaign \citep{Liu_2013a, Liu_2013b}. 
This anomaly is due to fiber misidentification during the data reduction process. Following the \citet{Lena_2014}
tutorial for Gemini GMOS data reduction, we have manually corrected the 
misidentified fibers for these 3 objects, and reconstructed the relevant figures. This correction leads 
to differences in the central $\sim1\arcsec$ of the [OIII] flux maps, while the changes in
the non-parametric measurement ($V_{\rm med}$, $W_{80}$, $A$ and $K$) maps are negligible. 
As a result, none of the scientific conclusions in \citet{Liu_2013a, Liu_2013b} are affected by this correction. For completeness, we show the corrected maps for these objects in the Appendix (Figure \ref{kin_new}, Figure \ref{maps_new},  Figure \ref{maps_new2}).

\citet{Liu_2013a} detected very extended, galaxy-wide emission-line nebulae on 20-30 kpc scales in every case via the [OIII]$\lambda5007$\AA\ emission line, indicative of organized galaxy-wide outflows. However, these data do not contain any information about the quasar's host galaxies or any information about the triggering of the winds. In this paper, we investigate the nature of the quasar host galaxies from Sample I and how various wind components are related to host galaxy properties.

\subsection{Multi-wavelength supporting data}

In addition to the HST imaging data for Sample~I and II and the Gemini IFU data for Sample I, we use multi-wavelength archival imaging data for our targets for spectral energy distribution (SED) decomposition which we describe in Section 3.2. We cross-correlate the positions of our targets as measured from SDSS with catalogs from the UKIRT Infrared Deep Survey (UKIDSS) in the near-IR, the \textit{Wide-field Infrared Survey Explorer (WISE)} in the mid-IR and the \textit{Herschel Space Observatory} and the \textit{Spitzer Space Telescope} in the far-IR. By definition, all sources are also covered by the SDSS in the optical. We convert the magnitudes (given in AB, Vega or SDSS magnitudes) reported by these various surveys to $F_{\nu}$. The SDSS magnitudes are very close to AB magnitudes to 0.01 mag precision, but in the $u-$ and $z-$band we apply a SDSS-to-AB correction of -0.04 and +0.02 mag, respectively~\footnote{http://classic.sdss.org/dr7/algorithms/fluxcal.html\#sdss2ab}. Tables \ref{multi_lambda} and \ref{multi_lambda2} in the Appendix report on the multi-wavelength data available for the sources in our sample.  

\section{Analysis}
\subsection{Quasar host galaxies}

\subsubsection{2D Surface Brightness Profiles}

We use the publicly free available code GALFIT \citep{Peng_2002, Peng_2010} to perform two-dimensional fitting of the surface brightness profiles on the yellow-band images of targets from Sample I and Sample II. The light profiles of galaxies can generally be modeled well by one or multiple S\'ersic profiles \citep{Sersic_1968} of the form
\begin{equation}
\Sigma(r) = \Sigma_{e}e^{-\kappa[(r/r_{e})^{1/n}-1]}
\end{equation}
where $r_{e}$ is the effective radius of the galaxy, $\Sigma_{e}$ is the surface brightness at $r_e$, $n$ is the power-law index also known as S\'ersic index, and $\kappa$ is coupled to $n$ such that half of the total flux is always within $r_e$. Bulges and elliptical galaxies tend to be well described by a de Vaucouleurs profile \citep[$n \simeq 4$][]{Vaucouleurs_1948}, while disk components generally follow an exponential profile \citep[$n < 1$,][]{Patterson_1940,Graham_2013}.

For each object in Sample I and Sample II we first measure the point-spread function (PSF) by using a non-saturated star in the image. A cutout of these stars is used as the PSF image in the GALFIT fitting procedure. We then fit either single S\'ersic or a combination of two S\'ersic profiles to the yellow-band images of our targets. The only exception is SDSS~J0149-0048, where we use a combination of one S\'ersic component and an edge-on disk component, which describes the surface brightness profile well. In contrast to the mostly empirically derived S\'ersic profile, the edge-on disk profile assumes that the disk component is locally isothermal and self-gravitating \citep[for details see][]{Kruit_1981}. GALFIT then convolves the models with the PSF to take into account the seeing and performs minimization based on the Levenberg-Marquardt algorithm. The output includes the integrated magnitude of the target, the effective radius $r_e$, the S\'ersic index $n$, the axis ratio and the position angle. For the purposes of the analysis in this paper, we are mostly interested in the S\'ersic parameter $n$, which indicates if the galaxy is disk-dominated (i.e. $n \leq 1$) or bulge-dominated (i.e. $n \geq 1$) allowing us to make conclusions about the evolutionary state of the galaxy (i.e., late-type vs. early-type). 

All except two galaxies in both samples are dominated by a bulge-like component (i.e. $n_{\rm{pri}} >1$) and if a second component was necessary in the fit, this second, fainter component is disk-like (i.e. $n_{\rm{sec}} < 1$). The only exceptions are SDSS~J0149-0048, which is best fit by a primary edge-on disk component and a fainter disk-like S\'ersic component and SDSS~J0759+1339 where both S\'ersic components are disk-like ($n_{\rm{pri}}$ and $n_{\rm{sec}} < 1$). 

\subsubsection{Morphological disturbances}
\begin{table*}
\caption{Host galaxy properties: results of 2D surface brightness fitting to the yellow-band images using GALFIT and results of spectral energy distribution fitting}
\begin{center}
\begin{tabular}{l C{1.5cm} c c C{4cm} c c}
\hline\hline
Object Name & Component Type & $n_{\rm{pri}}$ & $n_{\rm{sec}}$ & Disturbance Signs & log(M$_{\rm{stellar}}$/M$_{\sun}$) &SFR in M$_{\sun}/$year \\
(1) & (2) & (3) & (4) & (5) & (6) & (7)\\
\hline
SDSS~J0149-0048   & ED/S       & NA        & 0.7         & --          & 10.8 & 66             \\
SDSS~J0210-1001   & S/S        & 1.1      & 0.3         & --      & 10.2  & --                  \\
SDSS~J0319-0019  & S/S          & 2         & 0.15           & -- &  11.3& 13                     \\
SDSS~J0319-0058  & S          & 1.7       & --           & --   &10.6    & --                     \\
SDSS~J0321+0016   & S/S        & 1.6       & --           & C  &11.2     & 19    \\
SDSS~J0759+1339   & S/S        & 0.8       & 0.4         & --       & 11.3&44                \\
SDSS~J0841+2042   & S          & 1.4       & --           & -- &10.9  &--                       \\
SDSS~J0842+3625  & S          & 1.5       & --           & TT+C  &10.1&15                   \\
SDSS~J0858+4417   & S/S        & 2.4       & 0.8         &  TT+C &10.6     &121\\
SDSS~J1039+4512   & S        & 1.4       & --         &  C   &10.6    & 37            \\
SDSS~J1040+4745   & S/S        & 2.6       & 2.2         & TT     &   10.7    & 39        \\
       &            &           &             &             & &           \\
SDSS~J0123+0044   & S/S        & 2.6       & 0.3         &  large-scale TT   &10.7  &--                  \\
SDSS~J0920+4531   & S          & 3.6       & --           & -- & 11.0 & --                       \\
SDSS~J1039+6430   & S          & 4.1       & --           & -- & 10.7 & --                        \\
SDSS~J1106+0357   & S          & 14.4      & --           & TT   &  10.6  &5                   \\
SDSS~J1243-0232   & S/S        & 4.5       & 0.1         & --   & 10.7  & --                    \\
SDSS~J1301-0058   & S          & 1.5       & --          &  C  & 10.3 & --           \\
SDSS~J1323-0159   & S          & 2.7       & --           & --     &  10.3 &  8                   \\
SDSS~J1413-0142   & S          & 2.2       & --           & --    &   10.5 &13                    \\
SDSS~J2358-0009   & S/S        & 14.6      & 0.5         & TT+C &10.6& --      \\
\hline                
\end{tabular}
\end{center}
\label{galfit}
\begin{tablenotes}
\item \textit{Notes:} (1) object name (2) component types used for GALFIT surface brightness fitting. S denotes single S\'ersic profile and SS and ED/S double S\'ersic and Edge-on Disk + S\'ersic profiles, respectively. (3) S\'ersic index of the primary, i.e. brighter, component (4) S\'ersic index of the secondary, i.e. fainter, component (5) Signs for deviation from a smooth potential in the yellow-band image with C meaning companion galaxy and TT meaning tidal tail (6) stellar mass derived using CIGALE (7) star formation rates derived using DecompIR, the typical uncertainty is 0.3 dex.
        \end{tablenotes}
\end{table*}
The processes that trigger quasars remain an unresolved issue and are not yet well understood. Quasars could be triggered by major or minor mergers \citep[e.g.][]{Silk_1998, Hopkins_2008}, close passages of galaxies that alter the gravitational potential \cite[e.g.][]{Hopkins_2008} or secular interactions within disk galaxies auch as disk instabilities or turbulence \citep{Crenshaw_2003, Orban_2011}. Our quasars are not in disks, so disk instabilities are unlikely. We investigate the presence of ongoing merger activity and disturbances by subtracting the smooth best-fit surface brightness profile derived by GALFIT from the yellow-band HST images to identify the presence of tidal tails, dust lanes and general deviations from a smooth potential. We present a summary of our results in Tab. \ref{galfit}. 

\citet{Zakamska_2006} already noticed that SDSS~J2358-0009 shows a large tidal tail and is probably interacting with a companion galaxy at a projected distance of $\sim 29$~kpc. Three more sources in Sample II show faint tidal debris or faint companions \citep{Zakamska_2006}. Out of the sources newly observed with the HST (Sample I), two (SDSS~J0842+3625, SDSS~J1040+4745) show prominent tidal tails. The tail SDSS~J0842+3625 is extending to a companion galaxy at a projected distance of 17~kpc with which it is most likely interacting. In SDSS~J1040+4745 we detect a double nucleus with a separation of 2.5~kpc, probably currently undergoing a merger. In addition to these two galaxies, we find faint diffuse emission in the yellow-band image for SDSS~J0858+4417, at 4.5~kpc distance from the nominal position of the galaxy (see residual images in Figure \ref{stamps}). 
Two additional sources (SDSS~J0321+0016 and SDSS~J1039+4512) have accompanying galaxies at 6.5 and 9.5~kpc, respectively, but neither of the quasar host galaxies shows prominent morphological disturbances and it is unclear if these galaxies are interacting or are by-chance projections. 

Additionally, in two sources (SDSS~J0210-1001 and SDSS~J0319-0019) we detect faint shells at projected distances of $\sim 8-9$~kpc in the blue HST images with no counterparts in the yellow-band images. We speculate that these shells might be old merger signatures that are now illuminated by the quasar. For consistency and further comparison with other works, we do not report these shells in Table \ref{galfit}, nor do we include them in our statistics.

This means that 9 out of 20 galaxies in our samples ($45 \pm 10$\%) show clear signs of merger/disturbance activity with 5 galaxies having close-by companions and six showing signs of recent merger activity with apparent tidal tails.

\subsection{Spectral Energy Distributions}

Since our targets are by design type-2 quasars where the central black hole is hidden behind obscuring material in the form of a torus or a smoother dust distribution, the optical emission of our quasars is due to host galaxy, emission lines and scattered light.

Although the light of the quasar cannot propagate directly toward the observer because of large amounts of circumnuclear obscuration, it can escape along less obscured directions. It can then scatter off the interstellar medium in the host galaxy and reach the observer \citep[see e.g.][]{Zakamska_2006}. Contribution to the UV from scattered light can be as high as 75\% \citep{Obied_2015}.

Additional contribution from the quasar can occur in the mid- and far-IR due to quasar-heated dust. The dust torus surrounding the central quasar reprocesses a large fraction of the direct quasar emission (X-rays, UV and optical) which is re-emitted as thermal dust emission in the mid-IR. This part of the SED therefore has to be decomposed carefully into re-processed dust emission from the quasar as well as re-processed emission from star formation in the host galaxy \citep{Drouart_2014, Kirkpatrick_2014}. 


We are making use of the Python package CIGALE \citep[Code Investigating GALaxy Emission,][]{Noll_2009} to perform SED fitting and reconstruct properties of both the host galaxies and their quasars. CIGALE builds a library of models 
for both the stellar (i.e. host galaxy) and quasar components assuming a assuming stellar population models \citep{Bruzual_2003, Maraston_2005}, star formation histories (SFH), dust templates \citep{Dale_2002, Dale_2014, Draine_2007, Casey_2012} and quasar models from \citet{Fritz_2006}. In CIGALE, the energy balance is always fulfilled.


The parameters chosen for the fitting are presented in Tab. \ref{CIGALE} and are based on previous work and experience with fitting intermediate redshift quasars with CIGALE \citep[][and references therein]{Ciesla_2015}. 
For the quasar models, we fix the parameters $\beta$, $\gamma$ and $\theta$ which parametrize the dust distribution within the torus. We choose typical values found in \citet{Fritz_2006} and by fixing these parameters we also avoid degeneracies in the model templates. 

In Figure \ref{SEDs} we present the multi-wavelength imaging data of the galaxies in our samples and the best-fitting SEDs as derived by CIGALE. For the analysis of this paper, we are mostly interested in the derived stellar masses for our quasar host galaxies, their quasar contribution to the far-IR emission and their star formation rates. 

The restframe optical and near-IR emission is well described by the models  and stellar masses have a typical uncertainty of only about 10\%. The galaxies span a mass range of $10^{10}-10^{11.3}$ M$_{\sun}$ (Table \ref{galfit}).

As Figure \ref{SEDs} shows, the dust models and theoretical quasar models from \citet{Fritz_2006} used in the CIGALE fits cannot well reproduce the hot dust emission at mid-IR and far-IR wavelengths. In particular, the mid-IR emission ($\sim 4 - 30 \mu$m) dominated by hot dust heated by the quasar, is underestimated. In fact, when fitting only the data at $\lambda<3.4\mu$m, i.e. the rest-frame optical and near-IR, the reduced $\chi^2$ of the fits declines by 20\%. Therefore, CIGALE's mean estimate of frac$_{\rm{AGN}} = 0.7\pm0.2$, the fraction of the infrared luminosity $L_{\rm{IR}}$ that is due to the quasar (or active galactic nucleus, AGN), should be considered as a lower limit which shows that the IR emission is most likely significantly dominated by reprocessed quasar light rather than star formation. We do not report star formation rates derived by CIGALE because the far-IR fits are poor.

\subsection{Modeling the IR emission}

Since the \citet{Fritz_2006} models provide an insufficient description of the far-IR emission, we attempt to describe the mid- and far-IR empirically, using the SED fitting procedure DecompIR \citep{Mullaney_2011} and single and two-temperature component modified blackbodies. The key distinction between quasar-heated and SF-heated dust is that the former is more compact and therefore heated to higher temperatures than the latter.

\begin{figure*}
\begin{center}
\includegraphics[trim=1.65cm 0 0 0cm, clip = true, scale=0.245]{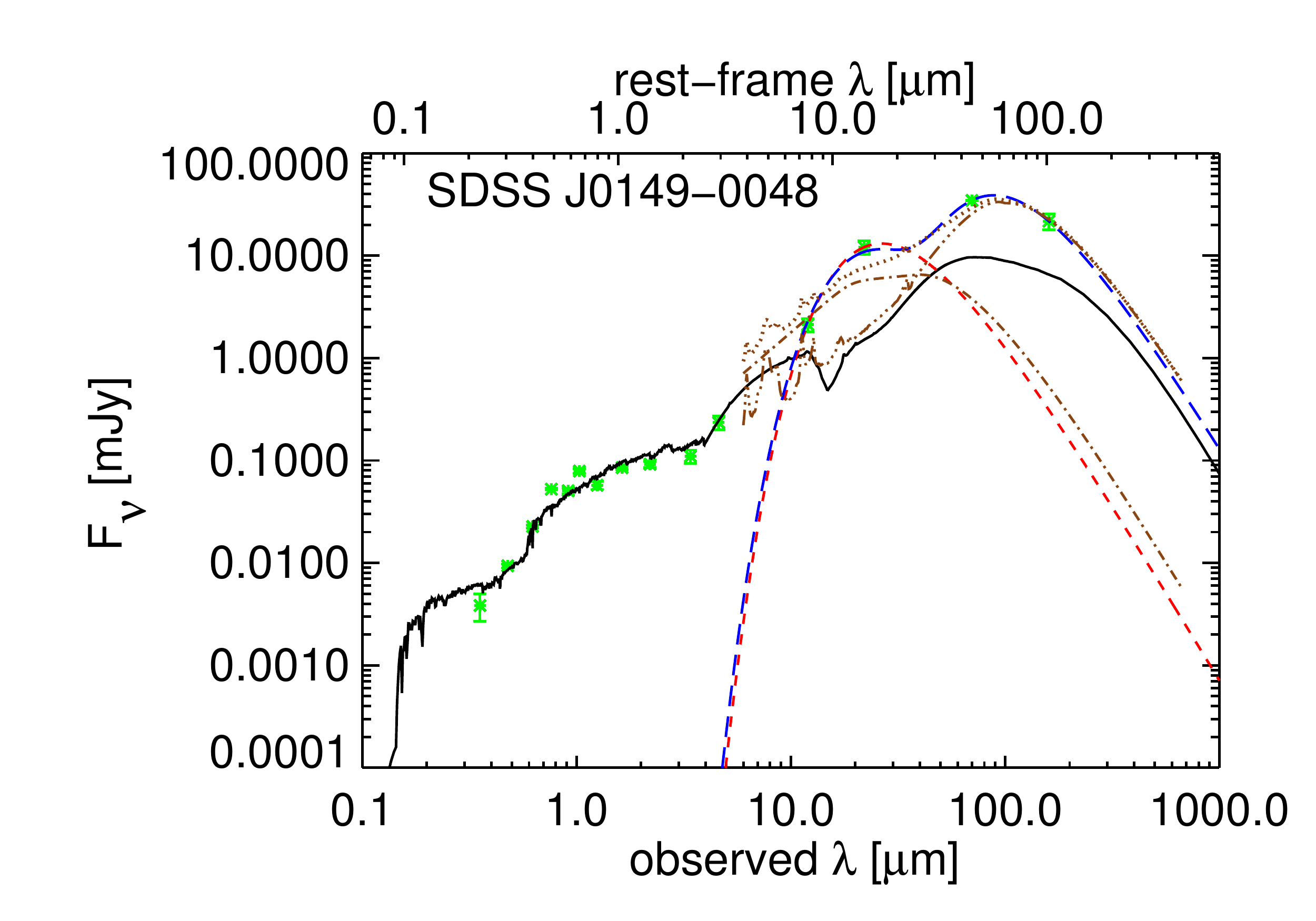}
\includegraphics[trim=1.65cm 0 0 0cm, clip = true, scale=0.245]{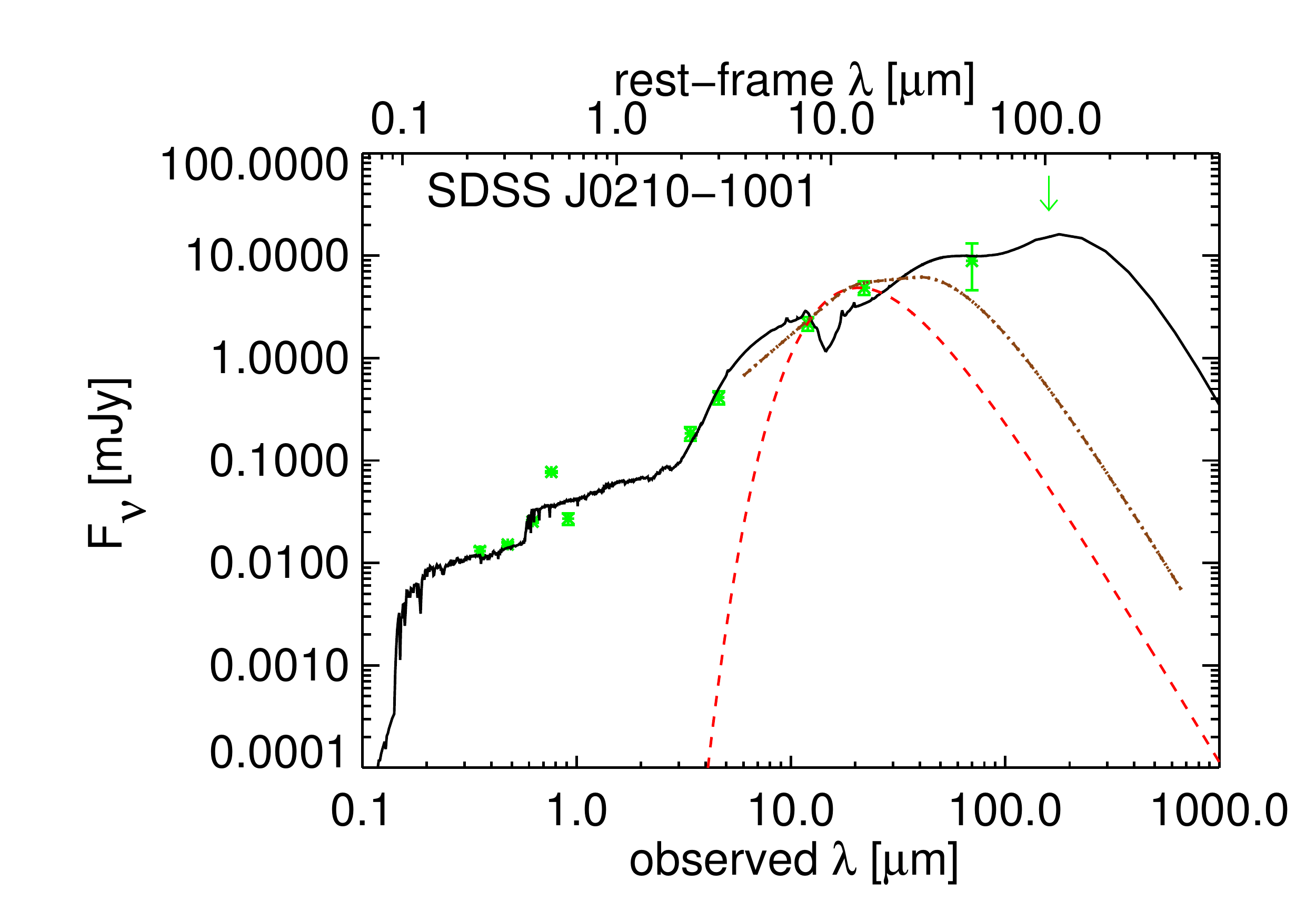}
\includegraphics[trim=1.65cm 0 0 0cm, clip = true, scale=0.245]{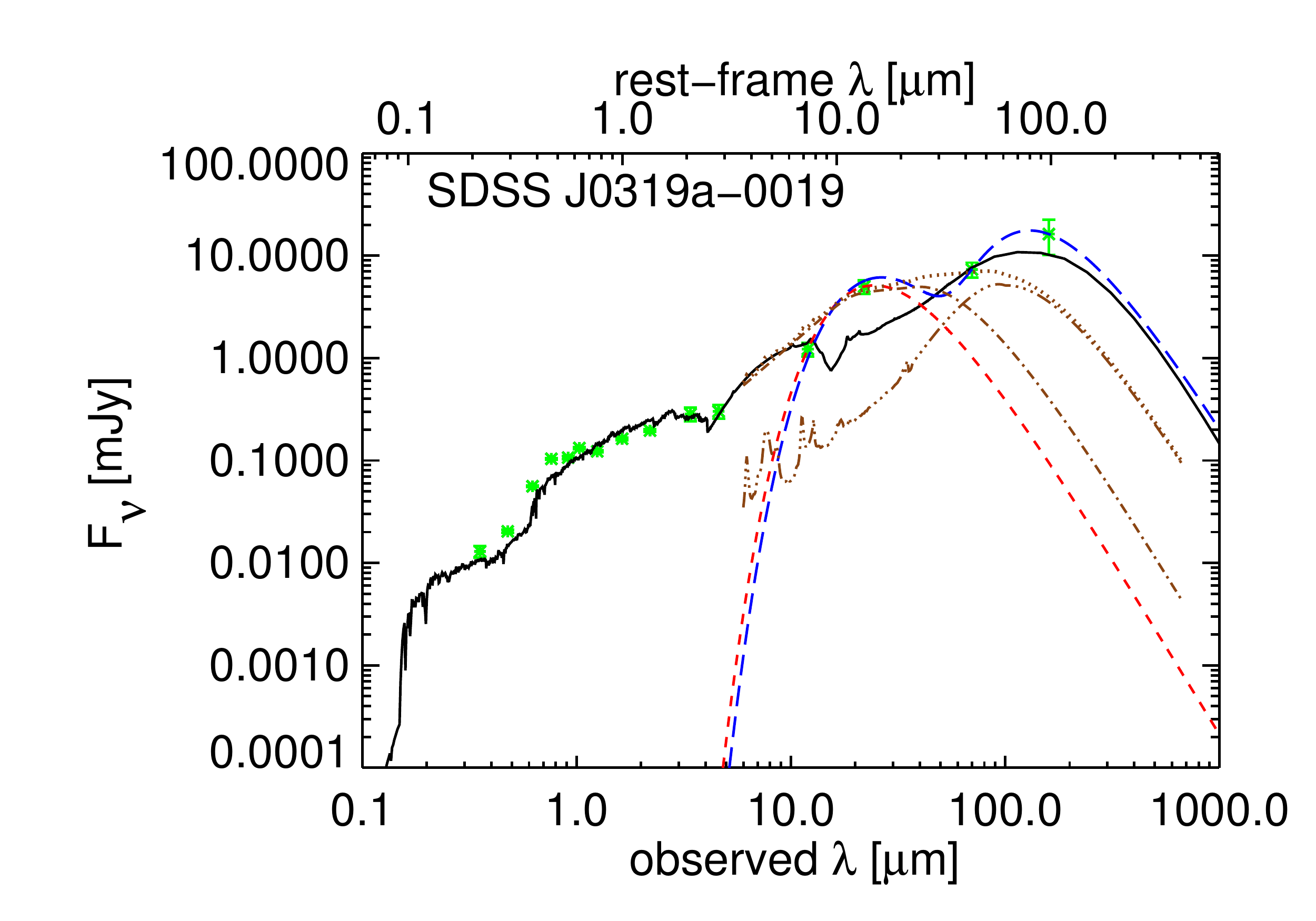}
\includegraphics[trim=1.65cm 0 0 0cm, clip = true, scale=0.245]{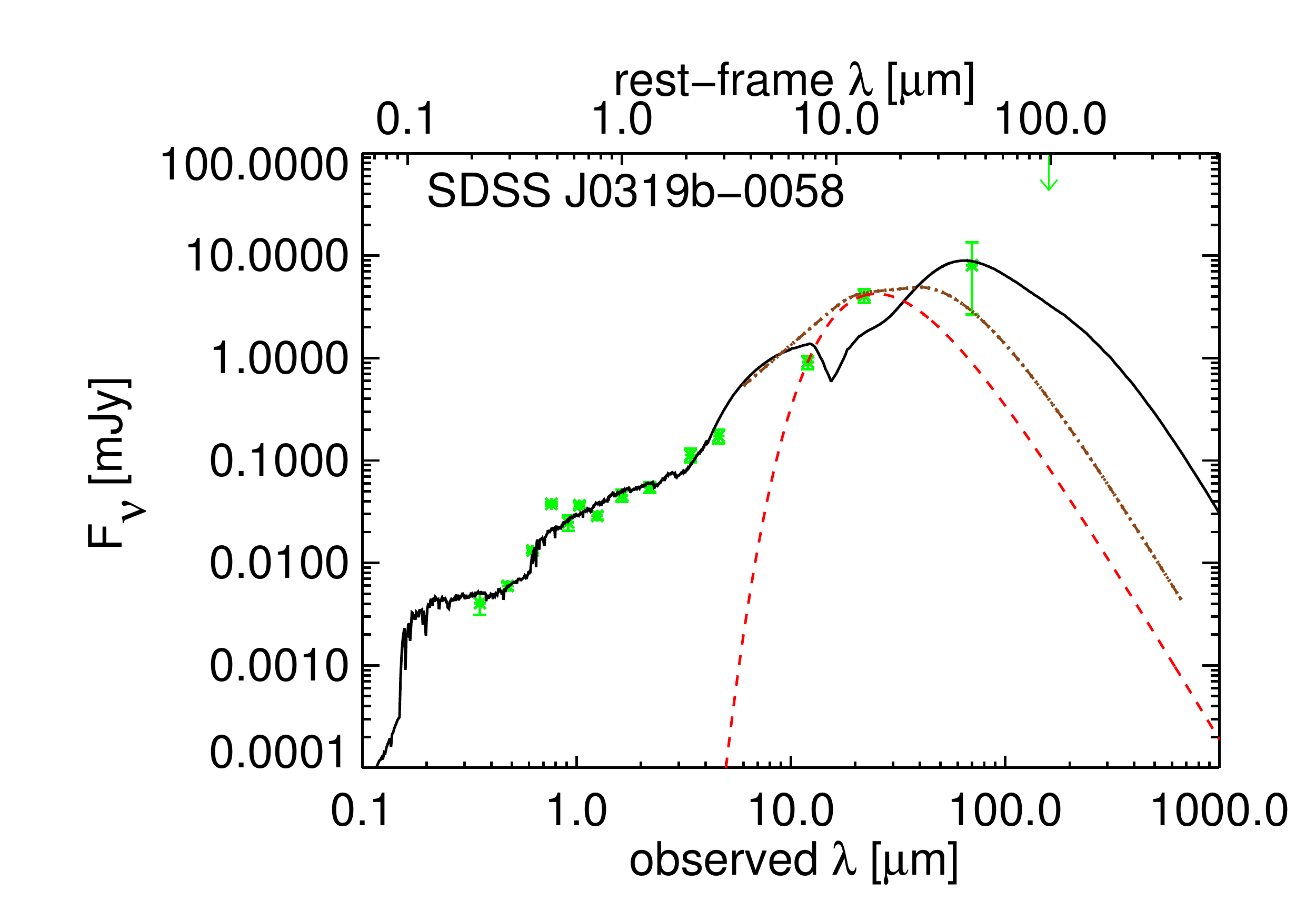}
\includegraphics[trim=1.65cm 0 0 0cm, clip = true, scale=0.245]{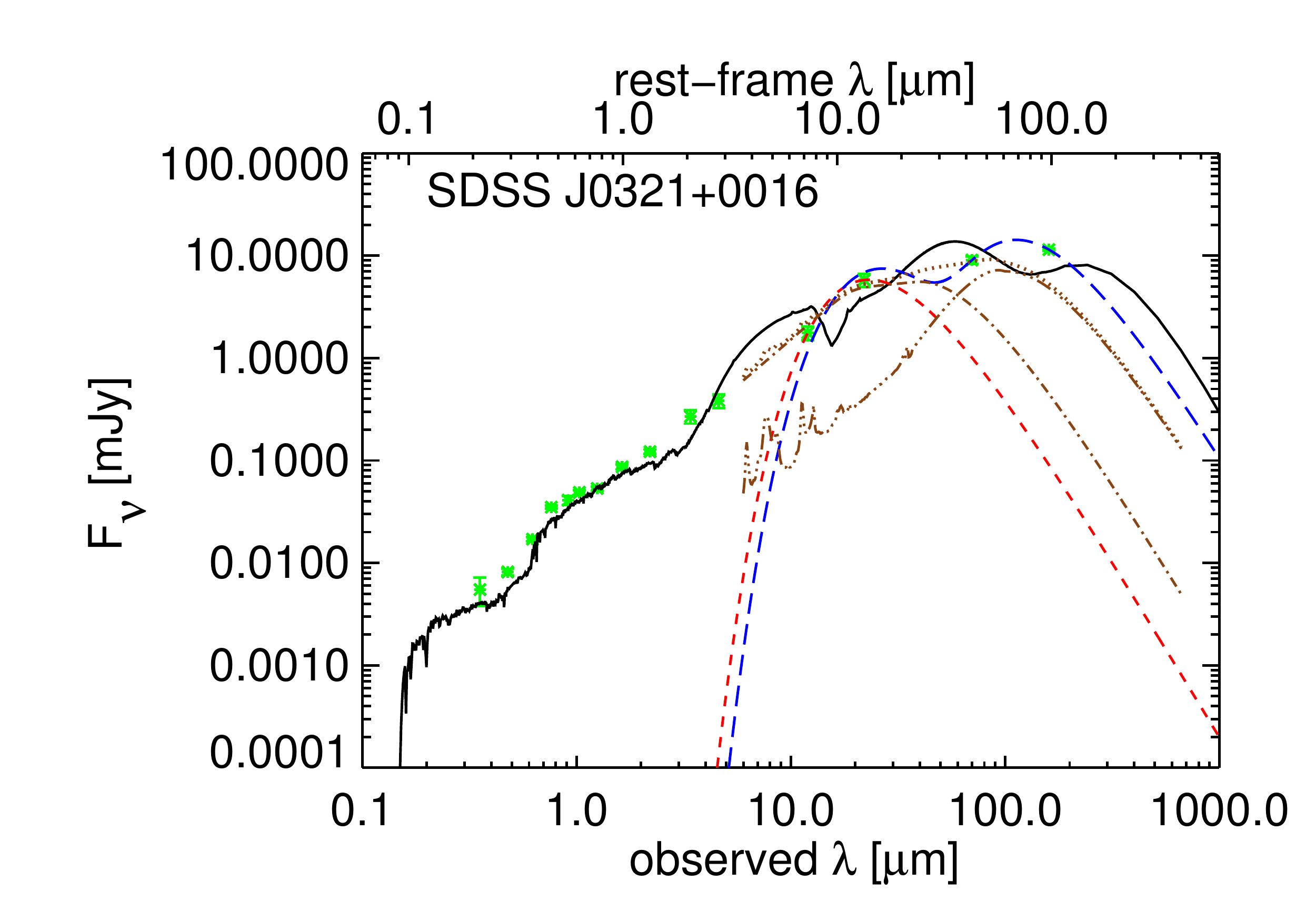}
\includegraphics[trim=1.65cm 0 0 0cm, clip = true, scale=0.245]{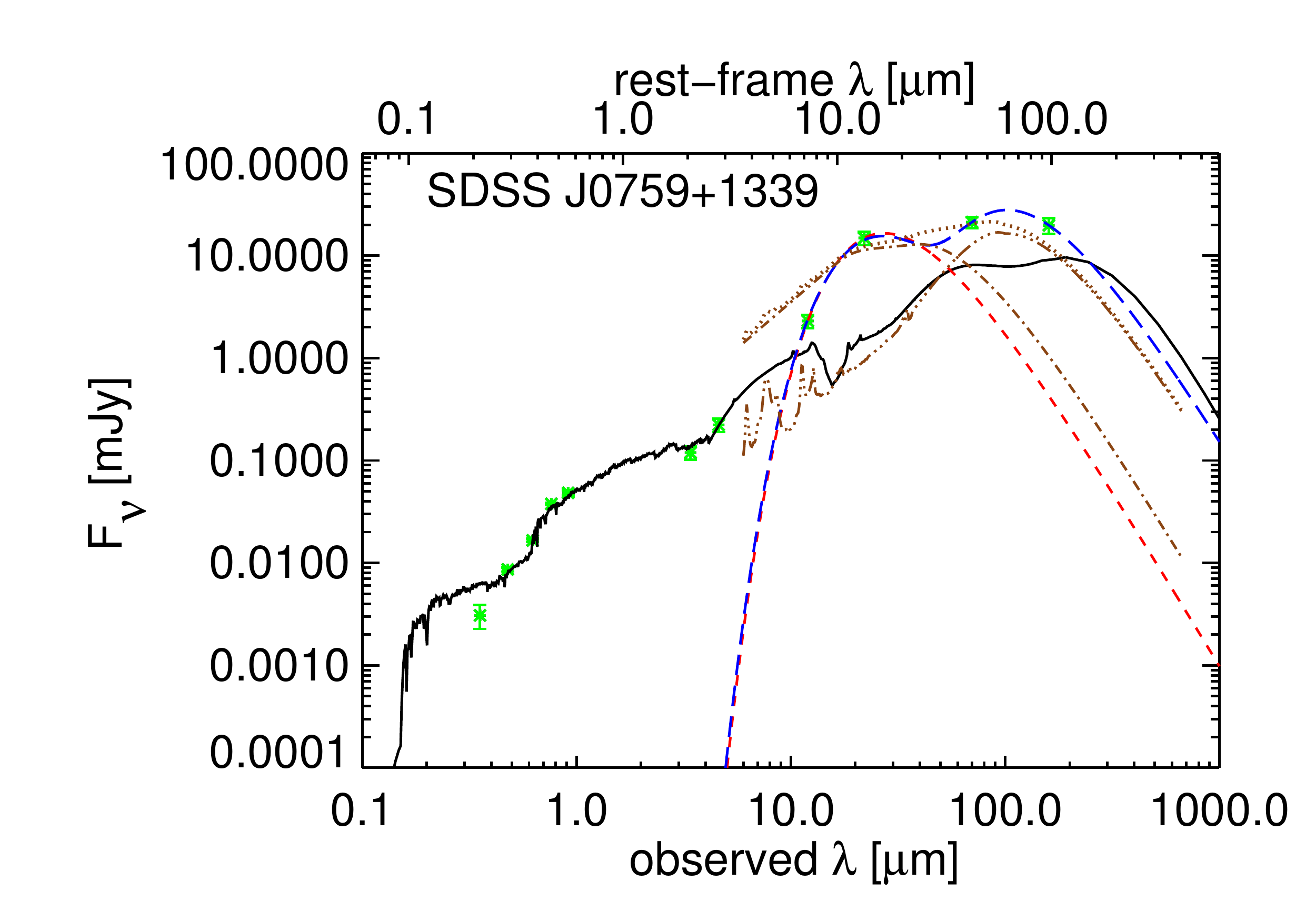}
\includegraphics[trim=1.65cm 0 0 0cm, clip = true, scale=0.245]{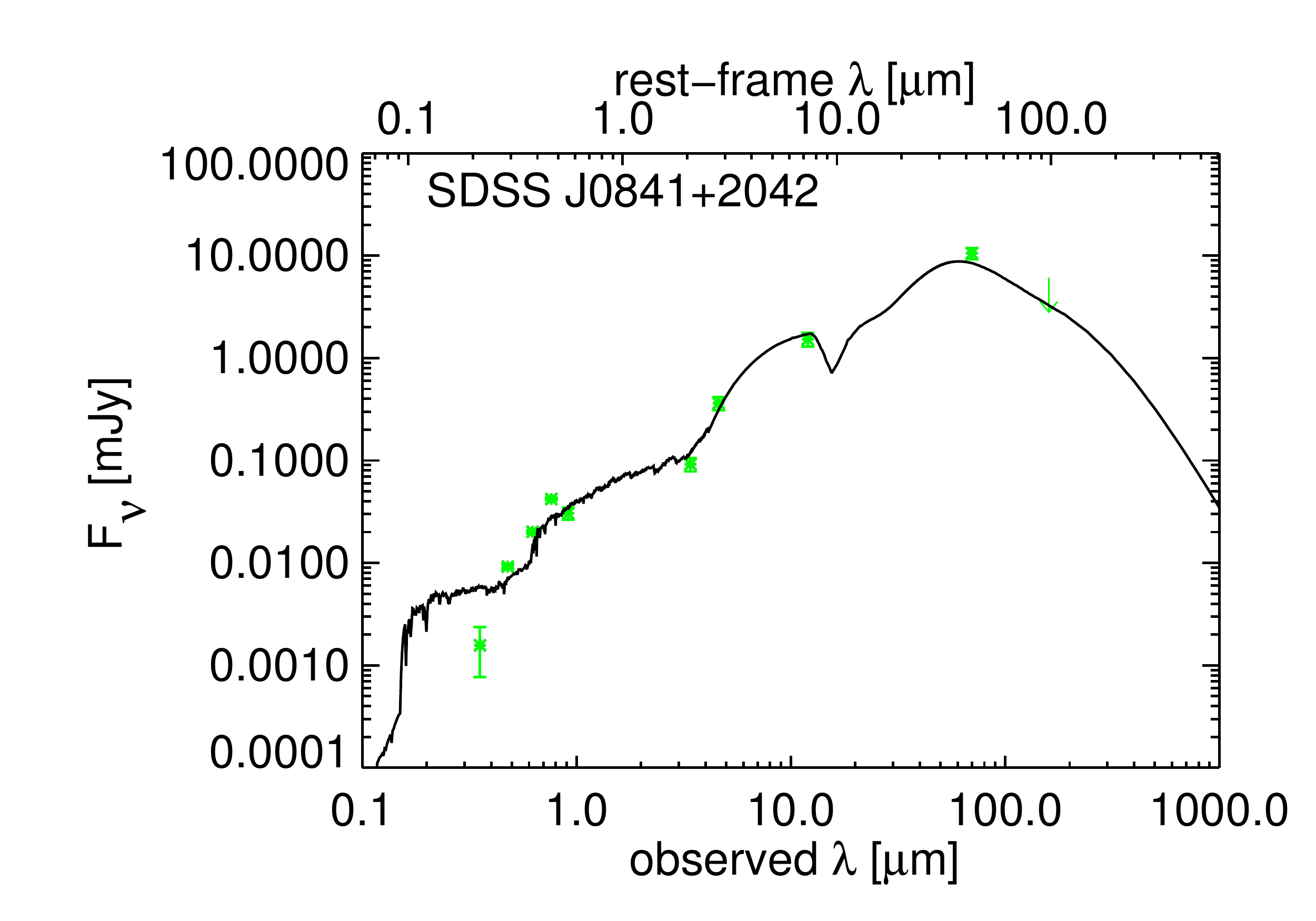}
\includegraphics[trim=1.65cm 0 0 0cm, clip = true, scale=0.245]{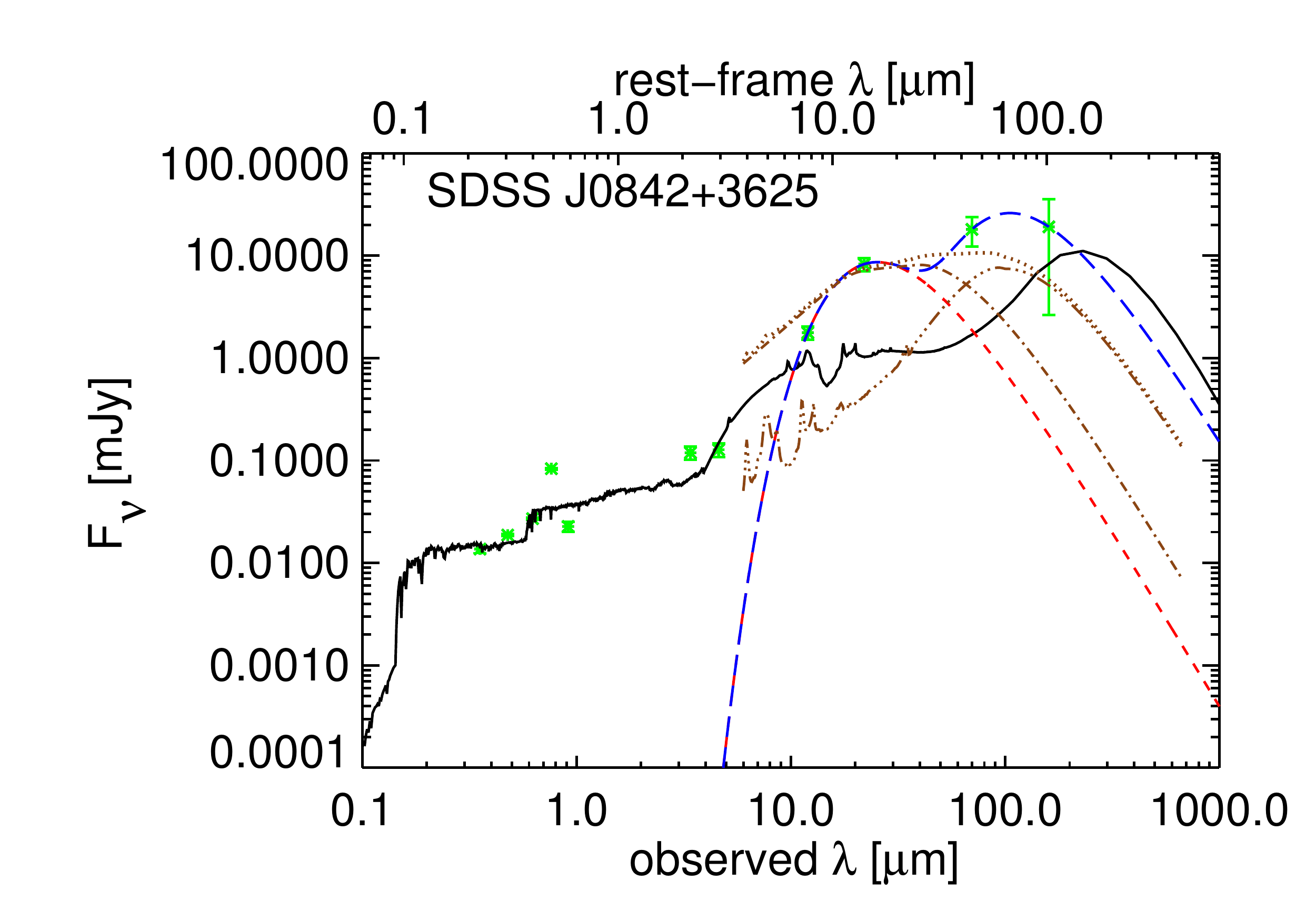}
\includegraphics[trim=1.65cm 0 0 0cm, clip = true, scale=0.245]{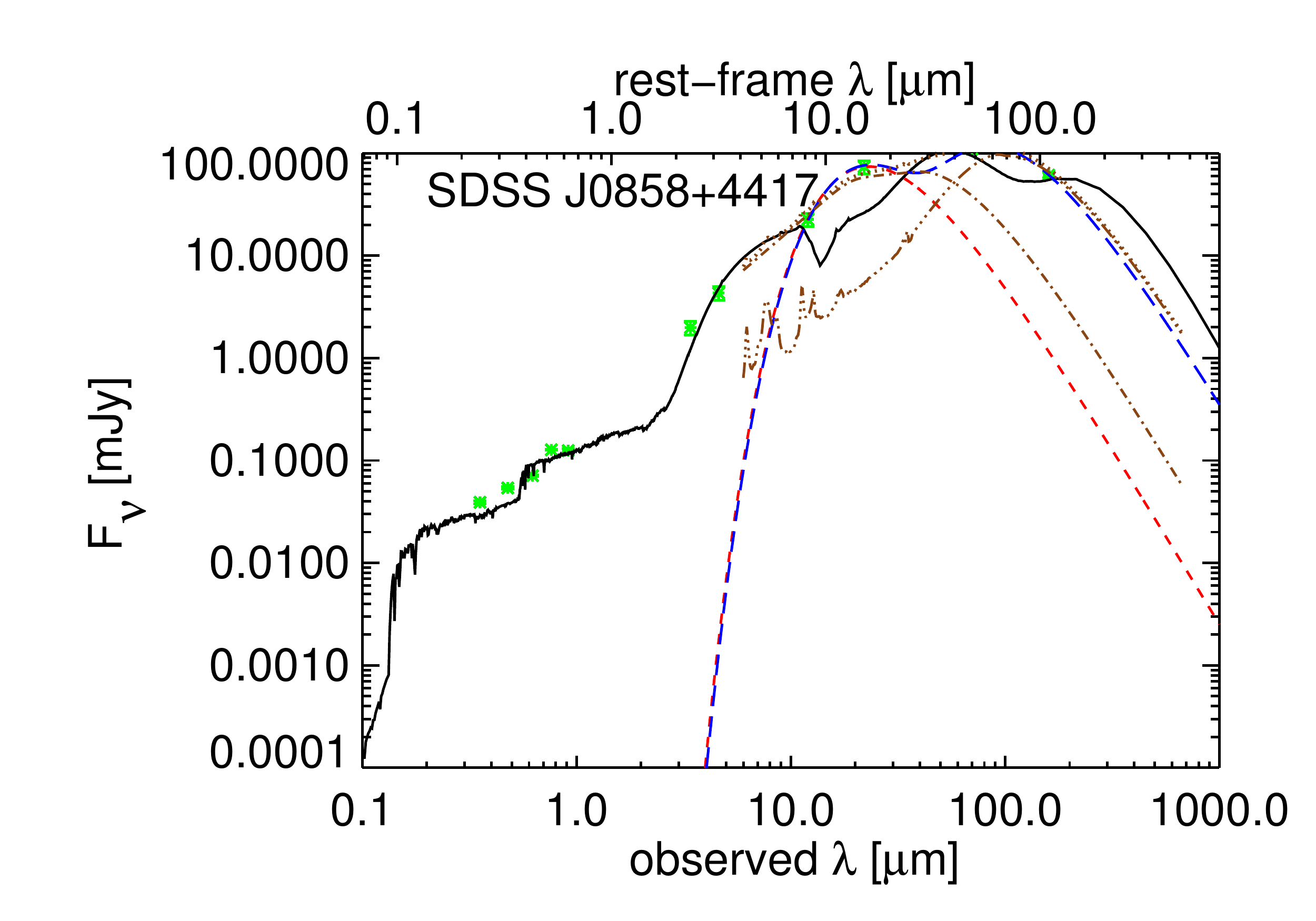}
\includegraphics[trim=1.65cm 0 0 0cm, clip = true, scale=0.245]{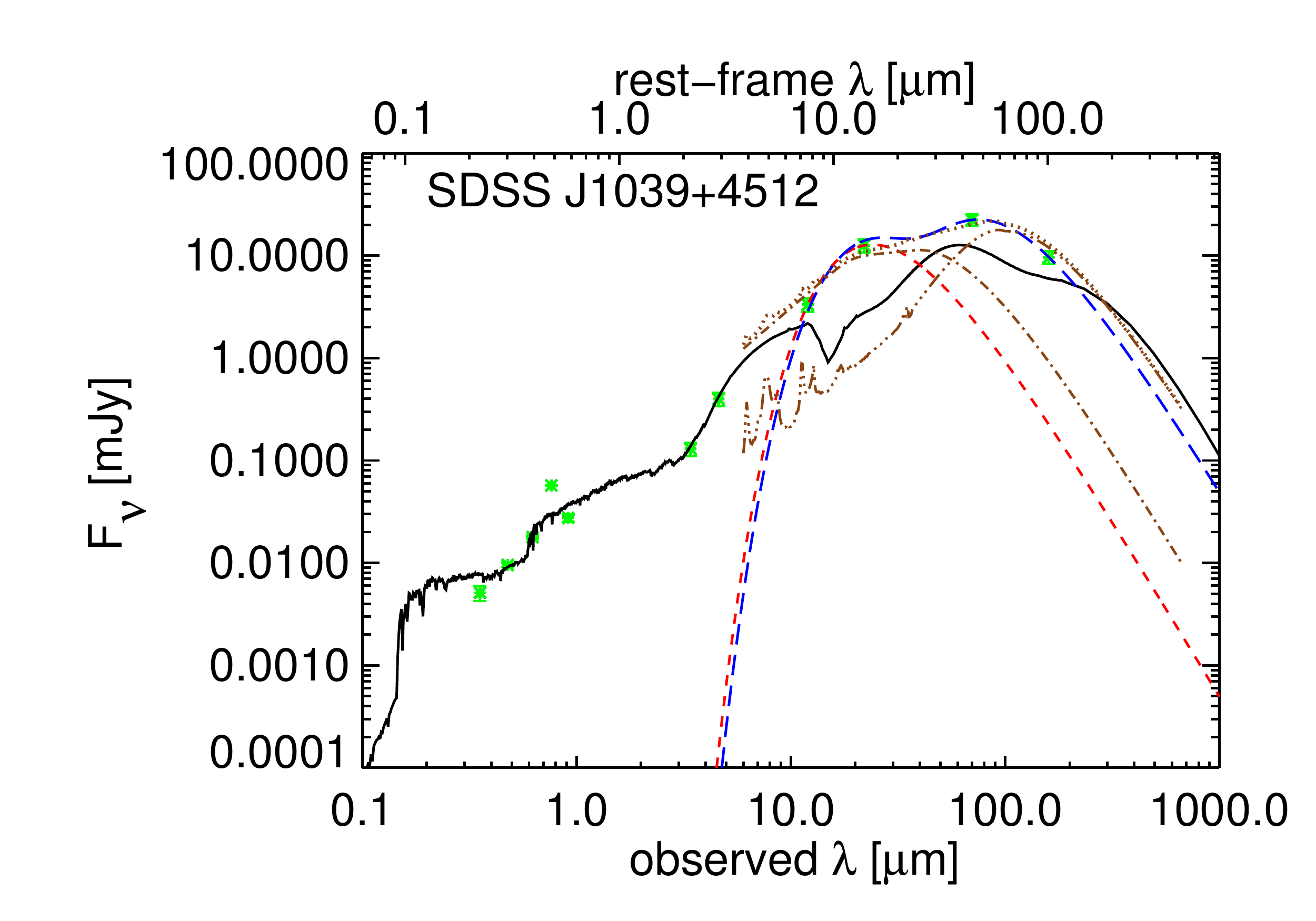}
\includegraphics[trim=1.65cm 0 0 0cm, clip = true, scale=0.245]{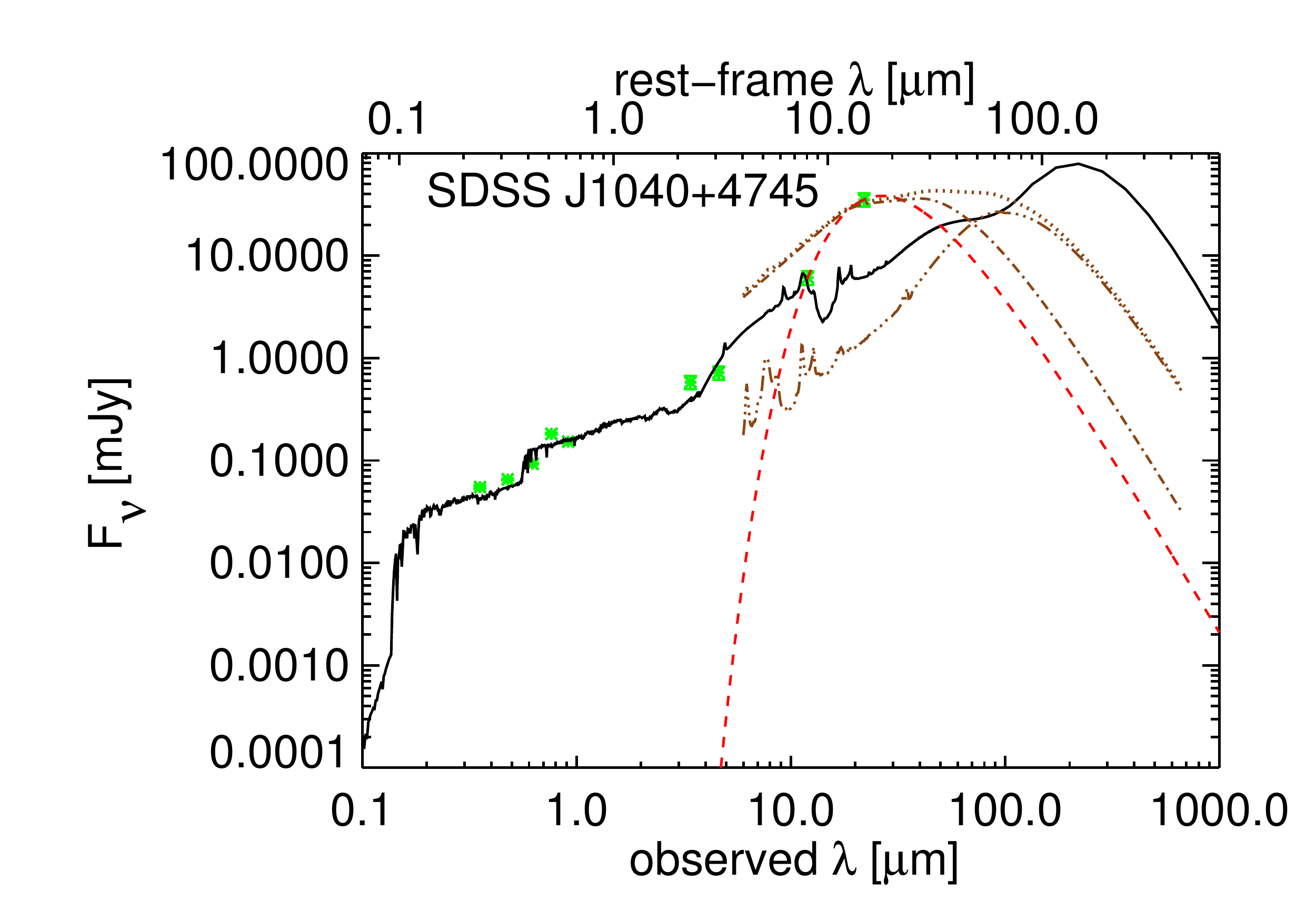}
\includegraphics[trim=1.65cm 0 0 0cm, clip = true, scale=0.245]{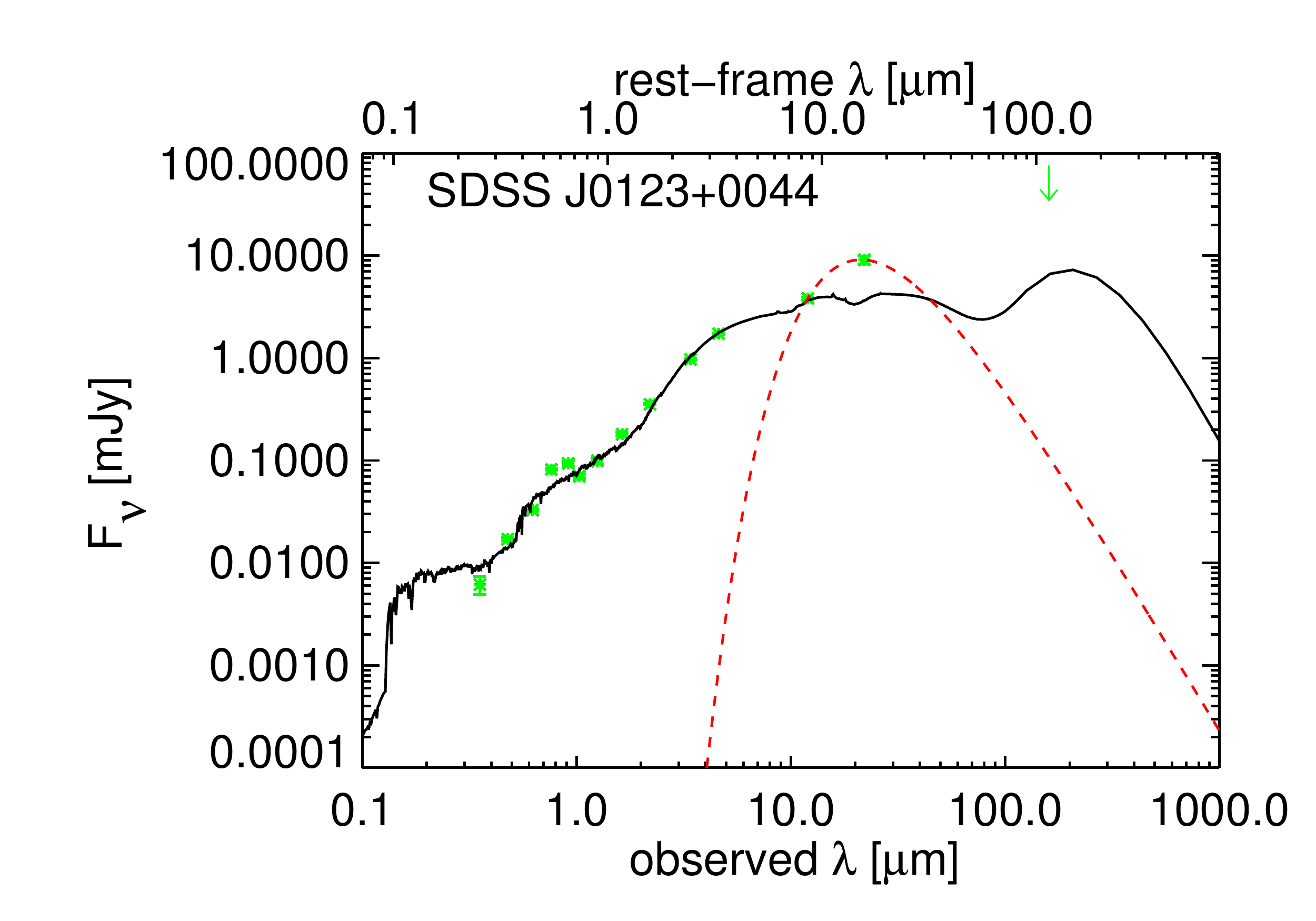}
\includegraphics[trim=1.65cm 0 0 0cm, clip = true, scale=0.245]{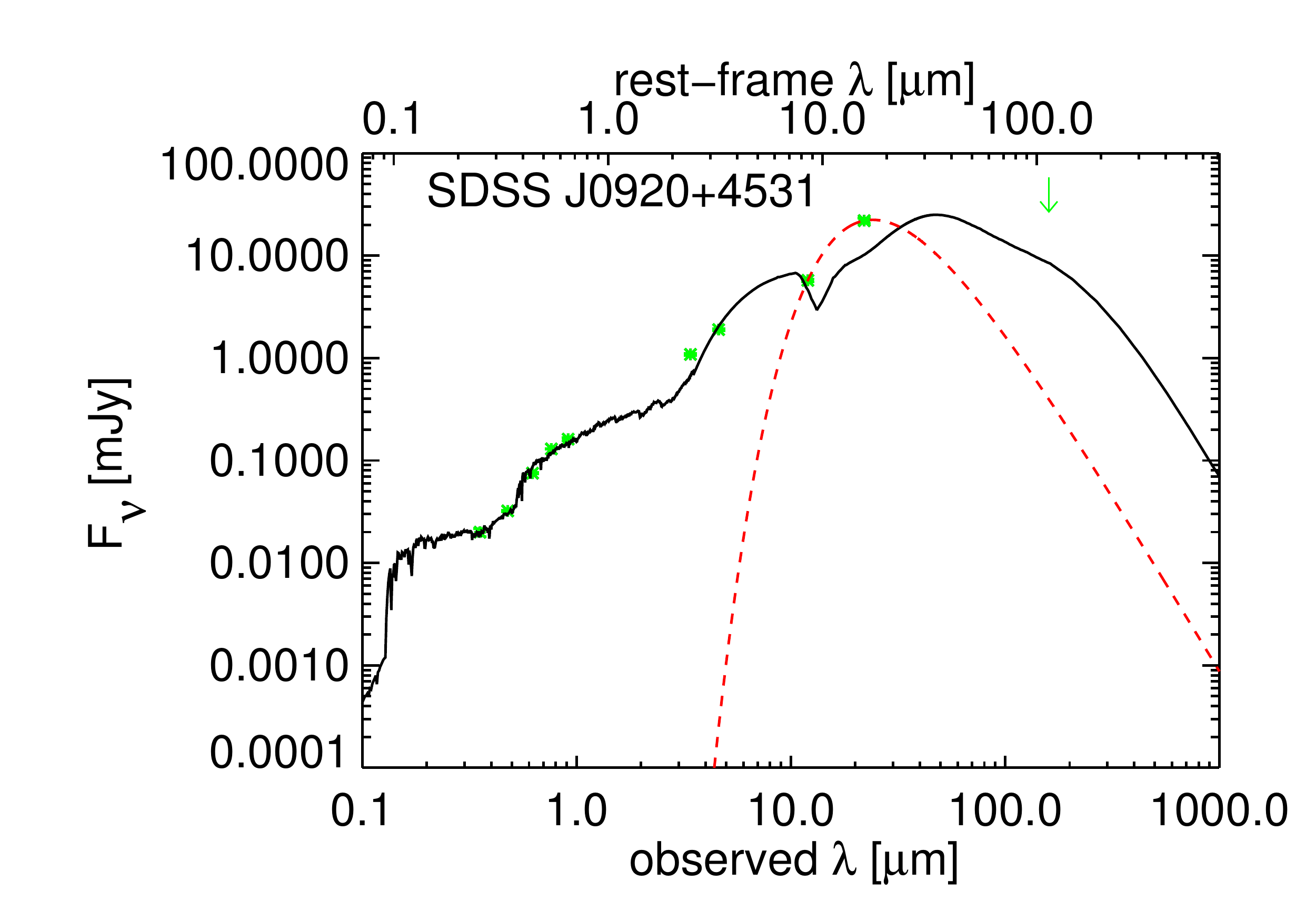}
\includegraphics[trim=1.65cm 0 0 0cm, clip = true, scale=0.245]{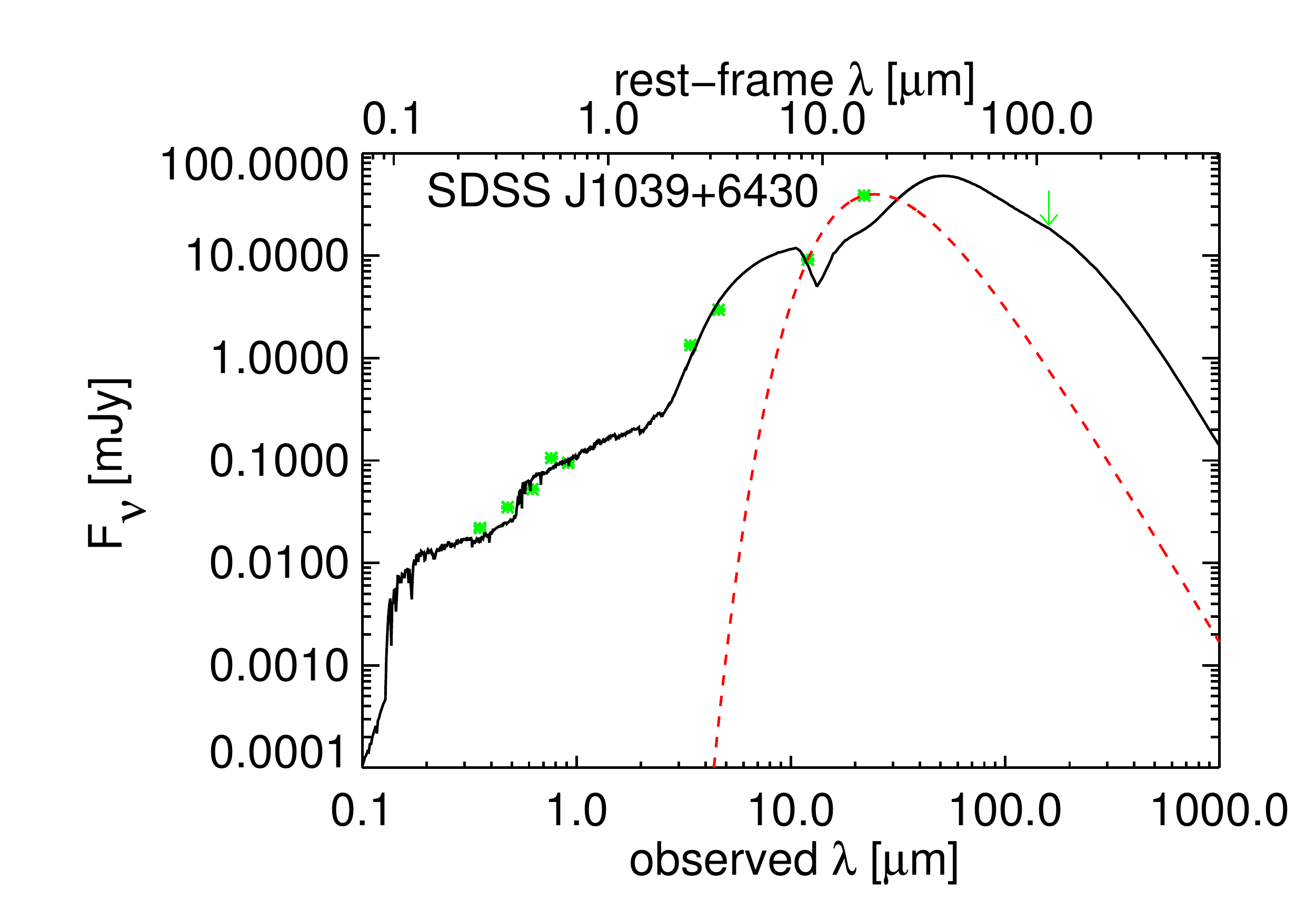}
\includegraphics[trim=1.65cm 0 0 0cm, clip = true, scale=0.245]{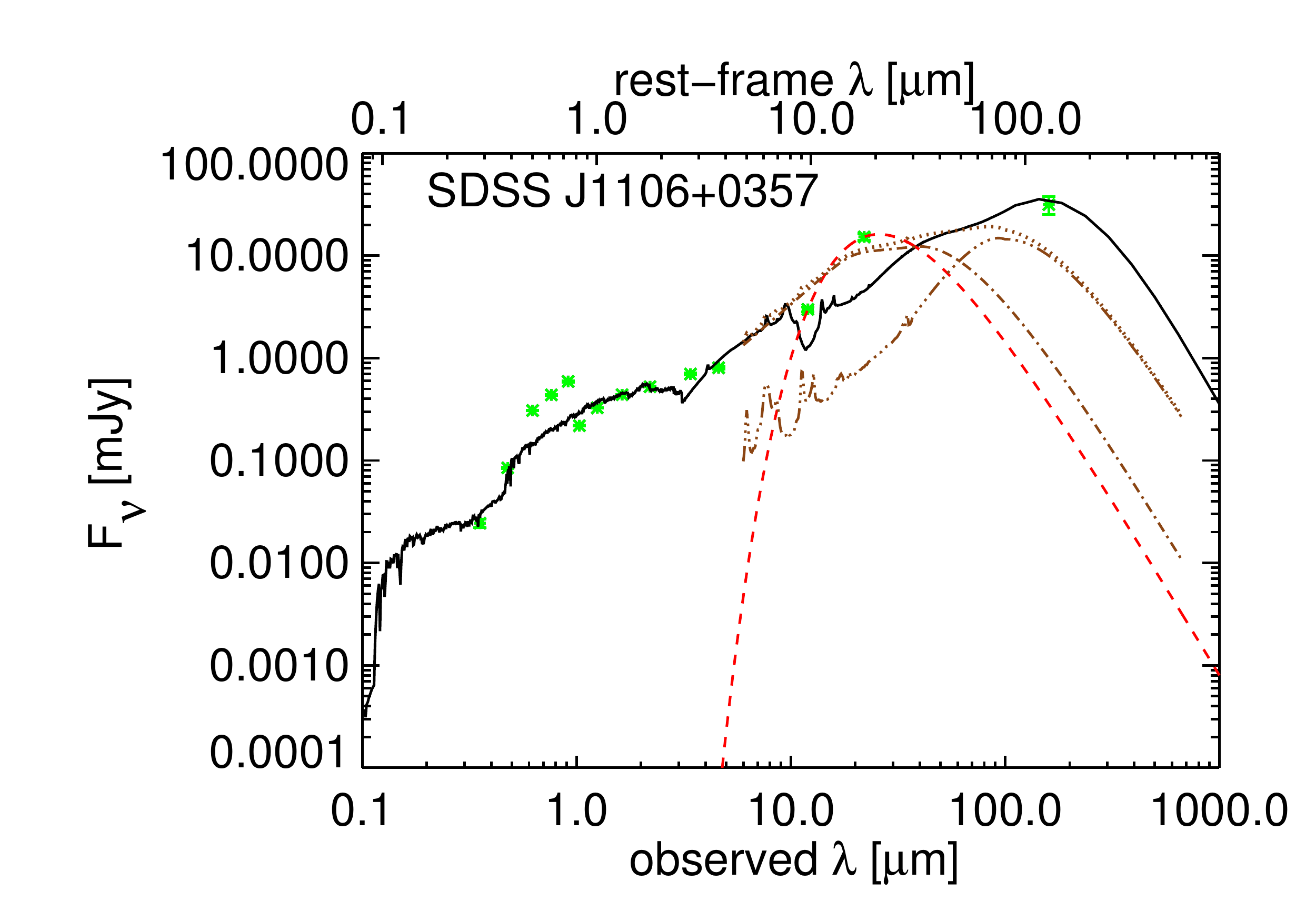}
\caption{Spectral energy distributions of the sources in our samples. Green asterisks show the broad-band photometry points as measured by SDSS, UKIDSS, \textit{WISE} and \textit{Herschel}. The black solid line shows the best fit model using the CIGALE fitting routine. The red and blue dashed and long dashed lines show the single modified and double modified black body fits to the far-IR data, respectively. The brown lines show results of the template fitting of the far-IR data using DecompIR, where the dashed-dotted brown line shows the quasar component, the dashed-double-dotted line shows the SF component and the dotted line shows the total far-IR emission.}
\label{SEDs}
\end{center}
\end{figure*}

DecompIR uses $\chi^2$ minimization to constrain the relative contribution of quasar emission and star-formation to the mid-far-IR emission in galaxies. It uses an empirical library derived from local starbursts and an empirical quasar template described by broken power-laws and a modified black body. Although this quasar template is empirically derived and is expected to show discrepancies on an object-to-object basis due to large variety of quasar SED's, it represents well the average mid-IR SED of quasars \citep{Dale_2014}. Due to limited IR coverage of some of our sources (i.e. less than 5 photometry points between rest-frame $6-1000\mu$m), we can only fit 13 out of our 20 sources using DecompIR (see Figure \ref{SEDs}). These empirical templates generally capture the far-IR emission better than CIGALE but fail at reproducing the steep rise with $\lambda$ at mid-IR wavelengths like in source SDSS~J0759+1339. These SED shapes are typical for very hot quasar-heated dust in heavily obscured type-2 quasars \citep{Assef_2015, Tsai_2015}. The average AGN fraction, i.e. the fraction of the infrared luminosity $L_{\rm{IR}}$ that is due to the quasar, is $frac_{\rm{AGN}} = 0.8 \pm 0.15$, consistent with the mid- and far-IR bolometric luminosity being almost entirely dominated by the quasar and with values derived using CIGALE. 

Assuming that star formation is not contributing significantly to the IR bolometric luminosity, we further attempt to describe the dust bump using a single- and two-temperature component modified blackbody. A blackbody fit is usually the most simplistic far-IR SED fit. Accounting for the galaxies' dust not being perfectly non-reflective and variations in opacity, the flux density should be modeled as a modified blackbody of the form 
\begin{equation}
S(\nu,T) \propto \frac{(1-e^{-\tau(\nu)})\nu^3}{e^{h\nu/kT}-1}
\end{equation}  
where $S(\nu,T)$ is the flux density at $\nu$ for a given temperature $T$,  $\tau(\nu)$ is the optical depth and is commonly represented as $\tau(\nu) = (\nu/\nu_{0})^\beta$, where $\beta$ is the spectral emissivity index and $\nu_{0}$ is the frequency where optical depth equals unity. In the optically thin case, appropriate for far-IR emission, this expression reduces to 
 \begin{equation}
S(\nu,T) \propto \frac{\nu^{\beta}\nu^3}{e^{h\nu/kT}-1}.
\end{equation}  
The spectral emissivity index $\beta$ ranges typically between $1-2$ and is often assumed to be 1.5 \citep{Hildebrand_1983, Dunne_2001}. 

For the one component fit, we assume $\beta = 1.5$ and fit a modified blackbody to the mid-IR data between $4\mu$m$<\lambda<50\mu$m. We find a mean temperature of $T=191K$. Such high temperatures are usually attributed to dust being heated by the quasar rather than by star formation \citep{Kirkpatrick_2014}. This single modified blackbody fit fails at reproducing the emission at far-IR wavelengths and we conclude that multiple dust components are needed to fully describe the data. We therefore also fit a sum of two modified blackbodies to the data at $\lambda > 4 \mu$m. We fix $\beta_1$ and $\beta_2$ to 1.5 and assume $T_1 = 200$K to reduce the number of free parameters. Since the far-IR hump in some of our sources is not as well covered, we can only fit this two-component model to 7 out of 20 sources. The mean temperature of the colder component is found to be $T_2 = 50$K. While this dust temperature could be reached by heating through star formation, it would be on the upper end of SF-heated dust temperatures \citep{Wylezalek_2013a, Drouart_2014} and represents likely a mix of contributions of colder components of quasar-heated dust and warmer components of star-formation heated dust. A two-component modified blackbody fit is only an approximation the dust emission in the far-IR where a range of temperatures contributes. 

Following the standard definition for the IR luminosity $L_{\rm{IR}}$ \citep{Kennicutt_1998}, we integrate $F_{\nu}$ over the wavelength range $8-1000\mu$m rest-frame to derive $F_{\rm{IR}}$
and compute $L_{\rm{IR}}=4\pi D_{L}^{2} F_{\rm{IR}}$ where $D_{\rm{L}}$ is the luminosity distance. The IR luminosities derived using the three different methods agree with each other within a factor of 1.8 on average. This agreement might be surprising since the single-component modified blackbody fit does not capture the longer wavelength data at $\sim > 100\mu$m, but this is due to the fact that emission at longer wavelength, i.e. lower frequencies, does not contribute as much to the total energy output $\nu F_{\nu}$.

CIGALE, DecompIR and blackbody fits show that the IR bolometric output is consistent with being almost entirely dominated by quasar-heated dust emission, where only the longest wavelength flux measurements can be used to estimate star formation rates. 

We therefore report star formation rates derived from IR decomposition performed through template fitting with DecompIR, which is only possible if data at $160 \mu$m are available (Table \ref{galfit}). \citet{Zakamska_2015} recently estimated star formation rates based on the $160 \mu$m alone following \citet{Symeonidis_2008} for a larger sample of type-2 quasars which our sources were part of. Although the $160 \mu$m flux densities are likely dominated by star formation, any contribution from the quasar was not accounted for. Therefore, these measurements can be taken as upper limits on the actual star formation rate. The measurements performed in this paper using DecompIR agree with the upper limit measurements from \citet{Zakamska_2015} within a factor of two on average. Since the quasar-star formation decomposition is a challenging task considering the limited data coverage, we adopt this factor as the typical uncertainty on the star formation rates we report.




\addtocounter{figure}{-1}
\begin{figure*}
\begin{center}
\includegraphics[trim=1.65cm 0 0 0cm, clip = true, scale=0.245]{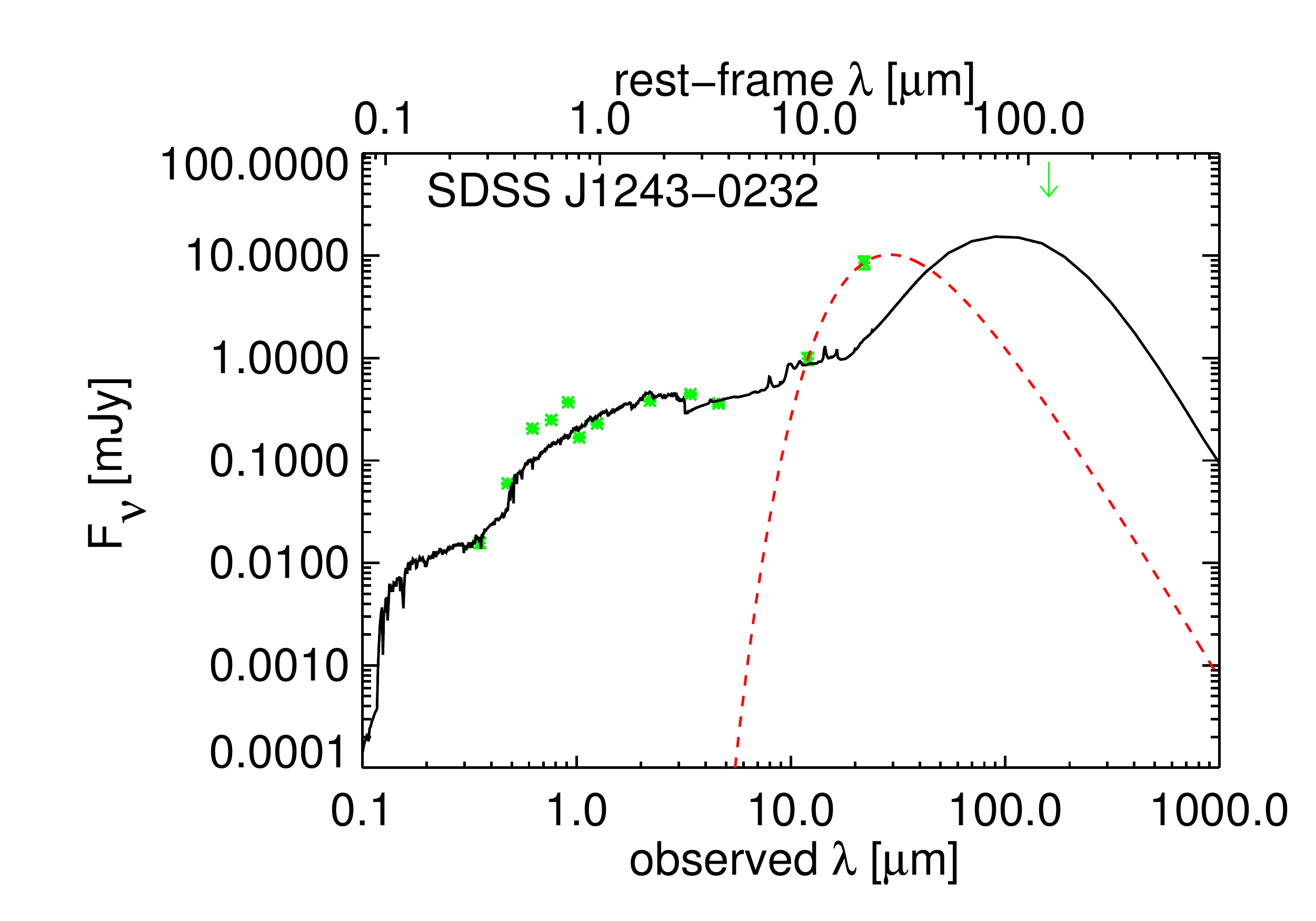}
\includegraphics[trim=1.65cm 0 0 0cm, clip = true, scale=0.245]{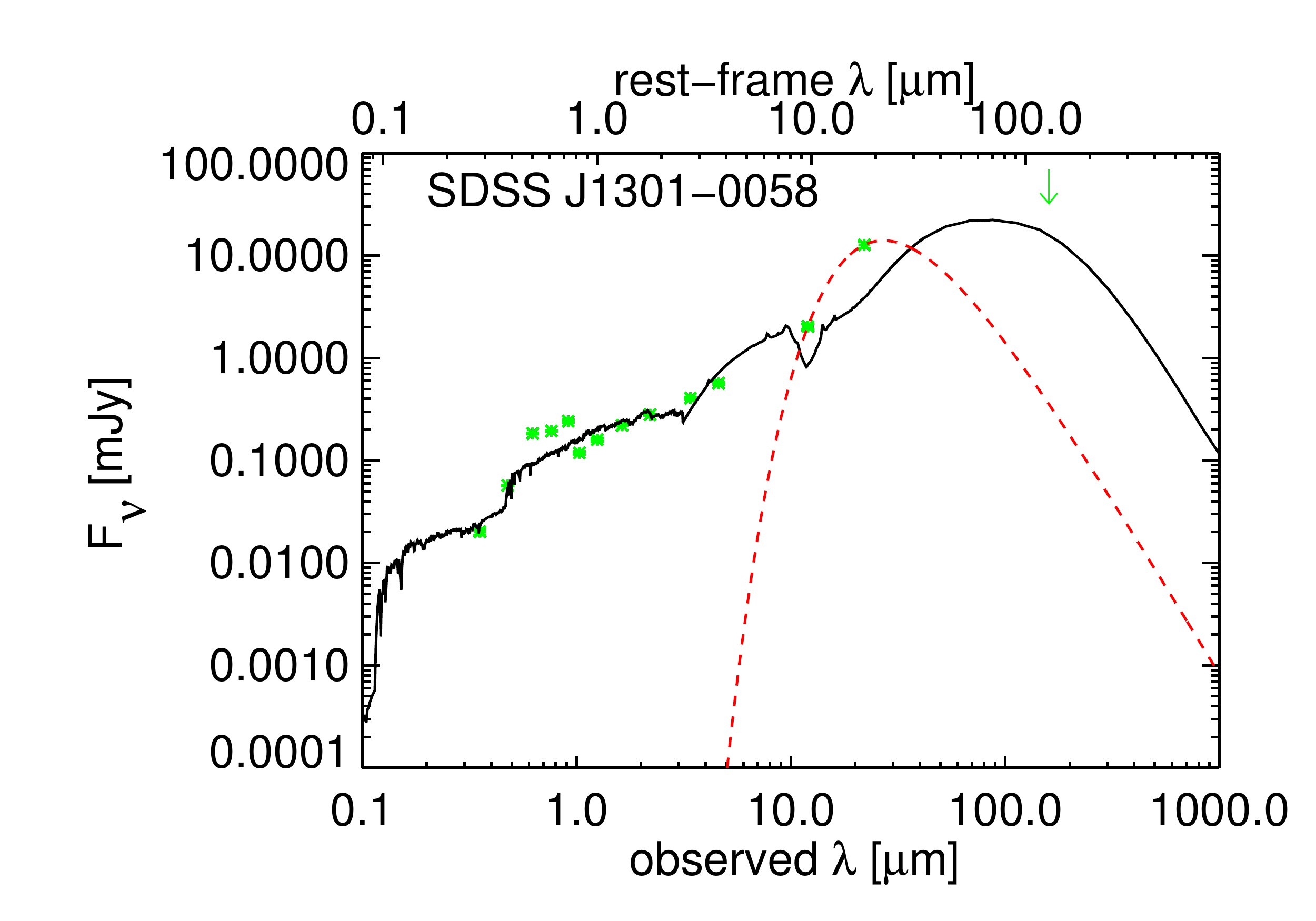}
\includegraphics[trim=1.65cm 0 0 0cm, clip = true, scale=0.245]{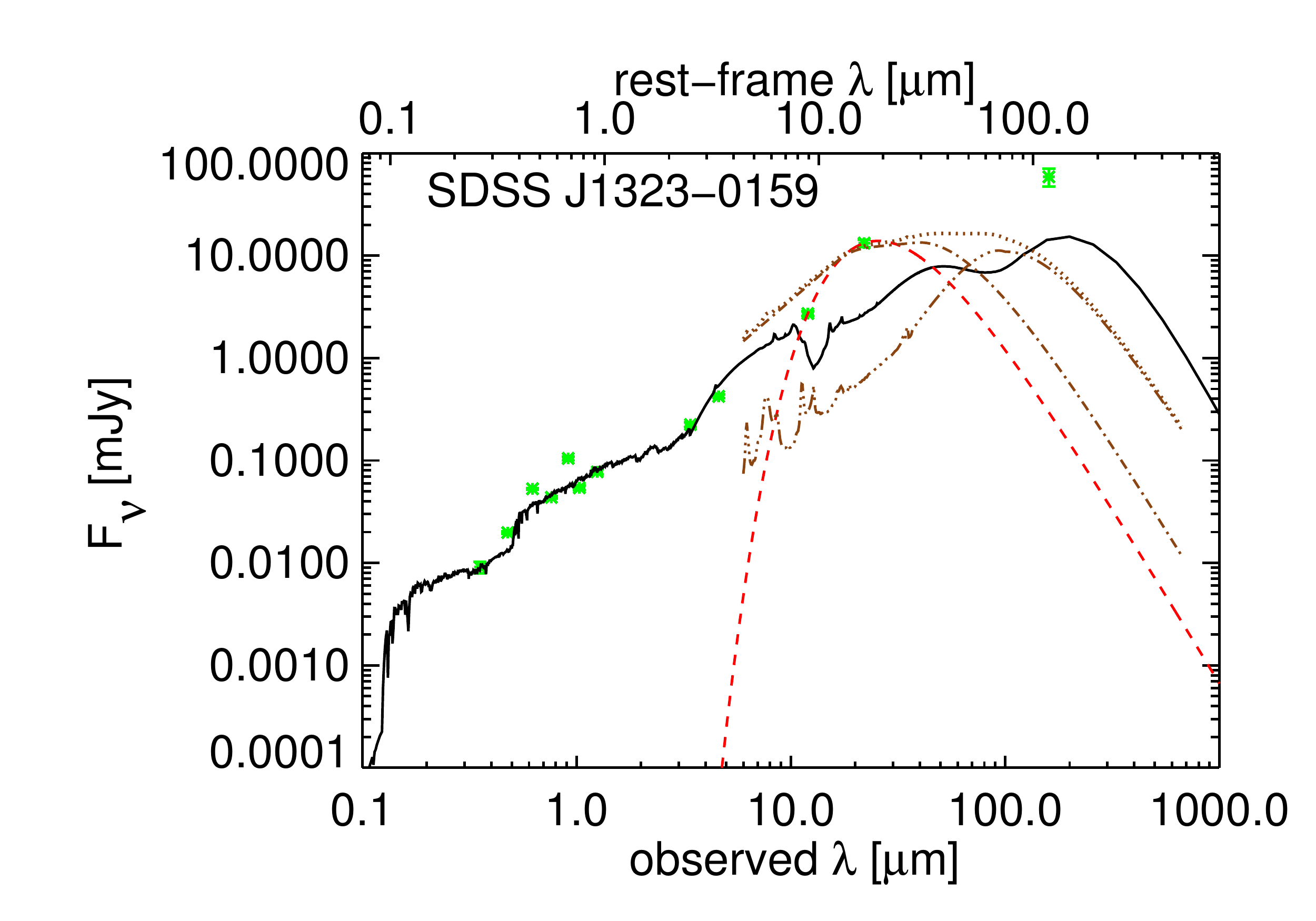}
\includegraphics[trim=1.65cm 0 0 0cm, clip = true, scale=0.245]{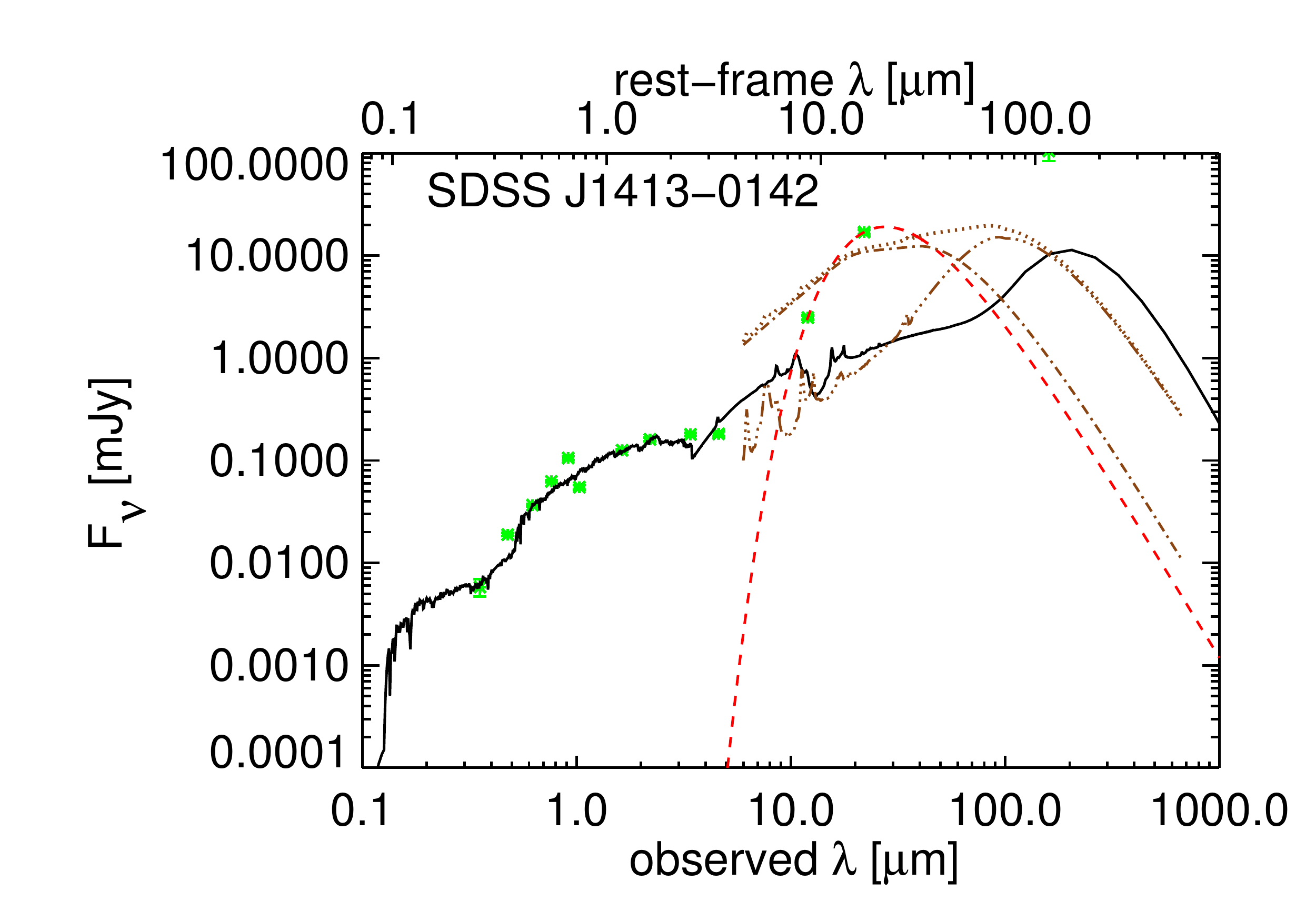}
\includegraphics[trim=1.65cm 0 0 0cm, clip = true, scale=0.245]{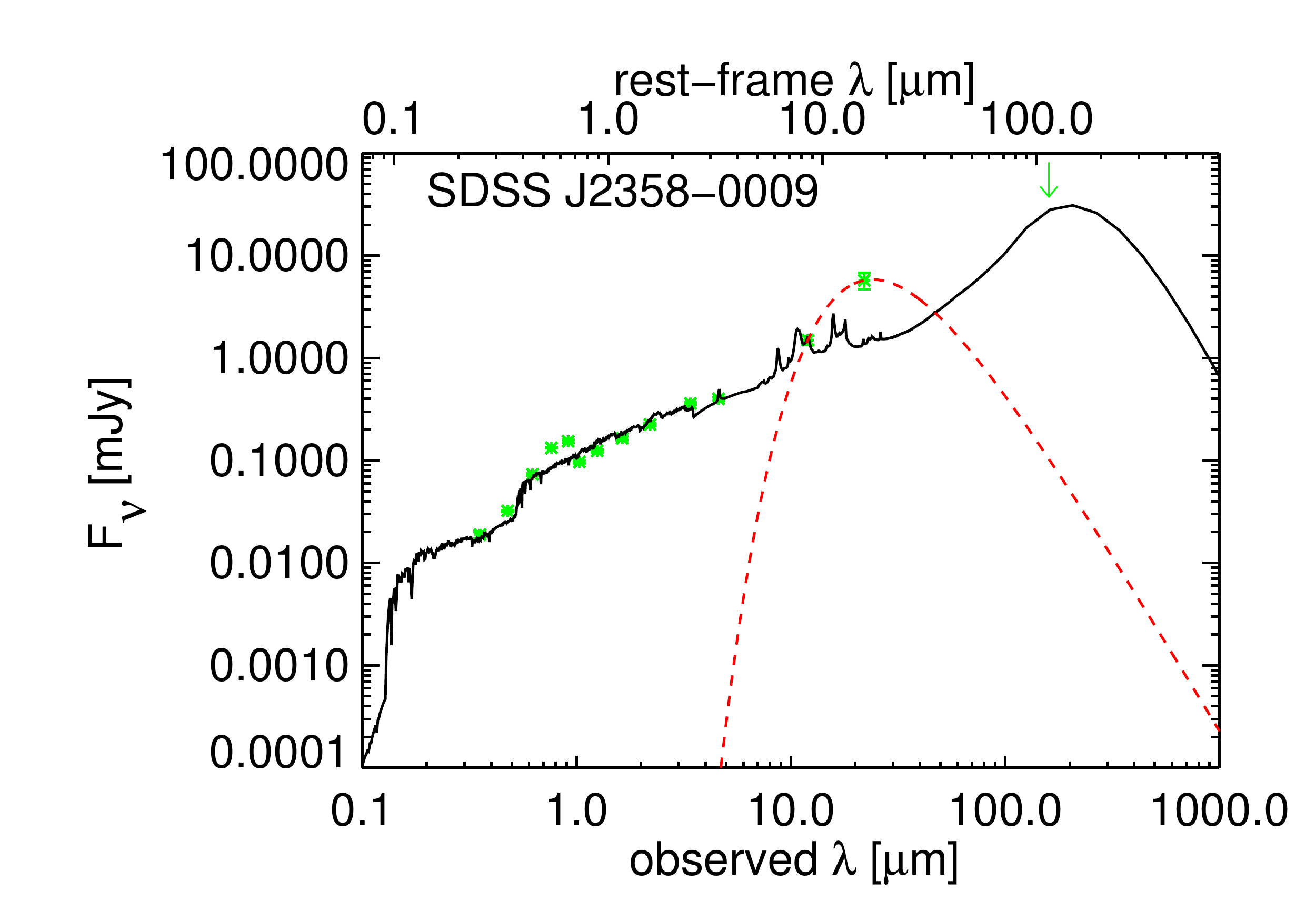}
\caption{Spectral energy distributions continued}
\end{center}
\end{figure*}

\begin{table*}
\caption{CIGALE parameters used for SED fitting.}
\begin{center}
\begin{tabular}{cc}
\hline\hline
Parameter & Values \\
\hline      
Star formation history (SFH) & delayed$^1$ \\
Metallicity & 0.02 \\
Interstellar Mass Function (IMF) & Salpeter \\
Stellar population models & \citet{Maraston_2005} \\
E-folding timescale of the main SPM (Gyr) & 1, 3, 5 \\
Age of the oldest stars in the galaxy (Gyr) & 1, 3, 5 \\
\hline
Dust attenutation & \citet{Calzetti_2000} \& \citet{Leitherer_2002} \\
Colour excess of stellar continuum light for the young population &  0.1, 0.2 \\
E(B$-$V)$_*$ reduction factor between old and young population & 0.44 \\
\hline
Dust template & \citet{Dale_2014} \\
IR power-law slope & 1.5, 2, 2.5 \\
\hline
Quasar emission & \citet{Fritz_2006} \\
Ratio of dust torus radii & 10, 60 \\
Optical depth at 9.7 $\mu$m & 0.1, 0.3 \\
$\beta$ & -0.5 \\
$\gamma$ & 0 \\
Opening angle of the dust torus & 40, 60 \\
Angle between quasar axis and line of sight $\psi$ & 0.001, 50.100, 89.990 \\
Fraction of $L_{\rm{IR}}$ due to the quasar & 0.1, 0.3, 0.5, 0.7, 0.9 \\          
\end{tabular}
\end{center}
\label{CIGALE}
\begin{tablenotes}
\item \textit{Notes:}  (1) $SFR(t) \propto t\ exp(-t/\tau)$
        \end{tablenotes}
\end{table*}

\subsection{Wind kinematics vs. quasar and host galaxy orientation}

The IFU observations of Sample I targeting the [OIII] emission line allow us for the first time to compare the velocity field of a quasar wind with the stellar light distribution of their host galaxies and scattered light observations which probe the opening and inclination angles of the quasar. 

Although the 2-D surface brightness fitting using GALFIT has shown that the galaxies in both Sample I and Sample II are on average bulge-dominated, close to round early-type galaxies with a mean ellipticity $\epsilon = 0.26$, many of them show some degree of ellipticity and GALFIT is able to derive their position angle, i.e. their orientation on the sky. At the same time, the blue-band HST images reveal the direction with respect to the plane of the sky where light from the quasar can escape along less obscured directions through scattering off dust in the interstellar medium which result in typical cone shaped scattering regions \citep[see ][]{Obied_2015}. In the quasar unification model, the scattering cones would be oriented perpendicular to the obscuring dust torus.

The [OIII] IFU kinematic data, in particular the velocity gradient, velocity dispersion and asymmetry maps, allow to draw conclusions about the geometry and nature of the wind/outflow. In a detailed discussion, \citet{Liu_2013b} show that the modest radial velocity differences, high velocity dispersions and uniformly negative asymmetry parameter are consistent with a model in which the ionized gas is outflowing quasi spherically, partially reddened by dust in the galaxy or a wide biconical outflow with cone opening angles $2\theta > 120^{\circ}$. A rotating disk model was excluded on the basis of the large velocity dispersions with a flat radial profile. Using energy considerations they concluded that the outflows cannot be powered by star formation alone but are likely powered by the central quasar. In this paper we have shown that these quasars do not reside in disk galaxies, ruling out the scenario in which velocity differences in the map are primarily due to reddening by the galaxy disk. 

For further analysis, we assume that the outflows are biconical with large opening angles which results in a velocity difference across the map, i.e. blue and redshifted sides. The direction of the velocity field, i.e. where and how the blue- and redshifted shifted semicircular sides are orientated, is an indication for the outflow axis. Combining the blue and yellow-band HST data with the Gemini IFU observations therefore allows for the first time to infer a quasi 3-D picture of the stellar distribution, quasar orientation and outflow axis.

To do so, we first determine the position angle $\vec{A}_{\rm{wind}}$ of the wind direction, i.e. the position angle of the velocity gradient in the [OIII] velocity maps. We then measure the angle between the position angle of the host galaxy, as determined by GALFIT, $\vec{A}_{\rm{host}}$, the position angle of the quasar as determined by the scattering cone analysis \citep{Obied_2015} $\vec{A}_{\rm{quasar}}$ and $\vec{A}_{\rm{wind}}$ (see Fig. \ref{cartoon} for a schematic representation). When measuring the angles between those axes, we always choose the acute angle, i.e. $\alpha_{\vec{A}_{\rm{wind}}, \vec{A}_{\rm{host}}}$, $\alpha_{\vec{A}_{\rm{wind}},\vec{A}_{\rm{quasar}}}$ and $\alpha_{\vec{A}_{\rm{quasar}},\vec{A}_{\rm{host}}}$ are all $< 90^{\circ}$. Fig. \ref{angles} shows the distribution of the measured angles. Although there is scatter, {$\vec{A}_{\rm{quasar}}$ and $\vec{A}_{\rm{host}}$ are preferentially parallel. {$\vec{A}_{\rm{quasar}}$ and $\vec{A}_{\rm{wind}}$ tend to be parallel, as well. As described in Section 3.1.1,  6 out of the 11 sources in Sample I have a second, disk-like (i.e. $n_{\rm{sec}} < 1$) component and we repeat the angle analysis using the position angle of this second component. The distribution of the angles between the quasar scattering cones and the host galaxies' secondary stellar component is almost flat and we measure no preferred orientation between the scattering cone direction and the disk-like component.

\begin{figure}
\begin{center}
\includegraphics[trim = 3cm 6cm 3cm 4cm, clip = true, scale = 0.45]{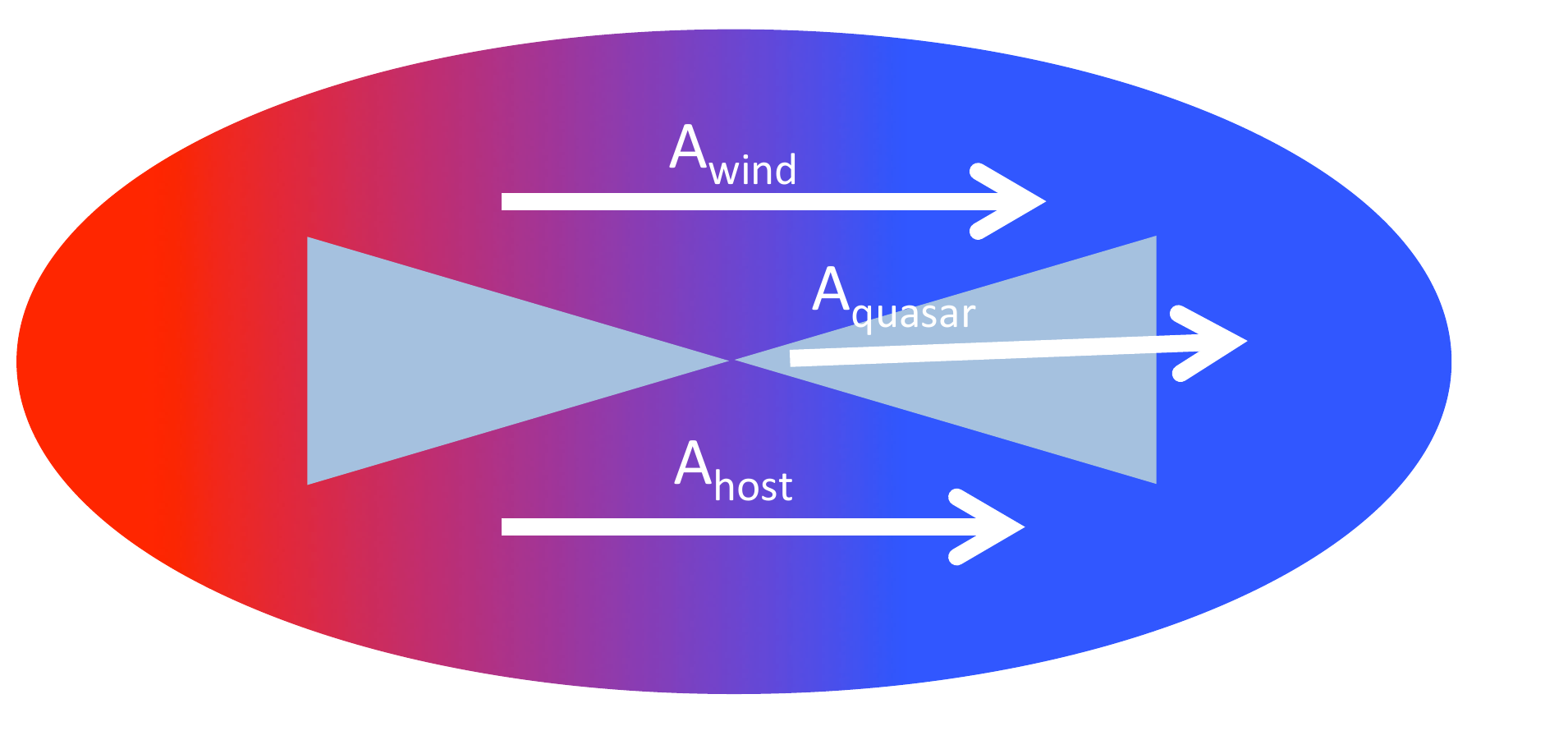}
\caption{Cartoon showing how the quasar wind $\vec{A}_{\rm{wind}}$, the quasar scattering cones $\vec{A}_{\rm{quasar}}$ and the host galaxy $\vec{A}_{\rm{host}}$ tend to be oriented with respect to another. The shape of the ellipse shows an extreme version of the host galaxy shape and the blue and red sides show how the receding/approaching sides of the wind are oriented with respect to the host galaxy and the quasar scattering regions (shown in light blue in the center of the cartoon).}
\label{cartoon}
\end{center}
\end{figure}

\begin{figure*}
\begin{center}
\includegraphics[trim=0cm 0cm 0cm 1cm, clip = true, scale = 0.4]{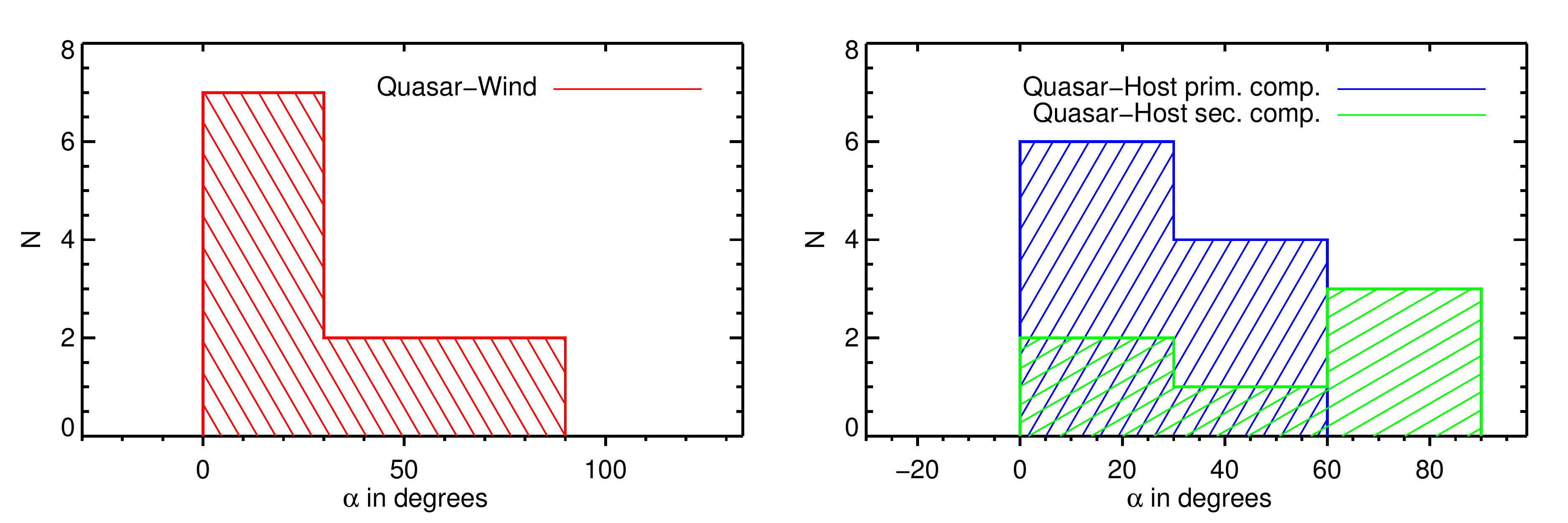} 
\caption{\textit{Left:} Distribution of measured angles between the quasar scattering cones and the wind axis (red) and \textit{Right:}  the quasar scattering regions and the orientation of the host galaxy's primary and secondary component (blue and green). The quasar scattering cones are preferentially aligned with the direction of the wind and preferentially aligned with the host galaxy axis. The cartoon in Fig. \ref{cartoon} visualizes these trends.}
\label{angles}
\end{center}
\end{figure*}

\section{Discussion}

\subsection{Quasar host galaxies}

The [OIII]$\lambda\lambda$4959,5007 emission line is one of the best tracers for quasar bolometric luminosity for optically selected obscured quasars \citep{Reyes_2008}. While the large sample of [OIII] selected luminous type-2 quasars presented in \citet{Reyes_2008} has been shown to constitute at least half of the entire (type-1 and type-2) quasar population and to be representative of `typical' black hole activity, in this paper we have focused on only the most luminous type-2 quasars with $L_{\rm{[OIII]}} > 10^{9}$~L$_{\sun}$. The quasars in our samples in fact belong to the 10\% most luminous objects in a given volume \citet{Reyes_2008} at $0.3 < z < 0.5$. At the same time, the space density of such luminous quasars peaks at around $1.5 < z < 2.5$ \citep[e.g. ][]{Croom_2009}, the epoch when quasars were most active. We therefore investigate whether peculiarities of the host galaxies or host galaxy environments might be responsible for this late low redshift outbreak of luminous quasar activity that we are observing for the sources in our samples. 

\subsubsection{Merger Fraction}

In Sec. 3.1.2 we have shown that a significant fraction of about half of our objects  ($45\pm10$\%) show signs of recent or ongoing merger activity in the form of tidal tails or close-by companions. In order to investigate if this is an unusual high fraction, a careful comparison with quiescent non-active galaxies is needed.

\citet{Bessiere_2012} compare a complete sample of 20 type-2 quasars with $L_{\rm{[OIII]}} > 10^{8.5}$ L$_{\sun}$ at $0.3 < z < 0.4$ with a sample of early-type quiescent galaxies from \citet{Ramos_Almeida_2012} and find a similar rate of interaction signatures between these two samples ($75\pm20$\% vs. $67\pm14$\%). The detected features in the type-2 quasar sample are, however, about 2 magnitudes brighter (r-band) in terms of surface brightness compared to the features seen in the quiescent galaxy sample. This lead the authors to the conclusion that the mergers in the two samples might have different progenitors (gas rich vs. gas poor encounters, minor vs. major mergers) or that they may be viewing the interactions at different stages.

Similarly to the analysis in this paper, \citet{Villar_Martin_2012} analyzed a sample of 42 type-2 quasars at $0.3 < z < 0.4$ with  $L_{\rm{[OIII]}} > 10^{8}$ L$_{\sun}$ that were observed with ACS/WFC on the HST. Visual inspection showed that most (72\%) of these quasars are hosted by elliptical galaxies. A significant fraction of 59\% show signs of morphological disturbances, such as tails, shells or double nuclei. The merger fractions of these less luminous quasars studied in \citet{Ramos_Almeida_2012} and \citet{Villar_Martin_2012} are similar or even higher to the merger fraction found in our work ($45\pm10$\%).

The average surface brightness limit of merger signatures for type-2 quasars and ellipticals probed in \citet{Ramos_Almeida_2012} are $\mu_{r, type-2} = 23.4$~mag~arcsec$^{-2}$ $\mu_{r, ell} = 24.3$~mag~arcsec$^{-2}$, respectively, which is about two magnitudes fainter than the faintest signatures we observe. If we instead re-compute the merger fraction for the ellipticals in \citet{Ramos_Almeida_2012} considering only galaxies that have merger signatures with $\mu_{r, ell} = 22.3$~mag~arcsec$^{-2}$ (the surface brightness limit of our faintest signatures), we find that only about 15\% of elliptical galaxies show merger signatures equally bright. Matching at fixed surface brightness is indeed the proper way to compare as tidal signatures become fainter with age but do not disappear entirely. 

While quiescent elliptical galaxies have most likely been assembled through minor, gas-poor mergers \citep{Dokkum_2005}, the quasar host galaxies in our sample are highly star forming (see Section 4.1.2) and are hosting a luminous quasar, therefore requiring larger gas supplies. Simulations and observations have shown that merger signatures of such gas-rich encounters are brighter than gas-poor merger signatures \citep{Kennicutt_1996, Duc_2013}. Only 15\% of quiescent ellipticals show merger signatures brighter or as bright as the quasar host galaxies of our Samples I and II, whereas the merger fraction in our Samples is $45\pm10$\%. This is therefore a clear indication that mergers do play a significant role in triggering the most luminous quasars at $z \sim 0.5$.

\subsubsection{Stellar Masses and Star Formation Rates}

\begin{figure}
\begin{center}
\includegraphics[trim = 7cm 0cm 1cm 0cm, clip =true, scale  = 0.25]{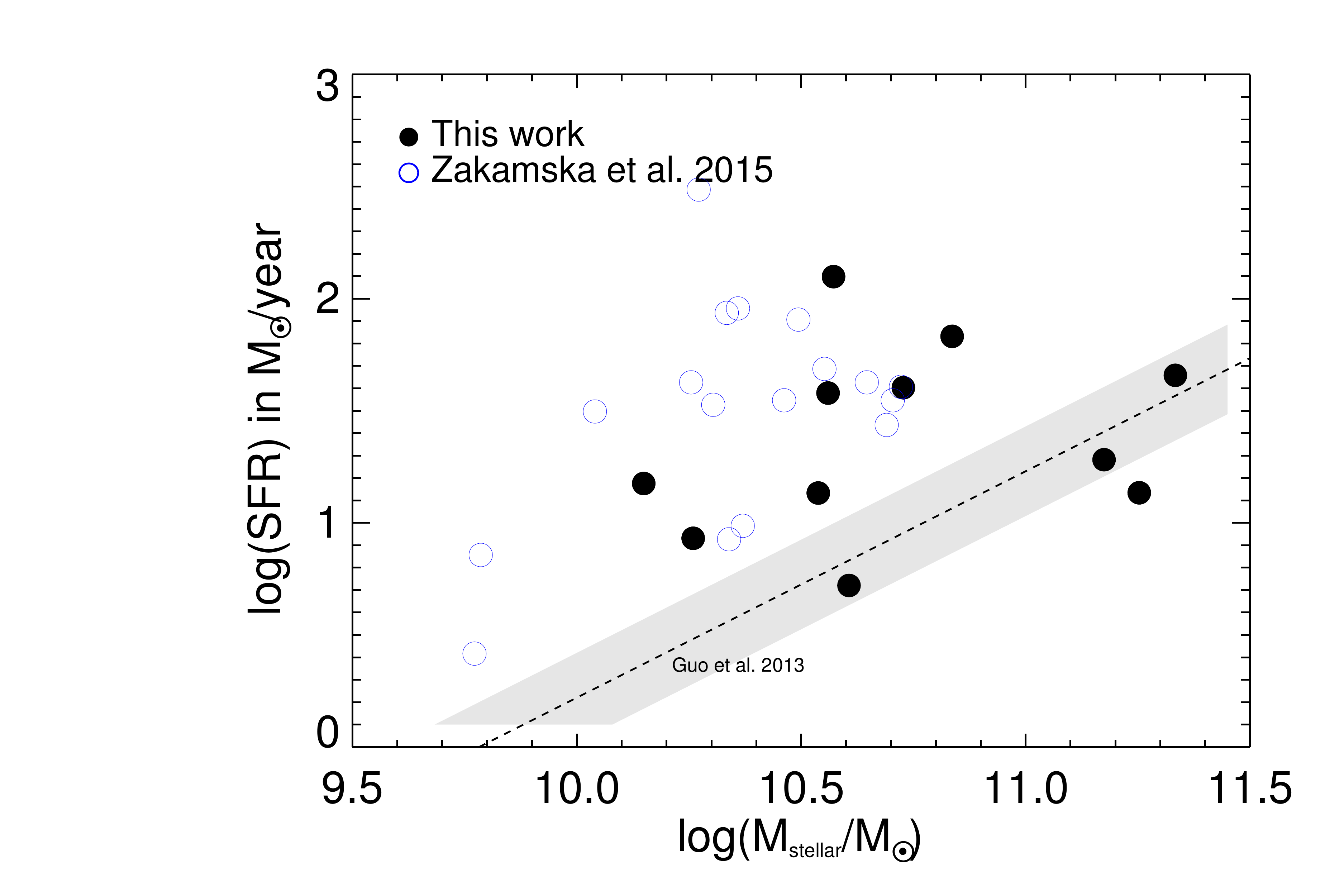}
\includegraphics[trim = 7cm 0cm 1cm 0cm, clip =true, scale  = 0.25]{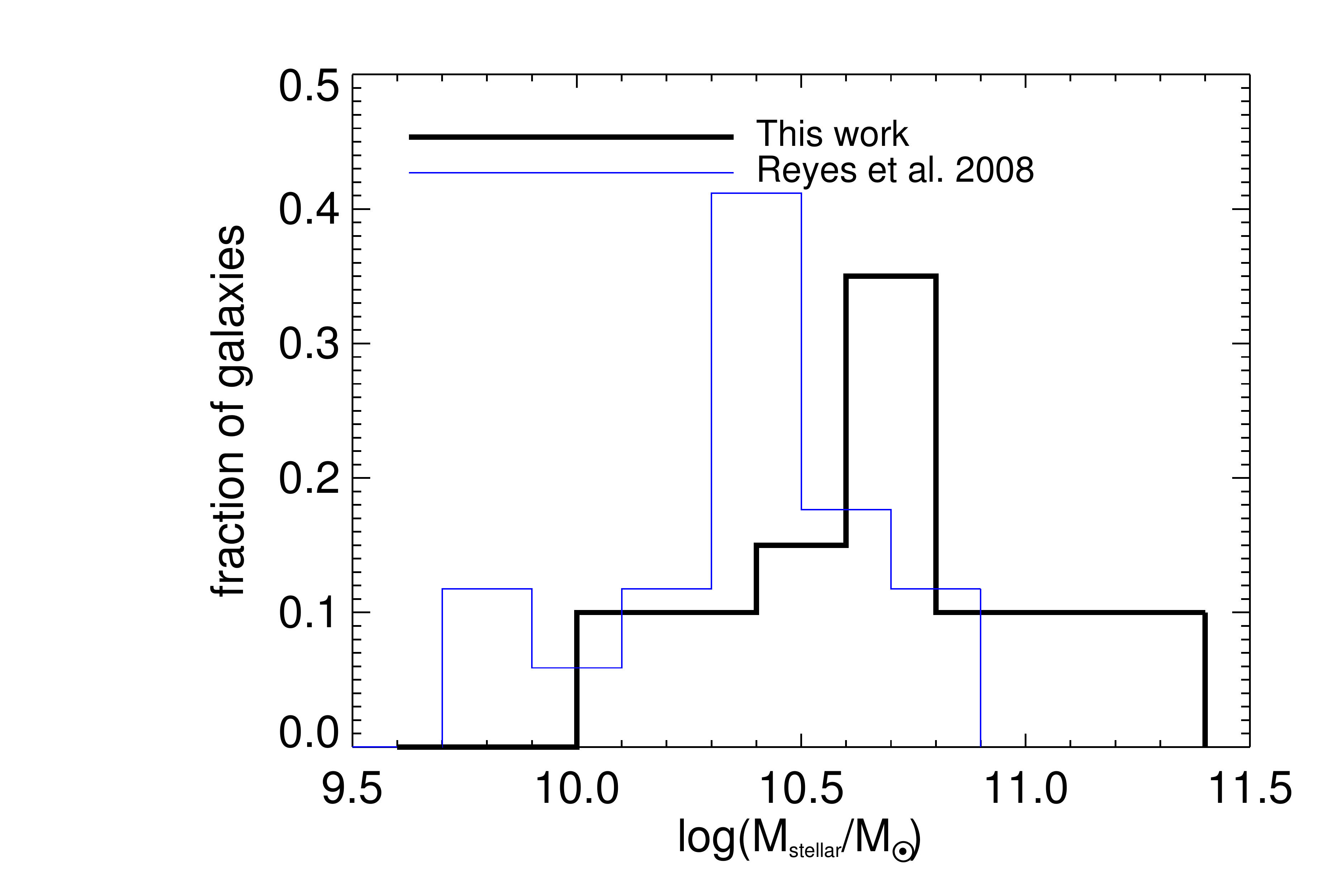}
\caption{\textit{Upper panel:} SFR-M$_{\rm{stellar}}$ relation for sources in Sample I and Sample II for which we were able to derive reliable SFR (filled back circles). For comparison (open blue circles) we also show type-2 quasars from \citet{Zakamska_2015} with $L_{\rm{[OIII]}} < 10^{9}$ L$_{\sun}$ and available SFRs from broad band data at $160 \mu$m. The dashed black line shows the SFR-mass relation for star forming galaxies at $z \sim 0.7$ and the grey shaded region its typical scatter. Although we have shown that type-2 quasar hosts are typically massive elliptical galaxies, they show high star formation rates even compared to regular star forming galaxies, indicative of large gas supplies available. \textit{Lower Panel:} Histogram of stellar masses for all sources in Sample I and Sample II (black histogram) compared to stellar masses in all type-2 quasars from \citet{Reyes_2008} with $L_{\rm{[OIII]}} < 10^{9}$ L$_{\sun}$ and available near-IR data necessary for deriving reliable stellar masses. The luminous quasars of Sample I and II are significantly more massive than less luminous quasars. The difference in average stellar mass can, however, be solely explained by their presumed difference in black hole mass, if accreting at the same Eddington ratio.}
\label{ms_sf}
\end{center}
\end{figure}

We now investigate to what extent the host galaxies of the sources in Sample I and Sample II are peculiar in terms of their stellar mass and star formation rate. In Section 3.1.1 we have shown that 90\% of the host galaxies in our samples are dominated by a bulge-like component and can therefore be generally classified as elliptical galaxies. \citet{Ilbert_2010} measured the galaxy stellar mass function for a sample of almost 200,000 galaxies from $z =0.2$ to $z=2$ as a function of redshift, morphological and spectral type from the COSMOS 2~deg$^{2}$ field. They show that typical mass for red and blue elliptical galaxies at $0.2 < z < 0.4$ is log($M^{*}/M_{\sun}$)~$= 10.9$ and log($M^{*}/M_{\sun}$)~$= 11.1$, respectively. Blue elliptical galaxies are classified as galaxies showing a bulge-dominated morphology using HST/ACS images but show signs of non-negligible star formation rates (similar to SFR in our quasars) based on their NUV-to-optical colors. At $0.2 < z < 0.4$ and log($M^{*}/M_{\sun}$)$ \sim10.9$ only about 30\% of all elliptical galaxies are classified as blue ellipticals.

Similar results have been found by \citet{Mortlock_2015} combining photometry from the Ultra Deep Survey (UDS), Cosmic Assembly Near-infrared Deep Extragalactic Legacy Survey (CANDELS) UDS and CANDELS the Great Observatories Origins Deep Survey-South (GOODS-S) surveys, albeit at slightly higher redshift. At $1 < z < 1.5$ they find that log($M^{*}/M_{\sun}$)$ =10.8$ for galaxies with a S\'{e}rsic index of $> 2.5$ and at $z \sim 1$, the fraction of massive log($M^{*}/M_{\sun}$)$ > 10$, elliptical galaxies which are red is $\sim$ 80\%. The average stellar mass of all sources in Sample I and Sample II is log($M_{\rm{stellar}}/M_{\sun}$)~$= 10.7\pm 0.3$ which compared to the $M^{*}$ values found by \citet{Ilbert_2010} and \citet{Mortlock_2015} shows that the host galaxies of these powerful quasars are close to $M^{*}$ of the elliptical galaxies at these redshifts. 

In Figure \ref{ms_sf} we show the relation between the stellar masses of our quasars' host galaxies as measured by CIGALE (see Section 3.2) and SFR for all sources in our sample for which we were able to derive reliable SFR using DecompIR (see Section 3.3). For reference we also show the best fit to the mass-SFR relation at $z \sim 0.7$ derived by \citet{Guo_2013} using a sample of almost 13,000 SF galaxies from COSMOS. The mass-SFR relation, often dubbed the `main sequence of star formation' is a well established relation between the SFRs and stellar masses of galaxies and holds both at low and high redshift \citep[e.g. ][]{Brinchmann_2004, Daddi_2007}. Galaxies that lie above that relation are starbursts, while quiescent galaxies lie below that relation.

At $0.2 < z < 0.4$ and $10.5 <$ log($M^{*}/M_{\sun}$)~$= 11.3$ nearly 100\% of quiescent galaxies have an elliptical morphology while only about 30\% of all elliptical galaxies are non-quiescent. The host galaxies of the quasars in Sample I and II are found to lie within the scatter of the mass-SFR relation or even significantly above, with an average star formation rate of $35 M_{\sun}$/year. Similar average SFR for quasar hosts have recently been measured by \citet{Zakamska_2015}, who utilize infrared spectroscopic and photometric data from \textit{Spitzer} and \textit{Herschel} to derive SFR for $\sim 300$ obscured and unobscured quasars at $z < 1$. 





So while the SFR for our luminous quasars of Sample I and Sample II may be typical (and even slightly higher) for the average type-2 quasar samples at $z <1$, it is notable that these host galaxies do not populate the SFR-M$_{\rm{stellar}}$ plane as would be expected based on their morphology, mass and redshift. An average elliptical galaxy at $z \sim 0.6$ and log($M_{\rm{stellar}}/M_{\sun}$)~$= 10.7$ would lie significantly below the SFR-mass relation, with SFR $< 1$M$_{\sun}$/year \citep{Ilbert_2010}. An average star-forming galaxy with log($M_{\rm{stellar}}/M_{\sun}$)~$= 10.7$ would lie on the SFR-mass relation with SFR $\sim 10$M$_{\sun}$/year \citet{Guo_2013}. Only 30\% of elliptical galaxies are star-forming with SFR $\sim 25$M$_{\sun}$/year and the hosts of our extremely luminous quasars (Sample I and Sample II) seem to belong to this blue elliptical population and may even populate the most star-forming tail of that blue elliptical population. 

\subsubsection{Eddington Ratios}

In addition to comparing the host galaxies of the sources in our samples to the overall population of elliptical galaxies, we also explore how the host galaxies' properties of the powerful quasars in our sample compare to the overall type-2 quasar host galaxy population. To this purpose, we utilize the type-2 quasar sample from \citet{Reyes_2008} of which 107 sources have 160$\mu$m fluxes from the \textit{Herschel} where star formation rates can be derived similarly to the sources in our samples. We cross-correlate this sample of 107 sources with the UKIDSS near-IR photometry source catalog and find matches for 45 sources. We follow the same procedure as described in Section 3.2 and use CIGALE to estimate stellar masses for this sample of 45 sources using SDSS and UKIDSS photometry. We then only consider sources with $L_{\rm{[OIII]}} < 10^{9}$ L$_{\sun}$, i.e. sources that are less luminous than the quasars in our Sample I and II, leaving us with 17 sources. In Figure \ref{ms_sf}, we show how the host galaxies of the less luminous quasars from \citet{Reyes_2008} compare with the luminous quasars studied in this paper. While the host galaxy masses of the less luminous quasars span a range of 9.6 $<$ log(M$_{\rm{stellar}}/M_{\sun}$) $<$ 10.7, the luminous quasars populate a significantly higher mass range of 10.1 $<$ log(M$_{\rm{stellar}}/M_{\sun}$) $<$ 11.4.

The bolometric luminosity of a quasar usually has a natural limit known as Eddington luminosity

\begin{equation}
L_{\rm{Edd}} = 3.28 \times 10^4\left(\frac{M_{\rm{BH}}}{M_{\sun}}\right) L_{\sun}
\end{equation}
where $\frac{M_{\rm{BH}}}{M_{\sun}}$ is the mass of the black hole in solar mass units and $L_{\sun}$ is the solar luminosity. The Eddington luminosity corresponds to the limit where force of radiation acting outward equals the gravitational force acting inward and is therefore a the maximum luminosity an object can radiate at. The ratio between the bolometric luminosity of a quasar and the Eddington luminosity is referred to as Eddington ratio $\lambda_E  = L_{\rm{Bol}}^{\rm{quasar}}/L_{\rm{Edd}}$. Typical Eddington ratios found for luminous quasars are about $\sim 0.1$ \citep{Kollmeier_2006, Lamastra_2009, Shen_2013}. 

We now explore if there are indications for a variations in $\lambda_E$ between our lower luminosity and high luminosity samples in Fig. \ref{ms_sf} which might be an indication for differences in the mass accretion rate or efficiency of the or the conversion of gravitational energy into radiation. $L_{\rm{Edd}}$ is a function of black hole mass (equation 4) which in turn can be estimated from the black hole mass-galaxy mass relation \citep[e.g. ][]{Salviander_2015}. At the same time we can use the [OIII] line luminosities to estimate the bolometric luminosity of the quasars. \citet{Lamastra_2009} derived [OIII] bolometric correction factors $C_{\rm{[OIII]}}$ using a sample of type-2 quasars selected from the SDSS which allow to convert [OIII] line luminosities to bolometric luminosities. They showed that in the range of $8.4 < \rm{log(}\rm{L_{\rm{[OIII]}}}/\rm{L_{\sun}}) < 10.4$, the line luminosities can be converted to bolometric luminosities using a constant conversion factor where $L_{\rm{bol}} = C_{\rm{[OIII]}}\times L_{\rm{[OIII]}}$, i.e. the bolometric luminosity is proportional to the [OIII] line luminosity, resulting in a bolometric luminosity ratio between the two samples is $\sim 6.2$. 

The mean stellar masses for the lower luminosity \citet{Reyes_2008} sample and our high luminosity quasar sample are log(M$_{\rm{stellar}}\rm{)}= 10.3$ and 10.6, respectively. Extrapolating from the relation found in \citet{Salviander_2015}, the average black hole mass ratio is $\sim 6.3$, which is in astonishing agreement with the bolometric luminosity ratio and shows that if the black hole masses are correlated with stellar masses the Eddington ratio does not change significantly between the lower luminosity and higher luminosity sample. 

This is in agreement with results from \citet{Bian_2006}. In this paper the authors use reliable stellar velocity dispersions for a sample of type-2 quasars from the SDSS \citep{Zakamska_2003} to derive black hole masses and Eddington luminosities. Using [OIII] line luminosities they calculate bolometric luminosities to infer Eddington ratios and study the correlation between Eddington ratios and [OIII] luminosities. No significant correlation was found, consistent with our results. The difference in stellar masses between the lower luminosity sample and our high luminosity quasar sample that we observe in Figure \ref{ms_sf} is therefore consistent with differences in black hole masses in agreement with the mass black hole-galaxy mass relation.

\subsection{Quasar and Host Galaxy Alignment}

In Section 3.4. we have measured the relative orientations between the host galaxies' position angles, the direction of the quasar scattering cones and the velocity field of the outflowing ionized gas. The quasar scattering cones tend to be parallel to the position angle of the wind $\vec{A}_{\rm{wind}}$, determined from the gradient of the blue- and redshifted shifted semicircular sides in the [OIII] velocity maps. Despite large scatter, the primary stellar component of the host galaxy seems to be aligned with the direction of the scattering regions. Figure \ref{cartoon} shows a schematic representation of these trends. 

While so far we have assumed that the scattering cones that we detect in the blue-band HST images are indeed radiation scattered off the interstellar medium, an alternative explanation could be that the quasar is just illuminating the stellar disc in those galaxies resulting in the detected blue emission. This can happen if radiation bicones from the quasar intercept the galactic disc at shallow angles and photo-ionize the gas in the disk \citep{Lena_2015}. While \citet{Obied_2015} argue against this scenario showing that the lateral surface brightness profiles of the scattered light are unlikely to be produced by an illuminated disc, we discuss this scenario on the basis of the relative orientations of the scattering regions, host galaxy and ionized wind.


We find here that the direction of the scattering regions coincides with the direction of the velocity field (see Figure \ref{cartoon}) consistent with a quasi spherical outflow illuminated by the quasar radiation that escapes the nuclear region. Although the same alignment between blue scattered light and velocity field would be expected in the case of an illuminated disk, the additional information about the high velocity dispersion of the ionized gas unlikely to be due to an rotating disk, rules out this possibility. In fact, in the only case where the host galaxy is clearly a disk (SDSS~0149-0048), the scattering cone is perpendicular to the stellar disk (see Figure \ref{stamps}). Additionally, we do not find any correlation between the secondary disc-like stellar components and the scattering regions (see Figure \ref{angles}), confirming our conclusion. 

Based on geometry and luminosity arguments, \citet{Obied_2015} determined which scattering cone is to be identified as the forward scattering one (if two cones are detected). In objects where only one cone is detected it is taken to be the forward-scattering one. Due to the strongly forward-scattering nature of dust, the backside cone is more likely to be not detected. In all but two cases where cone morphology is complex (SDSS~J0319-0058 and SDSS~J0321+0841) the position of the forward-scattering cone is also coincident with the blue-shifted side of the velocity field of the ionized gas, another confirmation that the blue radiation cones are indeed due to scattered light and that the ionized gas is in an organized radial outflow.

The remaining puzzle concerns the apparent alignment between the host galaxies' primary elliptical stellar component and the scattering regions. As mentioned earlier, the scatter for this angle measurement is large, but it is notable that for our 11 objects of Sample I, no angle between primary elliptical stellar component and scattering region is larger than 60 degrees. If any angle range ($0-30$, $30-60$ and $30-90$ degrees) was equally likely, then measuring no angle larger than 60 degrees for 11 sources would have a probability of only $\sim 1$\%. 

Similar alignments have been found between the optical morphologies of distant ($z > 0.6$) radio galaxies and their radio jets \citep{Chambers_1987, McCarthy_1987}. Several mechanisms have since been proposed to explain these observations. Some authors suggest that this `alignment effect' might be due to star formation triggered by the radio source as it expands into the ambient medium. This idea is supported by some observations which demonstrate unpolarized UV continuum with clumpy morphology and starburst-like spectral shape \citep{Rees_1989, Dey_1997}. However, most of the powerful $z \sim 1$ radio galaxies show strong polarization and broad emission lines, suggesting that most of the spatially extended continuum emission is scattered radiation which comes from a hidden AGN \citep{Cimatti_1997, Leyshon_1998}. 

While the origin of the `alignment effect' might be different in powerful high-redshift radio galaxies, the similarity to our observations is intriguing. Based on considerations presented above and in \citet{Obied_2015}, the scattering scenario is also the favored explanation for the alignment between our host galaxies and ionization cones; the scattered light  likly does partially contribute to the emission in the yellow-band image leading to a bias in the measured ellipticities. The ellipticities are indeed small, $\epsilon_{\rm{mean}} = 0.26$, and even a slight contamination by some scattered light could alter the axis ratios of the ellipses. Such contamination by scattered light is seen directly in some positive residual yellow-band images (see e.g. SDSS~J0319-0019 and SDSS~J1039+4512 in Figure \ref{stamps}).


\subsection{Unification tests}

In the standard orientation-based unification model for quasars \citep{Antonucci_1993}, a dusty torus surrounding the central black hole is responsible for type-1/type-2 dichotomy, which is produced by the different relative orientations of the torus with respect to the observer. In type-1 objects, where the quasar can be directly observed since its radiation is not obscured by the torus, characteristic UV, X-ray and optical signatures and broad emission lines from the accretion disk can be observed. In type-2 quasars, this direct view is obscured, and the quasar can be identified using indirect signatures. 

At the same time, the torus absorbs a significant fraction of the quasar radiation at X-ray, UV and optical wavelengths which is re-emitted in the mid-IR as thermal dust emission. This leads to a radial temperature gradient within the torus ranging from the sublimation temperature ($\sim$1500 K) at the inner surface of the torus to a few hundred Kelvin in the outer torus, so that the observed dust temperature probed by the IR radiation is expected to depend on the orientation of the quasar \citep{Pier_1992}. This standard unification model has succeeded in explaining many observational differences between type-1 and type-2 quasars. The detection of a hidden AGN in some type-2 quasars through polarization measurements is the strongest, but not only, evidence for this unification scheme \citep{Antonucci_1985, Vernet_2001}. 

Other observations appear to be in conflict with some key predictions of the unification model, where for example the environments of type-1 and type-2 quasars are expected to be identical, but type-2 quasar samples are more clustered than type-1 samples \citep{Koulouridis_2006, Dipompeo_2014}. Motivated by such observations, a dynamical model for quasars has been proposed in in which obscured, type-2 quasars evolve into type-1 quasars as their merger-triggered, dust-obscured quasars become more powerful and clear out the environment \citep{Hopkins_2006}.

Combining measurements of the quasar host galaxies as probed by the yellow-band and multi-wavelength observations presented in this paper, measurements of quasar scattering regions as probed by blue-band observations presented in \citet{Obied_2015} and measurements of the quasar wind as probed by [OIII] kinematic measurements presented in \citet{Liu_2013a, Liu_2013b}, we now attempt to test the classical unification model for quasars using a number novel methods.

\subsubsection{OIII luminosity vs. opening angle}

\begin{figure}
\begin{center}
\includegraphics[trim = 0.5cm 1.3cm 0.4cm 0cm, clip = true, scale = 0.31]{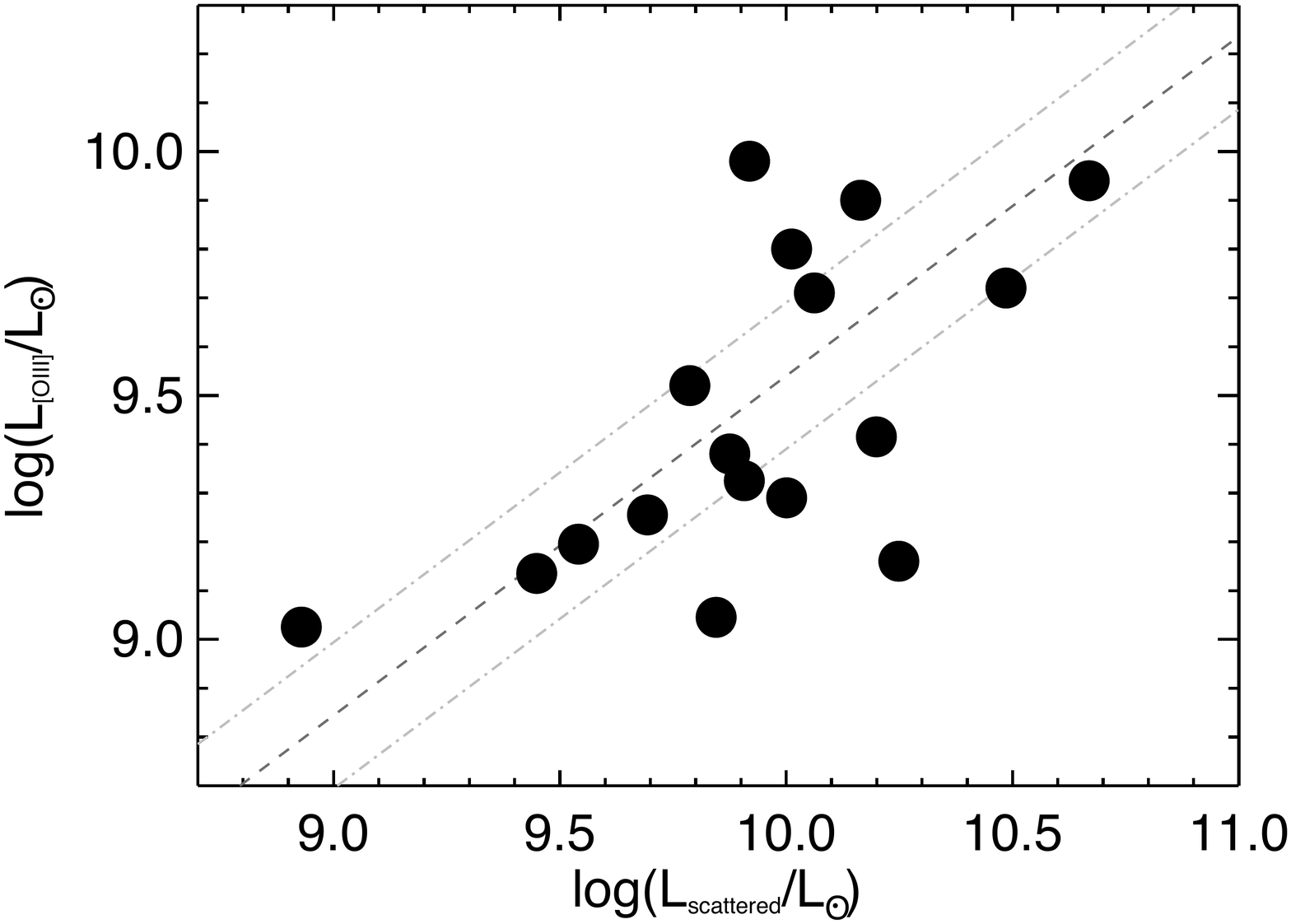}
\caption{[O III] line luminosity for sources in Sample I and II \citep{Zakamska_2006, Liu_2013a} as a function of scattered luminosity \citep{Obied_2015}. The dashed line shows a simple linear fit minimizing the vertical offsets. The positive correlation at 99.8\% significance confirms that outflowing ionized gas as measured by [OIII] is tightly related to the quasar scattered light.}
\label{OIII_scatter}
\end{center}
\end{figure}

In type-2 quasars the direct view to the nucleus is obscured, presumably by a torus-like dusty structure on scales of up to 100 pc. But the light can still escape along other unobscured directions, scatter off surrounding material and reach the observer. The scattered component can be identified either via its polarization signature or its morphology in imaging observations, as illumination of extended material by a light source blocked along some directions produces a characteristic conical shape in the plane of the sky \citep{Zakamska_2006, Borguet_2008}. 


\citet{Obied_2015} present modeling of such giant scattering regions of the objects in our Sample I and Sample II and present measurements for the cones' opening angles (which in turn are a probe for the geometry of the torus such as its inner radius and height). [OIII] line luminosities are available from \citet{Reyes_2008}.

Figure \ref{OIII_scatter} demonstrates that [OIII] and scattered light luminosity are strongly correlated at 99.8\% significance. While several possible variables can drive such a correlation, one possibility is that both are directly related to the opening angle of obscuration. We directly test this hypothesis in Figure \ref{OIII_opening}, where we show the correlation between the opening angles $2\theta$ of the scattering cones for the sources in Sample I and Sample II \citep{Obied_2015} and [OIII] luminosities from \citet{Reyes_2008} \citep[see also][]{Liu_2013a, Zakamska_2006}. Using a Spearman rank correlation test, we find a significant correlation between $2\theta$ and log(L$_{\rm{[OIII]}}$/L$_{\sun}$) at 96\% significance. 

The volume of an ionization cone with half opening angle $\theta$ is proportional to $h^3 (1-\rm{cos}(\theta))$ where $h$ is the height of the cone. The height $h$ corresponds to the spatial extent of the cone as measured by \citet{Obied_2015}. We further assume that the the cone is filled with ionized gas so that $L_{\rm{[OIII]}}$ is proportional to the emitting volume. 

Dividing the sources shown in Figure \ref{OIII_opening} into two subsamples with $2\theta < 50$~degrees and $2\theta > 50$~degrees, the median properties of these two subsamples are $2\theta_1 = 37$~degrees, $2\theta_2 = 70$~degrees and log(L$_{\rm{[OIII]}}$/L$_{\sun}$)$_1$ = 9.3 and log(L$_{\rm{[OIII]}}$/L$_{\sun}$)$_2$ = 9.6. The distribution of spatial extents is very broad (7~kpc$\pm$3~kpc) so that we assume the same median cone for both subsamples. Following the previously described assumptions, then
\begin{equation}
\frac{L_{\rm{[OIII]},1}}{L_{\rm{[OIII]},2}} = \frac{1-\rm{cos}(\theta_1 )}{1-\rm{cos}(\theta_2)}.
\end{equation}
We find $\frac{L_{\rm{[OIII]},1}}{L_{\rm{[OIII]},2}} \sim 3$ and $\frac{1-\rm{cos}(\theta_1 )}{1-\rm{cos}(\theta_2)} \sim 3.5$, which, considering the crude assumptions made, is remarkably consistent and shows that the geometry of the ionization cones is indeed correlated with the amount of ionized gas in the galaxy. 

Although the slope of the correlation is consistent with the volume of the cones being correlated to the amount of ionized gas, the scatter of the correlation is fairly large and on the order of 0.2 dex. \citet{Obied_2015} show that a significant amount of the quasar radiation could be leaking through a patchy, clumpy torus which could lead to such a broadening of the correlation presented in Figure \ref{OIII_opening}.

A similar correlation between [OIII] luminosity and opening angle has been found in powerful radio galaxies where opening angles have been estimated from the linear projected sizes of double radio galaxies and radio quasars \citep{Arshakian_2005}. This result has been interpreted in the context of the receding torus model in which the ionizing radiation from the AGN evaporates the circumnuclear dust forming the inner wall of the torus leading to a positive correlation between AGN power and the inner radius and therefore between AGN power and opening angle \citep{Lawrence_1991}. 

Our observations show for the first time that such a correlation can also arise because larger opening angles allow for more ionized gas due to the larger volumes of ionization cones. Our observations do not exclude a scenario in which larger opening angles are caused by higher luminosity AGN and it is therefore possible that both effects contribute to the correlation presented in Fig. \ref{OIII_opening}
 

\begin{figure}
\begin{center}
\includegraphics[trim = 0.5cm 6cm 0.5cm 6cm, clip = true, scale = 0.4]{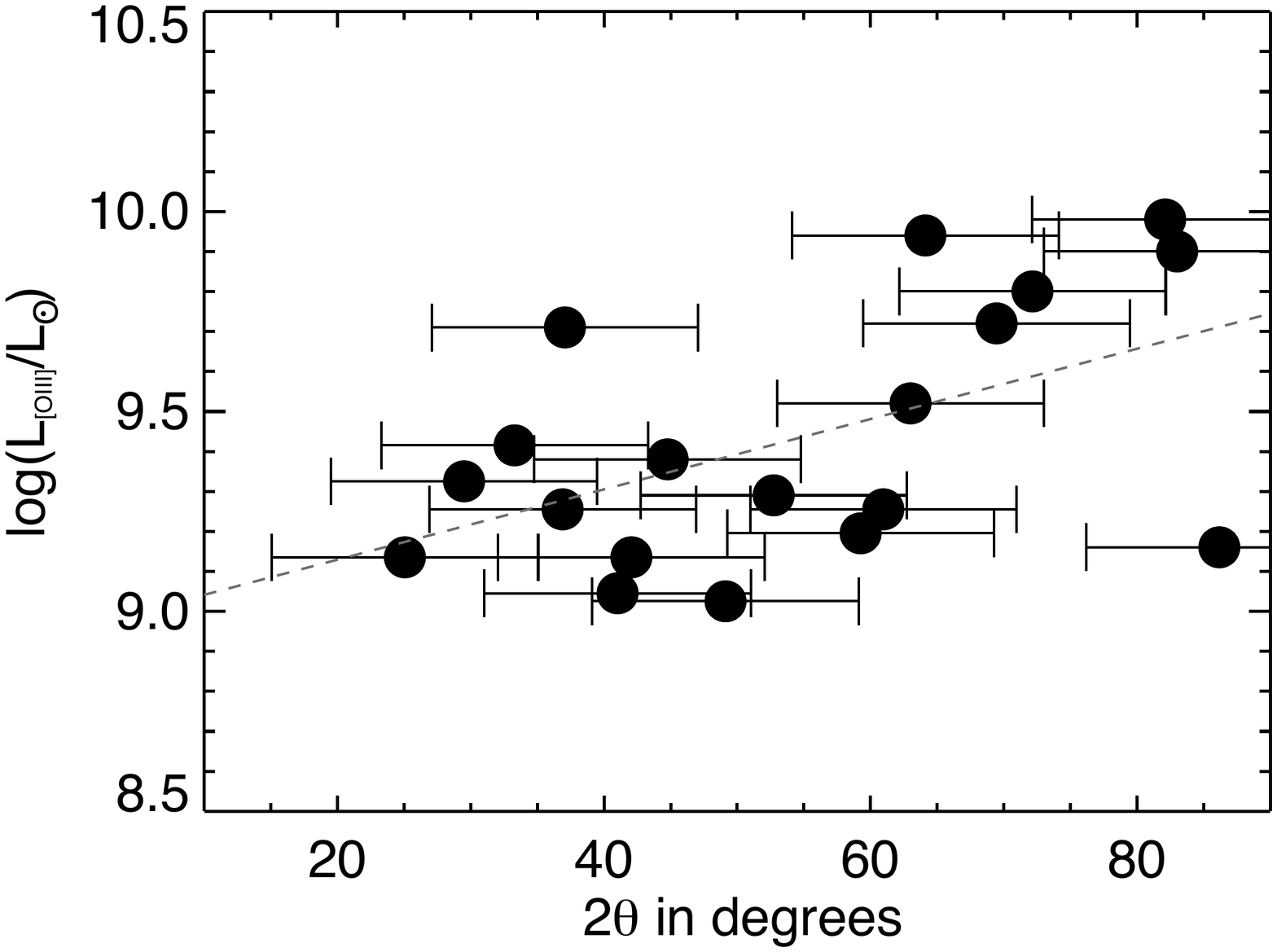}
\caption{[OIII] line luminosity for sources in Sample I and II \citep{Zakamska_2006, Liu_2013a} as a function of full opening angle $2\theta$ of the quasar scattering regions \citep{Obied_2015}. The dashed line shows a simple linear fit to the data. A Spearman rank test confirms a significant correlation at 96\% significance.}
\label{OIII_opening}
\end{center}
\end{figure}

\subsubsection{IR color vs. Inclination}

In type-2 quasars the direct emission from the quasar is obscured by dust resulting in absorption of the blue quasar spectrum and re-emission of this absorbed radiation as thermal emission by heated dust particles. The mid-IR SED (above $\sim 2\mu$m rest-frame) of type-2 quasars is therefore is dominated by red, thermal dust emission. In the standard quasar unification model the observed dust temperature probed by the IR radiation is dependent on the inclination angle of the system \citep{Pier_1992} as the hotter inner surface of the torus becomes less visible with increasing inclination angle (with respect to the line of sight).

The IR color can be used to estimate the maximum temperature of the dust visible to the observer. We have estimated dust temperatures using a modified black body model in Section 3.3, but this single component model is best for estimating the typical temperature at which the bulk of the dust radiates, while hotter, less significant dust components can be missed. Therefore, a simple color measurement can be more powerful than full modeling.

Inclination angles $i$ measured by \citet{Obied_2015} are defined relative to the line of sight. Figure \ref{IR_inclin} shows the [W1]-[W4] color (AB), for \textit{WISE} bands at $3.4\mu$m and $22\mu$m, probing rest-frame wavelengths of $\sim 2\mu$m and $\sim 15\mu$m, respectively, as a function of inclination angle $i$. For this analysis, we excluded source SDSS~J0841+2042, which is undetected at W4. In addition to the model inclination angles, we also show measurements by at least two human classifiers who classified the ionization cone images as low, medium and high inclination ($\sim 40, 60$ and 80 degrees, respectively). For both sets of angles, we observe a positive correlation with inclination angle at 50\% and 98\% significance, respectively. 

Despite large scatter and partly low significance, this tentative correlation is consistent with predictions from the unified model, in which smaller inclination angles should result in bluer mid-IR colors, as hotter parts of the inner torus become visible. It is becoming increasingly clear that the obscuring material is in fact not a smooth dusty torus \citep{Schartmann_2008, Nikutta_2009, Markowitz_2014}, but that quasar obscuration is clumpy, such that the observer has a reasonable chance of seeing the hotter inner sides of dusty clumps although the nominal line of sight (i.e., inclination angle of the system) would suggest otherwise. \citet{Obied_2015} shows that this might indeed be the case in $30 - 50$\% of the objects and explains the tentativeness of the correlation presented in Figure \ref{IR_inclin}.

\begin{figure}
\begin{center}
\includegraphics[trim = 0cm 0cm 0cm 0cm, clip = true, scale = 0.35]{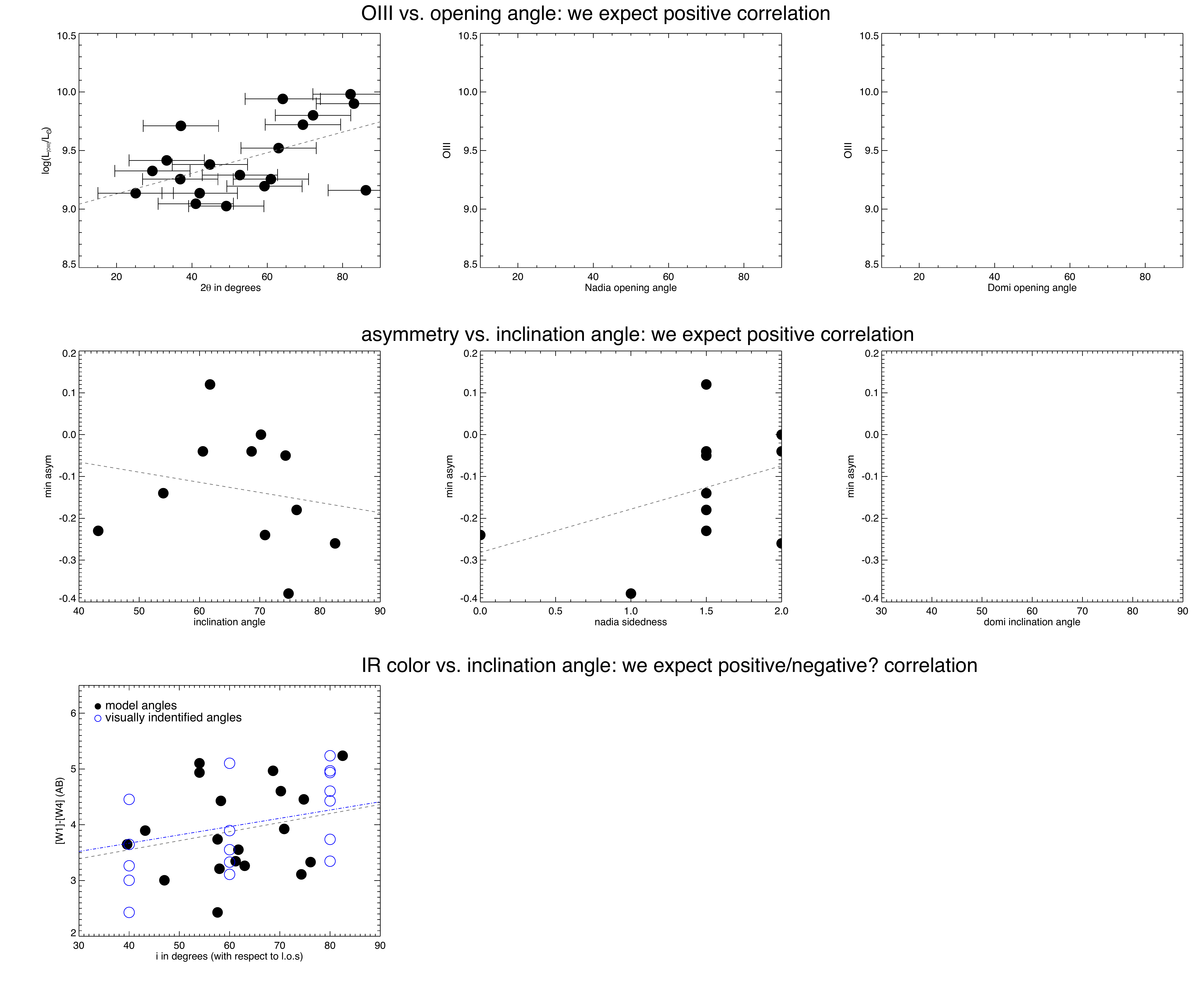}
\caption{WISE [W1]-[W4] color as a function of model inclination angles of the scattering regions \citep{Obied_2015} and inclination angles measured by at least two human classifiers. The dashed and dotted-dashed lines show a simple linear fit to the data. A Spearman rank tests confirms a correlation (50\% and 98\% significance, for model and human classified angles, respectively), indicating that quasars with higher inclination show redder colors, in agreement with expectations from the unified model for quasars.}
\label{IR_inclin}
\end{center}
\end{figure}

\section{Conclusions}

In this paper we have combined HST yellow-band observations of 20 powerful quasar host galaxies at $0.2 < z < 0.6$ with geometric measurements of quasar illumination presented in \citet{Obied_2015}, ionized wind diagnostics from \citet{Liu_2013a, Liu_2013b} and multi-wavelength archival observations from the optical to far-IR. We study the properties of the host galaxies as probed by the yellow-band observations and SED modeling to assess to which extent these host galaxies might be special and why they are hosting some of the most powerful quasars at their redshifts. We further study the 3D alignments between the quasars' host galaxies, quasar illumination and quasar winds. The main results are:

\begin{itemize}
\item 2-dimensional surface brightness fitting shows that all but two host galaxies are dominated by bulge-like elliptical components. If present, secondary components are faint and disk-like.
\item About half of our objects show signs of recent or ongoing merger activity in the form of tidal tails or close-by companions. Only about 15\% of quiescent elliptical galaxies at similar redshift show tidal signatures at the same surface brightness limit, suggesting that galaxy interactions play a significant role in triggering powerful quasars at $z \sim 0.5$.
\item SED fitting performed using CIGALE shows that the host galaxies span a mass range of $10^{10} - 10^{11.3}$~M$_{\sun}$, with an average mass of $10^{10.7\pm0.3}$~M$_{\sun}$. This is similar to M$^{*}$ in the redshift range $0.4 < z <0.6$, showing that these host galaxies are not unusual in terms of stellar mass.
\item Modeling of the IR emission demonstrates that the bolometric output of our objects is dominated by the quasar, with AGN fractions of $80 \pm15$\%. Nonetheless, the far-infrared emission is dominated by star formation. The star formation rates of $35$~M$_{\sun}$~year$^{-1}$ derived using DecompIR are unusually high compared to regular elliptical galaxies at $z \sim 0.5$. These host galaxies seem to be extremely gas-rich.
\item Less luminous type-2 quasars reside in less massive host galaxies. We find that this difference in stellar mass is consistent with the hypothesis that all type-2 quasars accrete with similar accretion rates and efficiencies, but with less luminous type-2 quasars powered by less massive black holes in agreement with the correlation between stellar mass and black hole mass.
\item No significant alignment is found between faint disk-like stellar components and ionized gas observations from \citet{Liu_2013b}, confirming that the gas fills the volume of the galaxy and is not restricted to a disk.
\item The direction of the quasar scattering cones probed by \citet{Obied_2015} and outflowing direction measured by \citet{Liu_2013b} are coincident, showing that the outflowing gas is photoionized by the quasar along the openings in circumnuclear obscuration. Furthermore, in most cases the forward-scattering cone is co-spatial with the blue-shifted [OIII] component, confirming that the [OIII] gas is indeed participating in an organized radial outflow.
\item We detect a tentative alignment between host galaxy position angle and quasar wind/scattering regions which is likely biased by some residual scattered light in the yellow-band images.
\item Novel quasar unification tests using a combination of various data sets (quasar scattering cone geometrical measurements, host galaxy properties and ionized gas information) show that the observed trends agree with expectations from the orientation-based unification model for quasars. In particular, we detect a positive correlation between cone opening angles and [OIII] luminosities with a slope that is consistent with the volume defined by the cone opening angle being filled with ionized gas. A weaker correlation is found between IR color and inclination angle, showing redder colors for more inclined systems. A likely explanation for the tentativeness of the correlation is that the torus is clumpy, smearing out the effects that would be expected for a smooth torus. 
\end{itemize}
These results show that the host galaxies of some of the most powerful quasars at $z \sim 0.5$ are massive elliptical galaxies. They are unusual in that they are gas rich and highly star forming (they are among the 30\% of blue, star forming ellipticals in this redshift range).

There is compelling evidence for significant (by a factor of two) size evolution of elliptical galaxies between $z=1$ and $z = 0.5$ \citep{Dokkum_2008, Wel_2008, Hopkins_2009} and continuous growth through accretion of companions or mergers has been proposed as one of the most likely explanations.  An unusually high fraction of host galaxies in our samples show bright signatures of recent or ongoing mergers. We propose that these luminous quasars are most likely triggered through minor, gas-rich mergers, possibly connected to this late-time size growth of elliptical galaxies. 

\section*{Acknowledgements}

Based on observations made with the NASA/ESA Hubble Space Telescope, obtained at the Space Telescope Science Institute, which is operated by the Association of Universities for Research in Astronomy, Inc., under NASA contract NAS 5-26555. These observations are associated with program GO-13307 and GO-9905. Support for program GO-13307 was provided by NASA through grant HST-GO-13307.01-A from the STScI. D.W. acknowledges support by Akbari-Mack Postdoctoral Fellowship. G.O. acknowledges support by the Provost's Undergraduate Research Award at Johns Hopkins University. 





\bibliographystyle{mnras}
\bibliography{bib_hst_project}

\begin{thebibliography}{}
\makeatletter
\relax
\def\mn@urlcharsother{\let\do\@makeother \do\$\do\&\do\#\do\^\do\_\do\%\do\~}
\def\mn@doi{\begingroup\mn@urlcharsother \@ifnextchar [ {\mn@doi@}
  {\mn@doi@[]}}
\def\mn@doi@[#1]#2{\def\@tempa{#1}\ifx\@tempa\@empty \href
  {http://dx.doi.org/#2} {doi:#2}\else \href {http://dx.doi.org/#2} {#1}\fi
  \endgroup}
\def\mn@eprint#1#2{\mn@eprint@#1:#2::\@nil}
\def\mn@eprint@arXiv#1{\href {http://arxiv.org/abs/#1} {{\tt arXiv:#1}}}
\def\mn@eprint@dblp#1{\href {http://dblp.uni-trier.de/rec/bibtex/#1.xml}
  {dblp:#1}}
\def\mn@eprint@#1:#2:#3:#4\@nil{\def\@tempa {#1}\def\@tempb {#2}\def\@tempc
  {#3}\ifx \@tempc \@empty \let \@tempc \@tempb \let \@tempb \@tempa \fi \ifx
  \@tempb \@empty \def\@tempb {arXiv}\fi \@ifundefined
  {mn@eprint@\@tempb}{\@tempb:\@tempc}{\expandafter \expandafter \csname
  mn@eprint@\@tempb\endcsname \expandafter{\@tempc}}}

\bibitem[\protect\citeauthoryear{{Abazajian} et~al.,}{{Abazajian}
  et~al.}{2009}]{Abazajian_2009}
{Abazajian} K.~N.,  et~al., 2009, \mn@doi [\apjs]
  {10.1088/0067-0049/182/2/543}, \href
  {http://adsabs.harvard.edu/abs/2009ApJS..182..543A} {182, 543}

\bibitem[\protect\citeauthoryear{{Antonucci}}{{Antonucci}}{1993}]{Antonucci_1993}
{Antonucci} R.,  1993, \mn@doi [\araa] {10.1146/annurev.aa.31.090193.002353},
  \href {http://adsabs.harvard.edu/abs/1993ARA%26A..31..473A} {31, 473}

\bibitem[\protect\citeauthoryear{{Antonucci} \& {Miller}}{{Antonucci} \&
  {Miller}}{1985}]{Antonucci_1985}
{Antonucci} R.~R.~J.,  {Miller} J.~S.,  1985, \mn@doi [\apj] {10.1086/163559},
  \href {http://adsabs.harvard.edu/abs/1985ApJ...297..621A} {297, 621}

\bibitem[\protect\citeauthoryear{{Arav}, {Moe}, {Costantini}, {Korista}, {Benn}
   \& {Ellison}}{{Arav} et~al.}{2008}]{Arav_2008}
{Arav} N.,  {Moe} M.,  {Costantini} E.,  {Korista} K.~T.,  {Benn} C.,
  {Ellison} S.,  2008, \mn@doi [\apj] {10.1086/588651}, \href
  {http://adsabs.harvard.edu/abs/2008ApJ...681..954A} {681, 954}

\bibitem[\protect\citeauthoryear{{Arshakian}}{{Arshakian}}{2005}]{Arshakian_2005}
{Arshakian} T.~G.,  2005, \mn@doi [\aap] {10.1051/0004-6361:20042341}, \href
  {http://adsabs.harvard.edu/abs/2005A%26A...436..817A} {436, 817}

\bibitem[\protect\citeauthoryear{{Assef} et~al.,}{{Assef}
  et~al.}{2015}]{Assef_2015}
{Assef} R.~J.,  et~al., 2015, \mn@doi [\apj] {10.1088/0004-637X/804/1/27},
  \href {http://adsabs.harvard.edu/abs/2015ApJ...804...27A} {804, 27}

\bibitem[\protect\citeauthoryear{{Bessiere}, {Tadhunter}, {Ramos Almeida}  \&
  {Villar Mart{\'{\i}}n}}{{Bessiere} et~al.}{2012}]{Bessiere_2012}
{Bessiere} P.~S.,  {Tadhunter} C.~N.,  {Ramos Almeida} C.,   {Villar
  Mart{\'{\i}}n} M.,  2012, \mn@doi [\mnras]
  {10.1111/j.1365-2966.2012.21701.x}, \href
  {http://adsabs.harvard.edu/abs/2012MNRAS.426..276B} {426, 276}

\bibitem[\protect\citeauthoryear{{Bian}, {Gu}, {Zhao}, {Chao}  \& {Cui}}{{Bian}
  et~al.}{2006}]{Bian_2006}
{Bian} W.,  {Gu} Q.,  {Zhao} Y.,  {Chao} L.,   {Cui} Q.,  2006, \mn@doi
  [\mnras] {10.1111/j.1365-2966.2006.10915.x}, \href
  {http://adsabs.harvard.edu/abs/2006MNRAS.372..876B} {372, 876}

\bibitem[\protect\citeauthoryear{{Borguet}, {Hutsem{\'e}kers}, {Letawe},
  {Letawe}  \& {Magain}}{{Borguet} et~al.}{2008}]{Borguet_2008}
{Borguet} B.,  {Hutsem{\'e}kers} D.,  {Letawe} G.,  {Letawe} Y.,   {Magain} P.,
   2008, \mn@doi [\aap] {10.1051/0004-6361:20077986}, \href
  {http://adsabs.harvard.edu/abs/2008A%26A...478..321B} {478, 321}

\bibitem[\protect\citeauthoryear{{Brinchmann}, {Charlot}, {White}, {Tremonti},
  {Kauffmann}, {Heckman}  \& {Brinkmann}}{{Brinchmann}
  et~al.}{2004}]{Brinchmann_2004}
{Brinchmann} J.,  {Charlot} S.,  {White} S.~D.~M.,  {Tremonti} C.,  {Kauffmann}
  G.,  {Heckman} T.,   {Brinkmann} J.,  2004, \mn@doi [\mnras]
  {10.1111/j.1365-2966.2004.07881.x}, \href
  {http://adsabs.harvard.edu/abs/2004MNRAS.351.1151B} {351, 1151}

\bibitem[\protect\citeauthoryear{{Brusa} et~al.,}{{Brusa}
  et~al.}{2015}]{Brusa_2015}
{Brusa} M.,  et~al., 2015, preprint, \href
  {http://adsabs.harvard.edu/abs/2015arXiv150301783B} {} (\mn@eprint {arXiv}
  {1503.01783})

\bibitem[\protect\citeauthoryear{{Bruzual} \& {Charlot}}{{Bruzual} \&
  {Charlot}}{2003}]{Bruzual_2003}
{Bruzual} G.,  {Charlot} S.,  2003, \mn@doi [\mnras]
  {10.1046/j.1365-8711.2003.06897.x}, \href
  {http://adsabs.harvard.edu/abs/2003MNRAS.344.1000B} {344, 1000}

\bibitem[\protect\citeauthoryear{{Calzetti}, {Armus}, {Bohlin}, {Kinney},
  {Koornneef}  \& {Storchi-Bergmann}}{{Calzetti} et~al.}{2000}]{Calzetti_2000}
{Calzetti} D.,  {Armus} L.,  {Bohlin} R.~C.,  {Kinney} A.~L.,  {Koornneef} J.,
   {Storchi-Bergmann} T.,  2000, \mn@doi [\apj] {10.1086/308692}, \href
  {http://adsabs.harvard.edu/abs/2000ApJ...533..682C} {533, 682}

\bibitem[\protect\citeauthoryear{{Casey}}{{Casey}}{2012}]{Casey_2012}
{Casey} C.~M.,  2012, \mn@doi [\mnras] {10.1111/j.1365-2966.2012.21455.x},
  \href {http://adsabs.harvard.edu/abs/2012MNRAS.425.3094C} {425, 3094}

\bibitem[\protect\citeauthoryear{{Chambers}, {Miley}  \& {van
  Breugel}}{{Chambers} et~al.}{1987}]{Chambers_1987}
{Chambers} K.~C.,  {Miley} G.~K.,   {van Breugel} W.,  1987, \mn@doi [\nat]
  {10.1038/329604a0}, \href {http://adsabs.harvard.edu/abs/1987Natur.329..604C}
  {329, 604}

\bibitem[\protect\citeauthoryear{{Ciesla} et~al.,}{{Ciesla}
  et~al.}{2015}]{Ciesla_2015}
{Ciesla} L.,  et~al., 2015, \mn@doi [\aap] {10.1051/0004-6361/201425252}, \href
  {http://adsabs.harvard.edu/abs/2015A%26A...576A..10C} {576, A10}

\bibitem[\protect\citeauthoryear{{Cimatti}, {Dey}, {Breugel}, {Hurt}  \&
  {Antonucci}}{{Cimatti} et~al.}{1997}]{Cimatti_1997}
{Cimatti} A.,  {Dey} A.,  {Breugel} W.~v.,  {Hurt} T.,   {Antonucci} R.,  1997,
  \apj, \href {http://adsabs.harvard.edu/abs/1997ApJ...476..677C} {476, 677}

\bibitem[\protect\citeauthoryear{{Crenshaw}, {Kraemer}  \& {Gabel}}{{Crenshaw}
  et~al.}{2003}]{Crenshaw_2003}
{Crenshaw} D.~M.,  {Kraemer} S.~B.,   {Gabel} J.~R.,  2003, \mn@doi [\aj]
  {10.1086/377625}, \href {http://adsabs.harvard.edu/abs/2003AJ....126.1690C}
  {126, 1690}

\bibitem[\protect\citeauthoryear{{Croom} et~al.,}{{Croom}
  et~al.}{2009}]{Croom_2009}
{Croom} S.~M.,  et~al., 2009, \mn@doi [\mnras]
  {10.1111/j.1365-2966.2009.15398.x}, \href
  {http://adsabs.harvard.edu/abs/2009MNRAS.399.1755C} {399, 1755}

\bibitem[\protect\citeauthoryear{{Croton} et~al.,}{{Croton}
  et~al.}{2006}]{Croton_2006}
{Croton} D.~J.,  et~al., 2006, \mn@doi [\mnras]
  {10.1111/j.1365-2966.2005.09675.x}, \href
  {http://adsabs.harvard.edu/abs/2006MNRAS.365...11C} {365, 11}

\bibitem[\protect\citeauthoryear{{Daddi} et~al.,}{{Daddi}
  et~al.}{2007}]{Daddi_2007}
{Daddi} E.,  et~al., 2007, \mn@doi [\apj] {10.1086/521818}, \href
  {http://adsabs.harvard.edu/abs/2007ApJ...670..156D} {670, 156}

\bibitem[\protect\citeauthoryear{{Dale} \& {Helou}}{{Dale} \&
  {Helou}}{2002}]{Dale_2002}
{Dale} D.~A.,  {Helou} G.,  2002, \mn@doi [\apj] {10.1086/341632}, \href
  {http://adsabs.harvard.edu/abs/2002ApJ...576..159D} {576, 159}

\bibitem[\protect\citeauthoryear{{Dale}, {Helou}, {Magdis}, {Armus},
  {D{\'{\i}}az-Santos}  \& {Shi}}{{Dale} et~al.}{2014}]{Dale_2014}
{Dale} D.~A.,  {Helou} G.,  {Magdis} G.~E.,  {Armus} L.,  {D{\'{\i}}az-Santos}
  T.,   {Shi} Y.,  2014, \mn@doi [\apj] {10.1088/0004-637X/784/1/83}, \href
  {http://adsabs.harvard.edu/abs/2014ApJ...784...83D} {784, 83}

\bibitem[\protect\citeauthoryear{{Dey}, {van Breugel}, {Vacca}  \&
  {Antonucci}}{{Dey} et~al.}{1997}]{Dey_1997}
{Dey} A.,  {van Breugel} W.,  {Vacca} W.~D.,   {Antonucci} R.,  1997, \apj,
  \href {http://adsabs.harvard.edu/abs/1997ApJ...490..698D} {490, 698}

\bibitem[\protect\citeauthoryear{{DiPompeo}, {Myers}, {Hickox}, {Geach}  \&
  {Hainline}}{{DiPompeo} et~al.}{2014}]{Dipompeo_2014}
{DiPompeo} M.~A.,  {Myers} A.~D.,  {Hickox} R.~C.,  {Geach} J.~E.,   {Hainline}
  K.~N.,  2014, \mn@doi [\mnras] {10.1093/mnras/stu1115}, \href
  {http://adsabs.harvard.edu/abs/2014MNRAS.442.3443D} {442, 3443}

\bibitem[\protect\citeauthoryear{{Draine} \& {Li}}{{Draine} \&
  {Li}}{2007}]{Draine_2007}
{Draine} B.~T.,  {Li} A.,  2007, \mn@doi [\apj] {10.1086/511055}, \href
  {http://adsabs.harvard.edu/abs/2007ApJ...657..810D} {657, 810}

\bibitem[\protect\citeauthoryear{{Drouart} et~al.,}{{Drouart}
  et~al.}{2014}]{Drouart_2014}
{Drouart} G.,  et~al., 2014, \mn@doi [\aap] {10.1051/0004-6361/201323310},
  \href {http://adsabs.harvard.edu/abs/2014A%26A...566A..53D} {566, A53}

\bibitem[\protect\citeauthoryear{{Duc} \& {Renaud}}{{Duc} \&
  {Renaud}}{2013}]{Duc_2013}
{Duc} P.-A.,  {Renaud} F.,  2013, in {Souchay} J.,  {Mathis} S.,   {Tokieda}
  T.,  eds,  Lecture Notes in Physics, Berlin Springer Verlag Vol. 861, Lecture
  Notes in Physics, Berlin Springer Verlag. p.~327 (\mn@eprint {arXiv}
  {1112.1922}), \mn@doi{10.1007/978-3-642-32961-6_9}

\bibitem[\protect\citeauthoryear{{Dunlop}, {McLure}, {Kukula}, {Baum}, {O'Dea}
  \& {Hughes}}{{Dunlop} et~al.}{2003}]{Dunlop_2003}
{Dunlop} J.~S.,  {McLure} R.~J.,  {Kukula} M.~J.,  {Baum} S.~A.,  {O'Dea}
  C.~P.,   {Hughes} D.~H.,  2003, \mn@doi [\mnras]
  {10.1046/j.1365-8711.2003.06333.x}, \href
  {http://adsabs.harvard.edu/abs/2003MNRAS.340.1095D} {340, 1095}

\bibitem[\protect\citeauthoryear{{Dunne} \& {Eales}}{{Dunne} \&
  {Eales}}{2001}]{Dunne_2001}
{Dunne} L.,  {Eales} S.~A.,  2001, \mn@doi [\mnras]
  {10.1046/j.1365-8711.2001.04789.x}, \href
  {http://adsabs.harvard.edu/abs/2001MNRAS.327..697D} {327, 697}

\bibitem[\protect\citeauthoryear{{Ferrarese} \& {Merritt}}{{Ferrarese} \&
  {Merritt}}{2000}]{Ferrarese_2000}
{Ferrarese} L.,  {Merritt} D.,  2000, \mn@doi [\apjl] {10.1086/312838}, \href
  {http://adsabs.harvard.edu/abs/2000ApJ...539L...9F} {539, L9}

\bibitem[\protect\citeauthoryear{{Floyd}, {Kukula}, {Dunlop}, {McLure},
  {Miller}, {Percival}, {Baum}  \& {O'Dea}}{{Floyd} et~al.}{2004}]{Floyd_2004}
{Floyd} D.~J.~E.,  {Kukula} M.~J.,  {Dunlop} J.~S.,  {McLure} R.~J.,  {Miller}
  L.,  {Percival} W.~J.,  {Baum} S.~A.,   {O'Dea} C.~P.,  2004, \mn@doi
  [\mnras] {10.1111/j.1365-2966.2004.08315.x}, \href
  {http://adsabs.harvard.edu/abs/2004MNRAS.355..196F} {355, 196}

\bibitem[\protect\citeauthoryear{{Fritz}, {Franceschini}  \&
  {Hatziminaoglou}}{{Fritz} et~al.}{2006}]{Fritz_2006}
{Fritz} J.,  {Franceschini} A.,   {Hatziminaoglou} E.,  2006, \mn@doi [\mnras]
  {10.1111/j.1365-2966.2006.09866.x}, \href
  {http://adsabs.harvard.edu/abs/2006MNRAS.366..767F} {366, 767}

\bibitem[\protect\citeauthoryear{{Gebhardt} et~al.,}{{Gebhardt}
  et~al.}{2000}]{Gebhardt_2000}
{Gebhardt} K.,  et~al., 2000, \mn@doi [\apjl] {10.1086/312840}, \href
  {http://adsabs.harvard.edu/abs/2000ApJ...539L..13G} {539, L13}

\bibitem[\protect\citeauthoryear{{Glikman}, {Simmons}, {Mailly}, {Schawinski},
  {Urry}  \& {Lacy}}{{Glikman} et~al.}{2015}]{Glikman_2015}
{Glikman} E.,  {Simmons} B.,  {Mailly} M.,  {Schawinski} K.,  {Urry} C.~M.,
  {Lacy} M.,  2015, \mn@doi [\apj] {10.1088/0004-637X/806/2/218}, \href
  {http://adsabs.harvard.edu/abs/2015ApJ...806..218G} {806, 218}

\bibitem[\protect\citeauthoryear{{Graham}}{{Graham}}{2013}]{Graham_2013}
{Graham} A.~W.,  2013, {Elliptical and Disk Galaxy Structure and Modern Scaling
  Laws}.
p.~91, \mn@doi{10.1007/978-94-007-5609-0_2}

\bibitem[\protect\citeauthoryear{{Guo}, {Zheng}  \& {Fu}}{{Guo}
  et~al.}{2013}]{Guo_2013}
{Guo} K.,  {Zheng} X.~Z.,   {Fu} H.,  2013, \mn@doi [\apj]
  {10.1088/0004-637X/778/1/23}, \href
  {http://adsabs.harvard.edu/abs/2013ApJ...778...23G} {778, 23}

\bibitem[\protect\citeauthoryear{{Hildebrand}}{{Hildebrand}}{1983}]{Hildebrand_1983}
{Hildebrand} R.~H.,  1983, \qjras, \href
  {http://adsabs.harvard.edu/abs/1983QJRAS..24..267H} {24, 267}

\bibitem[\protect\citeauthoryear{{Hopkins}, {Hernquist}, {Cox}, {Robertson}  \&
  {Springel}}{{Hopkins} et~al.}{2006}]{Hopkins_2006}
{Hopkins} P.~F.,  {Hernquist} L.,  {Cox} T.~J.,  {Robertson} B.,   {Springel}
  V.,  2006, \mn@doi [\apjs] {10.1086/499493}, \href
  {http://adsabs.harvard.edu/abs/2006ApJS..163...50H} {163, 50}

\bibitem[\protect\citeauthoryear{{Hopkins}, {Hernquist}, {Cox}  \& {Kere{\v
  s}}}{{Hopkins} et~al.}{2008}]{Hopkins_2008}
{Hopkins} P.~F.,  {Hernquist} L.,  {Cox} T.~J.,   {Kere{\v s}} D.,  2008,
  \mn@doi [\apjs] {10.1086/524362}, \href
  {http://adsabs.harvard.edu/abs/2008ApJS..175..356H} {175, 356}

\bibitem[\protect\citeauthoryear{{Hopkins}, {Bundy}, {Murray}, {Quataert},
  {Lauer}  \& {Ma}}{{Hopkins} et~al.}{2009}]{Hopkins_2009}
{Hopkins} P.~F.,  {Bundy} K.,  {Murray} N.,  {Quataert} E.,  {Lauer} T.~R.,
  {Ma} C.-P.,  2009, \mn@doi [\mnras] {10.1111/j.1365-2966.2009.15062.x}, \href
  {http://adsabs.harvard.edu/abs/2009MNRAS.398..898H} {398, 898}

\bibitem[\protect\citeauthoryear{{Ilbert} et~al.,}{{Ilbert}
  et~al.}{2010}]{Ilbert_2010}
{Ilbert} O.,  et~al., 2010, \mn@doi [\apj] {10.1088/0004-637X/709/2/644}, \href
  {http://adsabs.harvard.edu/abs/2010ApJ...709..644I} {709, 644}

\bibitem[\protect\citeauthoryear{{Kennicutt}}{{Kennicutt}}{1998}]{Kennicutt_1998}
{Kennicutt} Jr. R.~C.,  1998, \mn@doi [ApJ] {10.1086/305588}, \href
  {http://adsabs.harvard.edu/abs/1998ApJ...498..541K} {498, 541}

\bibitem[\protect\citeauthoryear{{Kennicutt}, {Schweizer}  \&
  {Barnes}}{{Kennicutt} et~al.}{1996}]{Kennicutt_1996}
{Kennicutt} Jr. R.~C.,  {Schweizer} F.,   {Barnes} J.~E.,  1996, {Galaxies:
  Interactions and Induced Star Formation}

\bibitem[\protect\citeauthoryear{{Kirkpatrick} et~al.,}{{Kirkpatrick}
  et~al.}{2014}]{Kirkpatrick_2014}
{Kirkpatrick} A.,  et~al., 2014, \mn@doi [\apj] {10.1088/0004-637X/789/2/130},
  \href {http://adsabs.harvard.edu/abs/2014ApJ...789..130K} {789, 130}

\bibitem[\protect\citeauthoryear{{Kollmeier} et~al.,}{{Kollmeier}
  et~al.}{2006}]{Kollmeier_2006}
{Kollmeier} J.~A.,  et~al., 2006, \mn@doi [\apj] {10.1086/505646}, \href
  {http://adsabs.harvard.edu/abs/2006ApJ...648..128K} {648, 128}

\bibitem[\protect\citeauthoryear{{Koulouridis}, {Plionis}, {Chavushyan},
  {Dultzin-Hacyan}, {Krongold}  \& {Goudis}}{{Koulouridis}
  et~al.}{2006}]{Koulouridis_2006}
{Koulouridis} E.,  {Plionis} M.,  {Chavushyan} V.,  {Dultzin-Hacyan} D.,
  {Krongold} Y.,   {Goudis} C.,  2006, \mn@doi [\apj] {10.1086/498421}, \href
  {http://adsabs.harvard.edu/abs/2006ApJ...639...37K} {639, 37}

\bibitem[\protect\citeauthoryear{{Lacy}, {Sajina}, {Petric}, {Seymour},
  {Canalizo}, {Ridgway}, {Armus}  \& {Storrie-Lombardi}}{{Lacy}
  et~al.}{2007}]{Lacy_2007}
{Lacy} M.,  {Sajina} A.,  {Petric} A.~O.,  {Seymour} N.,  {Canalizo} G.,
  {Ridgway} S.~E.,  {Armus} L.,   {Storrie-Lombardi} L.~J.,  2007, \mn@doi
  [\apjl] {10.1086/523851}, \href
  {http://adsabs.harvard.edu/abs/2007ApJ...669L..61L} {669, L61}

\bibitem[\protect\citeauthoryear{{Lamastra}, {Bianchi}, {Matt}, {Perola},
  {Barcons}  \& {Carrera}}{{Lamastra} et~al.}{2009}]{Lamastra_2009}
{Lamastra} A.,  {Bianchi} S.,  {Matt} G.,  {Perola} G.~C.,  {Barcons} X.,
  {Carrera} F.~J.,  2009, \mn@doi [\aap] {10.1051/0004-6361/200912023}, \href
  {http://adsabs.harvard.edu/abs/2009A%26A...504...73L} {504, 73}

\bibitem[\protect\citeauthoryear{{Lawrence}}{{Lawrence}}{1991}]{Lawrence_1991}
{Lawrence} A.,  1991, \mnras, \href
  {http://adsabs.harvard.edu/abs/1991MNRAS.252..586L} {252, 586}

\bibitem[\protect\citeauthoryear{{Leitherer}, {Calzetti}  \&
  {Martins}}{{Leitherer} et~al.}{2002}]{Leitherer_2002}
{Leitherer} C.,  {Calzetti} D.,   {Martins} L.~P.,  2002, \mn@doi [\apj]
  {10.1086/340902}, \href {http://adsabs.harvard.edu/abs/2002ApJ...574..114L}
  {574, 114}

\bibitem[\protect\citeauthoryear{{Lena}}{{Lena}}{2014}]{Lena_2014}
{Lena} D.,  2014, preprint, \href
  {http://adsabs.harvard.edu/abs/2014arXiv1409.8264L} {} (\mn@eprint {arXiv}
  {1409.8264})

\bibitem[\protect\citeauthoryear{{Lena} et~al.,}{{Lena}
  et~al.}{2015}]{Lena_2015}
{Lena} D.,  et~al., 2015, \mn@doi [\apj] {10.1088/0004-637X/806/1/84}, \href
  {http://adsabs.harvard.edu/abs/2015ApJ...806...84L} {806, 84}

\bibitem[\protect\citeauthoryear{{Letawe}, {Magain}, {Courbin}, {Jablonka},
  {Jahnke}, {Meylan}  \& {Wisotzki}}{{Letawe} et~al.}{2007}]{Letawe_2007}
{Letawe} G.,  {Magain} P.,  {Courbin} F.,  {Jablonka} P.,  {Jahnke} K.,
  {Meylan} G.,   {Wisotzki} L.,  2007, \mn@doi [\mnras]
  {10.1111/j.1365-2966.2007.11741.x}, \href
  {http://adsabs.harvard.edu/abs/2007MNRAS.378...83L} {378, 83}

\bibitem[\protect\citeauthoryear{{Leyshon} \& {Eales}}{{Leyshon} \&
  {Eales}}{1998}]{Leyshon_1998}
{Leyshon} G.,  {Eales} S.~A.,  1998, \mn@doi [\mnras]
  {10.1046/j.1365-8711.1998.29511145.x}, \href
  {http://adsabs.harvard.edu/abs/1998MNRAS.295...10L} {295, 10}

\bibitem[\protect\citeauthoryear{{Liu}, {Zakamska}, {Greene}, {Nesvadba}  \&
  {Liu}}{{Liu} et~al.}{2013a}]{Liu_2013a}
{Liu} G.,  {Zakamska} N.~L.,  {Greene} J.~E.,  {Nesvadba} N.~P.~H.,   {Liu} X.,
   2013a, \mn@doi [\mnras] {10.1093/mnras/stt051}, \href
  {http://adsabs.harvard.edu/abs/2013MNRAS.430.2327L} {430, 2327}

\bibitem[\protect\citeauthoryear{{Liu}, {Zakamska}, {Greene}, {Nesvadba}  \&
  {Liu}}{{Liu} et~al.}{2013b}]{Liu_2013b}
{Liu} G.,  {Zakamska} N.~L.,  {Greene} J.~E.,  {Nesvadba} N.~P.~H.,   {Liu} X.,
   2013b, \mn@doi [\mnras] {10.1093/mnras/stt1755}, \href
  {http://adsabs.harvard.edu/abs/2013MNRAS.436.2576L} {436, 2576}

\bibitem[\protect\citeauthoryear{{Lupton}, {Blanton}, {Fekete}, {Hogg},
  {O'Mullane}, {Szalay}  \& {Wherry}}{{Lupton} et~al.}{2004}]{Lupton_2004}
{Lupton} R.,  {Blanton} M.~R.,  {Fekete} G.,  {Hogg} D.~W.,  {O'Mullane} W.,
  {Szalay} A.,   {Wherry} N.,  2004, \mn@doi [\pasp] {10.1086/382245}, \href
  {http://adsabs.harvard.edu/abs/2004PASP..116..133L} {116, 133}

\bibitem[\protect\citeauthoryear{{Magorrian} et~al.,}{{Magorrian}
  et~al.}{1998}]{Magorrian_1998}
{Magorrian} J.,  et~al., 1998, \mn@doi [\aj] {10.1086/300353}, \href
  {http://adsabs.harvard.edu/abs/1998AJ....115.2285M} {115, 2285}

\bibitem[\protect\citeauthoryear{{Maraston}}{{Maraston}}{2005}]{Maraston_2005}
{Maraston} C.,  2005, \mn@doi [\mnras] {10.1111/j.1365-2966.2005.09270.x},
  \href {http://adsabs.harvard.edu/abs/2005MNRAS.362..799M} {362, 799}

\bibitem[\protect\citeauthoryear{{Markowitz}, {Krumpe}  \&
  {Nikutta}}{{Markowitz} et~al.}{2014}]{Markowitz_2014}
{Markowitz} A.~G.,  {Krumpe} M.,   {Nikutta} R.,  2014, \mn@doi [\mnras]
  {10.1093/mnras/stt2492}, \href
  {http://adsabs.harvard.edu/abs/2014MNRAS.439.1403M} {439, 1403}

\bibitem[\protect\citeauthoryear{{McCarthy}, {van Breugel}, {Spinrad}  \&
  {Djorgovski}}{{McCarthy} et~al.}{1987}]{McCarthy_1987}
{McCarthy} P.~J.,  {van Breugel} W.,  {Spinrad} H.,   {Djorgovski} S.,  1987,
  \mn@doi [\apjl] {10.1086/185000}, \href
  {http://adsabs.harvard.edu/abs/1987ApJ...321L..29M} {321, L29}

\bibitem[\protect\citeauthoryear{{Menci}, {Fiore}, {Puccetti}  \&
  {Cavaliere}}{{Menci} et~al.}{2008}]{Menci_2008}
{Menci} N.,  {Fiore} F.,  {Puccetti} S.,   {Cavaliere} A.,  2008, \mn@doi
  [\apj] {10.1086/591438}, \href
  {http://adsabs.harvard.edu/abs/2008ApJ...686..219M} {686, 219}

\bibitem[\protect\citeauthoryear{{Moe}, {Arav}, {Bautista}  \& {Korista}}{{Moe}
  et~al.}{2009}]{Moe_2009}
{Moe} M.,  {Arav} N.,  {Bautista} M.~A.,   {Korista} K.~T.,  2009, \mn@doi
  [\apj] {10.1088/0004-637X/706/1/525}, \href
  {http://adsabs.harvard.edu/abs/2009ApJ...706..525M} {706, 525}

\bibitem[\protect\citeauthoryear{{Mortlock} et~al.,}{{Mortlock}
  et~al.}{2015}]{Mortlock_2015}
{Mortlock} A.,  et~al., 2015, \mn@doi [\mnras] {10.1093/mnras/stu2403}, \href
  {http://adsabs.harvard.edu/abs/2015MNRAS.447....2M} {447, 2}

\bibitem[\protect\citeauthoryear{{Mullaney}, {Alexander}, {Goulding}  \&
  {Hickox}}{{Mullaney} et~al.}{2011}]{Mullaney_2011}
{Mullaney} J.~R.,  {Alexander} D.~M.,  {Goulding} A.~D.,   {Hickox} R.~C.,
  2011, \mn@doi [\mnras] {10.1111/j.1365-2966.2011.18448.x}, \href
  {http://adsabs.harvard.edu/abs/2011MNRAS.414.1082M} {414, 1082}

\bibitem[\protect\citeauthoryear{{Nesvadba}, {Lehnert}, {De Breuck}, {Gilbert}
  \& {van Breugel}}{{Nesvadba} et~al.}{2008}]{Nesvadba_2008}
{Nesvadba} N.~P.~H.,  {Lehnert} M.~D.,  {De Breuck} C.,  {Gilbert} A.~M.,
  {van Breugel} W.,  2008, \mn@doi [\aap] {10.1051/0004-6361:200810346}, \href
  {http://adsabs.harvard.edu/abs/2008A%26A...491..407N} {491, 407}

\bibitem[\protect\citeauthoryear{{Nikutta}, {Elitzur}  \& {Lacy}}{{Nikutta}
  et~al.}{2009}]{Nikutta_2009}
{Nikutta} R.,  {Elitzur} M.,   {Lacy} M.,  2009, \mn@doi [\apj]
  {10.1088/0004-637X/707/2/1550}, \href
  {http://adsabs.harvard.edu/abs/2009ApJ...707.1550N} {707, 1550}

\bibitem[\protect\citeauthoryear{{Noll}, {Burgarella}, {Giovannoli}, {Buat},
  {Marcillac}  \& {Mu{\~n}oz-Mateos}}{{Noll} et~al.}{2009}]{Noll_2009}
{Noll} S.,  {Burgarella} D.,  {Giovannoli} E.,  {Buat} V.,  {Marcillac} D.,
  {Mu{\~n}oz-Mateos} J.~C.,  2009, \mn@doi [\aap]
  {10.1051/0004-6361/200912497}, \href
  {http://adsabs.harvard.edu/abs/2009A%26A...507.1793N} {507, 1793}

\bibitem[\protect\citeauthoryear{{O'Dea} et~al.,}{{O'Dea}
  et~al.}{2002}]{Odea_2002}
{O'Dea} C.~P.,  et~al., 2002, \mn@doi [\aj] {10.1086/340076}, \href
  {http://adsabs.harvard.edu/abs/2002AJ....123.2333O} {123, 2333}

\bibitem[\protect\citeauthoryear{{Obied}, {Zakamska}, {Wylezalek}  \&
  {Liu}}{{Obied} et~al.}{2015}]{Obied_2015}
{Obied} G.,  {Zakamska} N.~L.,  {Wylezalek} D.,   {Liu} G.,  2015, \mn@doi
  [\mnras, submitted] {10.1111/j.1365-2966.2009.15062.x}, \href
  {http://adsabs.harvard.edu/abs/2009MNRAS.398..898H} {000, 000}

\bibitem[\protect\citeauthoryear{{Orban de Xivry}, {Davies}, {Schartmann},
  {Komossa}, {Marconi}, {Hicks}, {Engel}  \& {Tacconi}}{{Orban de Xivry}
  et~al.}{2011}]{Orban_2011}
{Orban de Xivry} G.,  {Davies} R.,  {Schartmann} M.,  {Komossa} S.,  {Marconi}
  A.,  {Hicks} E.,  {Engel} H.,   {Tacconi} L.,  2011, \mn@doi [\mnras]
  {10.1111/j.1365-2966.2011.19439.x}, \href
  {http://adsabs.harvard.edu/abs/2011MNRAS.417.2721O} {417, 2721}

\bibitem[\protect\citeauthoryear{{Patterson}}{{Patterson}}{1940}]{Patterson_1940}
{Patterson} F.~S.,  1940, Harvard College Observatory Bulletin, \href
  {http://adsabs.harvard.edu/abs/1940BHarO.914....9P} {914, 9}

\bibitem[\protect\citeauthoryear{{Peng}, {Ho}, {Impey}  \& {Rix}}{{Peng}
  et~al.}{2002}]{Peng_2002}
{Peng} C.~Y.,  {Ho} L.~C.,  {Impey} C.~D.,   {Rix} H.-W.,  2002, \mn@doi [\aj]
  {10.1086/340952}, \href {http://adsabs.harvard.edu/abs/2002AJ....124..266P}
  {124, 266}

\bibitem[\protect\citeauthoryear{{Peng}, {Ho}, {Impey}  \& {Rix}}{{Peng}
  et~al.}{2010}]{Peng_2010}
{Peng} C.~Y.,  {Ho} L.~C.,  {Impey} C.~D.,   {Rix} H.-W.,  2010, \mn@doi [\aj]
  {10.1088/0004-6256/139/6/2097}, \href
  {http://adsabs.harvard.edu/abs/2010AJ....139.2097P} {139, 2097}

\bibitem[\protect\citeauthoryear{{Pier} \& {Krolik}}{{Pier} \&
  {Krolik}}{1992}]{Pier_1992}
{Pier} E.~A.,  {Krolik} J.~H.,  1992, \mn@doi [\apj] {10.1086/172042}, \href
  {http://adsabs.harvard.edu/abs/1992ApJ...401...99P} {401, 99}

\bibitem[\protect\citeauthoryear{{Ramos Almeida} et~al.,}{{Ramos Almeida}
  et~al.}{2012}]{Ramos_Almeida_2012}
{Ramos Almeida} C.,  et~al., 2012, \mn@doi [\mnras]
  {10.1111/j.1365-2966.2011.19731.x}, \href
  {http://adsabs.harvard.edu/abs/2012MNRAS.419..687R} {419, 687}

\bibitem[\protect\citeauthoryear{{Rees}}{{Rees}}{1989}]{Rees_1989}
{Rees} M.~J.,  1989, \mnras, \href
  {http://adsabs.harvard.edu/abs/1989MNRAS.239P...1R} {239, 1P}

\bibitem[\protect\citeauthoryear{{Reyes} et~al.,}{{Reyes}
  et~al.}{2008}]{Reyes_2008}
{Reyes} R.,  et~al., 2008, \mn@doi [\aj] {10.1088/0004-6256/136/6/2373}, \href
  {http://adsabs.harvard.edu/abs/2008AJ....136.2373R} {136, 2373}

\bibitem[\protect\citeauthoryear{{Salviander}, {Shields}  \&
  {Bonning}}{{Salviander} et~al.}{2015}]{Salviander_2015}
{Salviander} S.,  {Shields} G.~A.,   {Bonning} E.~W.,  2015, \mn@doi [\apj]
  {10.1088/0004-637X/799/2/173}, \href
  {http://adsabs.harvard.edu/abs/2015ApJ...799..173S} {799, 173}

\bibitem[\protect\citeauthoryear{{Sanders}, {Soifer}, {Elias}, {Madore},
  {Matthews}, {Neugebauer}  \& {Scoville}}{{Sanders}
  et~al.}{1988}]{Sanders_1988}
{Sanders} D.~B.,  {Soifer} B.~T.,  {Elias} J.~H.,  {Madore} B.~F.,  {Matthews}
  K.,  {Neugebauer} G.,   {Scoville} N.~Z.,  1988, \mn@doi [\apj]
  {10.1086/165983}, \href {http://adsabs.harvard.edu/abs/1988ApJ...325...74S}
  {325, 74}

\bibitem[\protect\citeauthoryear{{Schartmann}, {Meisenheimer}, {Camenzind},
  {Wolf}, {Tristram}  \& {Henning}}{{Schartmann}
  et~al.}{2008}]{Schartmann_2008}
{Schartmann} M.,  {Meisenheimer} K.,  {Camenzind} M.,  {Wolf} S.,  {Tristram}
  K.~R.~W.,   {Henning} T.,  2008, \mn@doi [\aap] {10.1051/0004-6361:20078907},
  \href {http://adsabs.harvard.edu/abs/2008A%26A...482...67S} {482, 67}

\bibitem[\protect\citeauthoryear{{Sersic}}{{Sersic}}{1968}]{Sersic_1968}
{Sersic} J.~L.,  1968, {Atlas de galaxias australes}

\bibitem[\protect\citeauthoryear{{Shen}}{{Shen}}{2013}]{Shen_2013}
{Shen} Y.,  2013, Bulletin of the Astronomical Society of India, \href
  {http://adsabs.harvard.edu/abs/2013BASI...41...61S} {41, 61}

\bibitem[\protect\citeauthoryear{{Silk} \& {Rees}}{{Silk} \&
  {Rees}}{1998}]{Silk_1998}
{Silk} J.,  {Rees} M.~J.,  1998, \aap, \href
  {http://adsabs.harvard.edu/abs/1998A%26A...331L...1S} {331, L1}

\bibitem[\protect\citeauthoryear{{Somerville}, {Hopkins}, {Cox}, {Robertson}
  \& {Hernquist}}{{Somerville} et~al.}{2008}]{Somerville_2008}
{Somerville} R.~S.,  {Hopkins} P.~F.,  {Cox} T.~J.,  {Robertson} B.~E.,
  {Hernquist} L.,  2008, \mn@doi [\mnras] {10.1111/j.1365-2966.2008.13805.x},
  \href {http://adsabs.harvard.edu/abs/2008MNRAS.391..481S} {391, 481}

\bibitem[\protect\citeauthoryear{{Symeonidis}, {Willner}, {Rigopoulou},
  {Huang}, {Fazio}  \& {Jarvis}}{{Symeonidis} et~al.}{2008}]{Symeonidis_2008}
{Symeonidis} M.,  {Willner} S.~P.,  {Rigopoulou} D.,  {Huang} J.-S.,  {Fazio}
  G.~G.,   {Jarvis} M.~J.,  2008, \mn@doi [\mnras]
  {10.1111/j.1365-2966.2008.12899.x}, \href
  {http://adsabs.harvard.edu/abs/2008MNRAS.385.1015S} {385, 1015}

\bibitem[\protect\citeauthoryear{{Tsai} et~al.,}{{Tsai}
  et~al.}{2015}]{Tsai_2015}
{Tsai} C.-W.,  et~al., 2015, \mn@doi [\apj] {10.1088/0004-637X/805/2/90}, \href
  {http://adsabs.harvard.edu/abs/2015ApJ...805...90T} {805, 90}

\bibitem[\protect\citeauthoryear{{Urrutia}, {Lacy}  \& {Becker}}{{Urrutia}
  et~al.}{2008}]{Urrutia_2008}
{Urrutia} T.,  {Lacy} M.,   {Becker} R.~H.,  2008, \mn@doi [\apj]
  {10.1086/523959}, \href {http://adsabs.harvard.edu/abs/2008ApJ...674...80U}
  {674, 80}

\bibitem[\protect\citeauthoryear{{Vernet}, {Fosbury}, {Villar-Mart{\'{\i}}n},
  {Cohen}, {Cimatti}, {di Serego Alighieri}  \& {Goodrich}}{{Vernet}
  et~al.}{2001}]{Vernet_2001}
{Vernet} J.,  {Fosbury} R.~A.~E.,  {Villar-Mart{\'{\i}}n} M.,  {Cohen} M.~H.,
  {Cimatti} A.,  {di Serego Alighieri} S.,   {Goodrich} R.~W.,  2001, \mn@doi
  [\aap] {10.1051/0004-6361:20000076}, \href
  {http://adsabs.harvard.edu/abs/2001A%26A...366....7V} {366, 7}

\bibitem[\protect\citeauthoryear{{Villar-Mart{\'{\i}}n}, {Cabrera Lavers},
  {Bessiere}, {Tadhunter}, {Rose}  \& {de Breuck}}{{Villar-Mart{\'{\i}}n}
  et~al.}{2012}]{Villar_Martin_2012}
{Villar-Mart{\'{\i}}n} M.,  {Cabrera Lavers} A.,  {Bessiere} P.,  {Tadhunter}
  C.,  {Rose} M.,   {de Breuck} C.,  2012, \mn@doi [\mnras]
  {10.1111/j.1365-2966.2012.20652.x}, \href
  {http://adsabs.harvard.edu/abs/2012MNRAS.423...80V} {423, 80}

\bibitem[\protect\citeauthoryear{{Villforth} \& {Hamann}}{{Villforth} \&
  {Hamann}}{2015}]{Villforth_2015}
{Villforth} C.,  {Hamann} F.,  2015, \mn@doi [\aj]
  {10.1088/0004-6256/149/3/92}, \href
  {http://adsabs.harvard.edu/abs/2015AJ....149...92V} {149, 92}

\bibitem[\protect\citeauthoryear{{Villforth} et~al.,}{{Villforth}
  et~al.}{2014}]{Villforth_2014}
{Villforth} C.,  et~al., 2014, \mn@doi [\mnras] {10.1093/mnras/stu173}, \href
  {http://adsabs.harvard.edu/abs/2014MNRAS.439.3342V} {439, 3342}

\bibitem[\protect\citeauthoryear{{Wylezalek} et~al.,}{{Wylezalek}
  et~al.}{2013}]{Wylezalek_2013a}
{Wylezalek} D.,  et~al., 2013, \mn@doi [\mnras] {10.1093/mnras/sts264}, \href
  {http://adsabs.harvard.edu/abs/2013MNRAS.428.3206W} {428, 3206}

\bibitem[\protect\citeauthoryear{{York} et~al.,}{{York}
  et~al.}{2000}]{York_2000}
{York} D.~G.,  et~al., 2000, \mn@doi [\aj] {10.1086/301513}, \href
  {http://adsabs.harvard.edu/abs/2000AJ....120.1579Y} {120, 1579}

\bibitem[\protect\citeauthoryear{{Zakamska}}{{Zakamska}}{2015}]{Zakamska_2015}
{Zakamska} N.~L.,  2015, MNRAS, submitted

\bibitem[\protect\citeauthoryear{{Zakamska} \& {Greene}}{{Zakamska} \&
  {Greene}}{2014}]{Zakamska_2014}
{Zakamska} N.~L.,  {Greene} J.~E.,  2014, \mn@doi [\mnras]
  {10.1093/mnras/stu842}, \href
  {http://adsabs.harvard.edu/abs/2014MNRAS.442..784Z} {442, 784}

\bibitem[\protect\citeauthoryear{{Zakamska} et~al.,}{{Zakamska}
  et~al.}{2003}]{Zakamska_2003}
{Zakamska} N.~L.,  et~al., 2003, \mn@doi [\aj] {10.1086/378610}, \href
  {http://adsabs.harvard.edu/abs/2003AJ....126.2125Z} {126, 2125}

\bibitem[\protect\citeauthoryear{{Zakamska} et~al.,}{{Zakamska}
  et~al.}{2006}]{Zakamska_2006}
{Zakamska} N.~L.,  et~al., 2006, \mn@doi [\aj] {10.1086/506986}, \href
  {http://adsabs.harvard.edu/abs/2006AJ....132.1496Z} {132, 1496}

\bibitem[\protect\citeauthoryear{{de Vaucouleurs}}{{de
  Vaucouleurs}}{1948}]{Vaucouleurs_1948}
{de Vaucouleurs} G.,  1948, Annales d'Astrophysique, \href
  {http://adsabs.harvard.edu/abs/1948AnAp...11..247D} {11, 247}

\bibitem[\protect\citeauthoryear{{van Dokkum}}{{van
  Dokkum}}{2005}]{Dokkum_2005}
{van Dokkum} P.~G.,  2005, \mn@doi [\aj] {10.1086/497593}, \href
  {http://adsabs.harvard.edu/abs/2005AJ....130.2647V} {130, 2647}

\bibitem[\protect\citeauthoryear{{van Dokkum} et~al.,}{{van Dokkum}
  et~al.}{2008}]{Dokkum_2008}
{van Dokkum} P.~G.,  et~al., 2008, \mn@doi [\apjl] {10.1086/587874}, \href
  {http://adsabs.harvard.edu/abs/2008ApJ...677L...5V} {677, L5}

\bibitem[\protect\citeauthoryear{{van der Kruit} \& {Searle}}{{van der Kruit}
  \& {Searle}}{1981}]{Kruit_1981}
{van der Kruit} P.~C.,  {Searle} L.,  1981, \aap, \href
  {http://adsabs.harvard.edu/abs/1981A%26A....95..116V} {95, 116}

\bibitem[\protect\citeauthoryear{{van der Wel}, {Holden}, {Zirm}, {Franx},
  {Rettura}, {Illingworth}  \& {Ford}}{{van der Wel} et~al.}{2008}]{Wel_2008}
{van der Wel} A.,  {Holden} B.~P.,  {Zirm} A.~W.,  {Franx} M.,  {Rettura} A.,
  {Illingworth} G.~D.,   {Ford} H.~C.,  2008, \mn@doi [\apj] {10.1086/592267},
  \href {http://adsabs.harvard.edu/abs/2008ApJ...688...48V} {688, 48}

\makeatother
\end{thebibliography}




\appendix

\section{Improved Gemini GMOS IFU reductions}

\begin{figure*}[H]
\centering
\includegraphics[scale=0.28,clip=clip,trim=0mm 0mm 3.5cm 0mm]{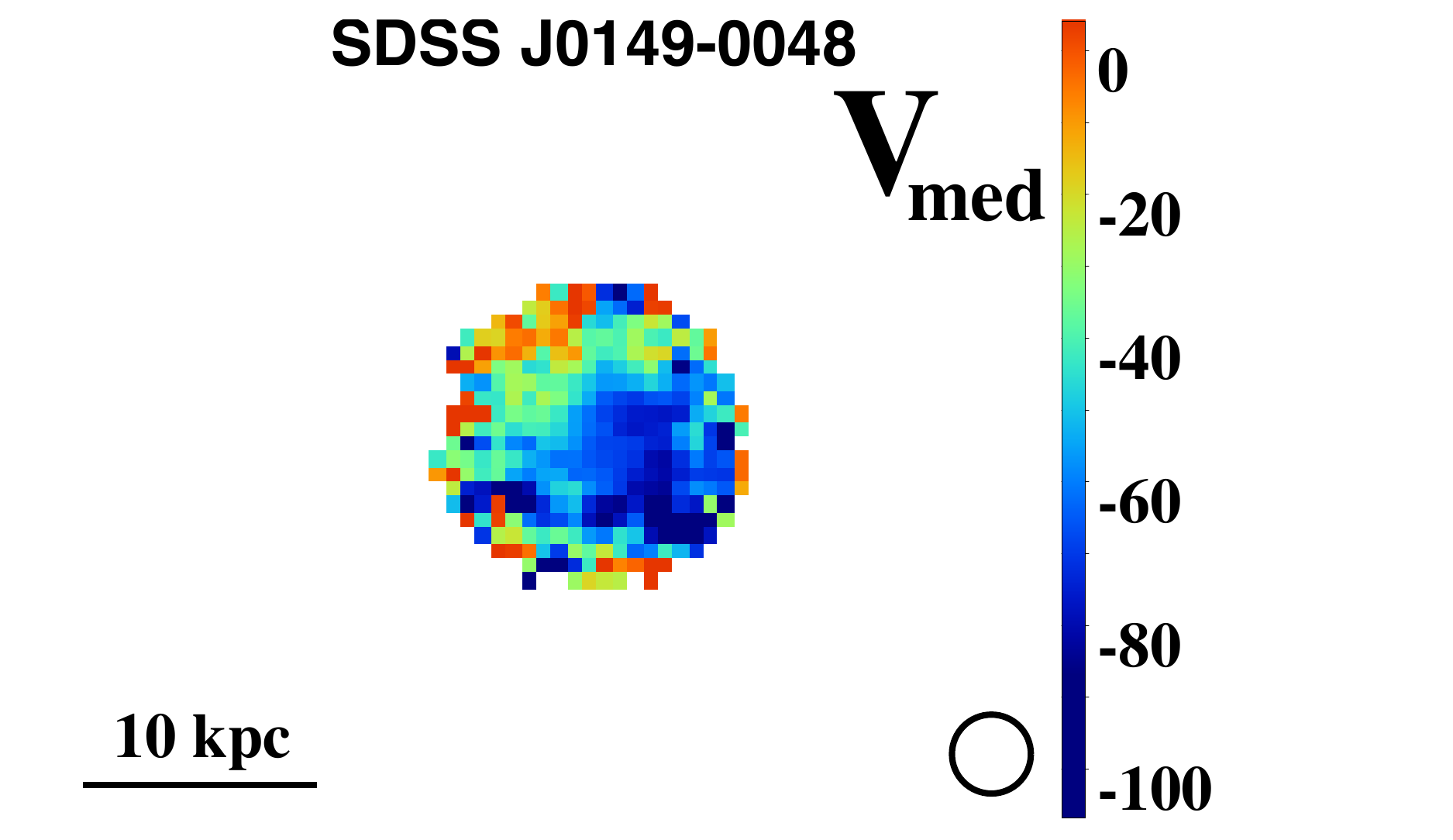}%
\includegraphics[scale=0.28,clip=clip,trim=0mm 0mm 3.5cm 0mm]{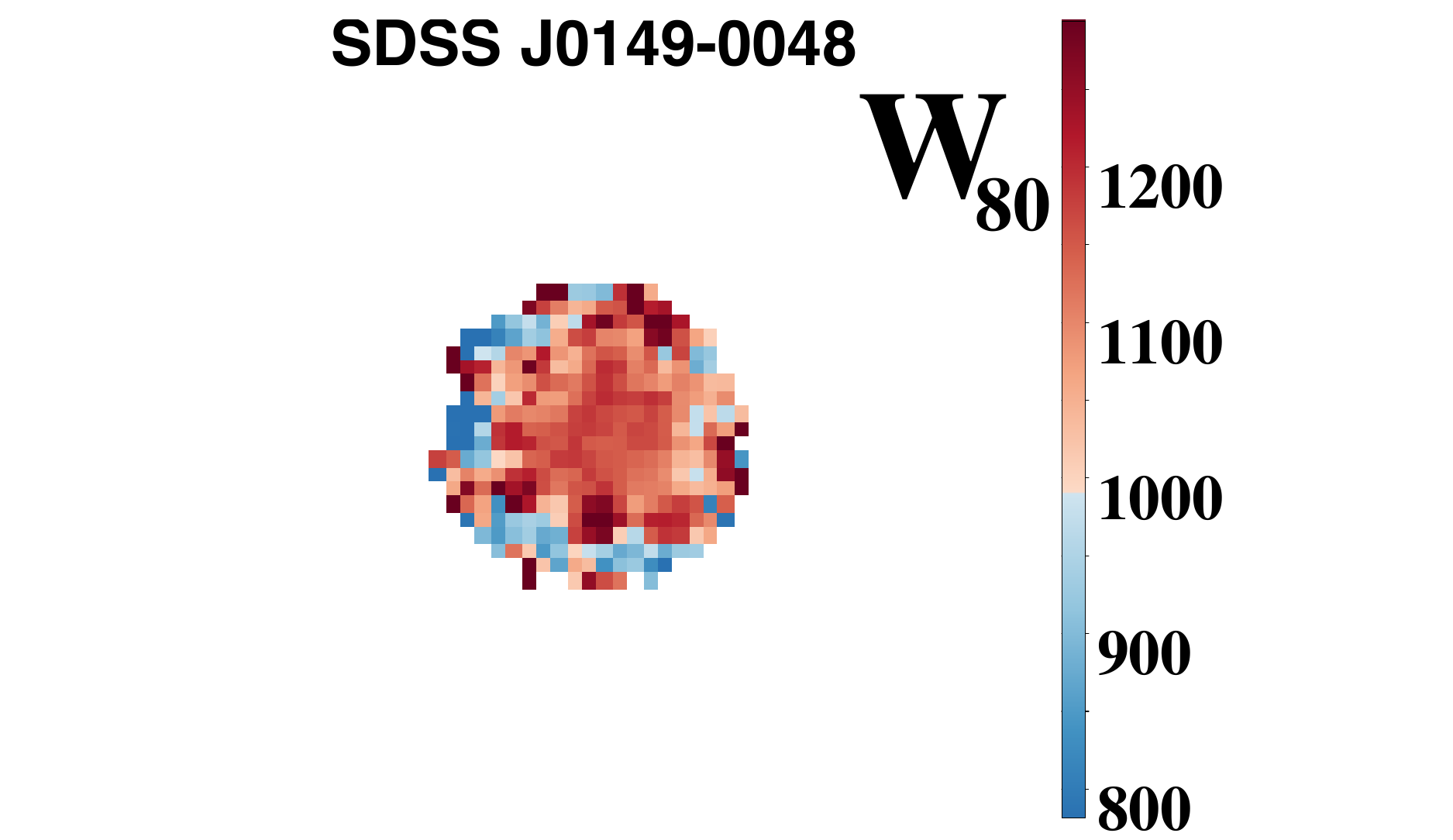}%
\includegraphics[scale=0.28,clip=clip,trim=0mm 0mm 3.5cm 0mm]{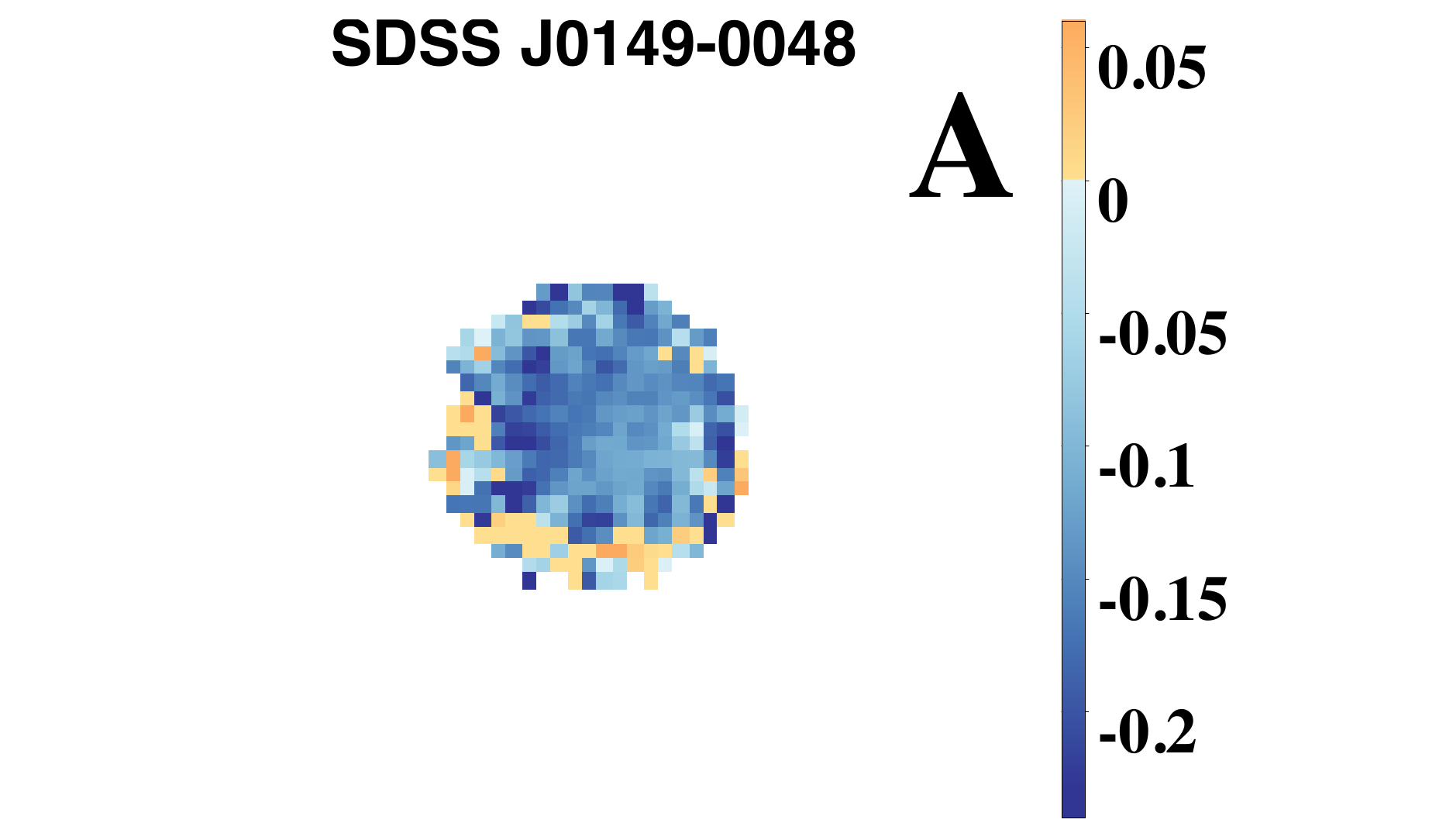}%
\includegraphics[scale=0.28,clip=clip,trim=0mm 0mm 3.5cm 0mm]{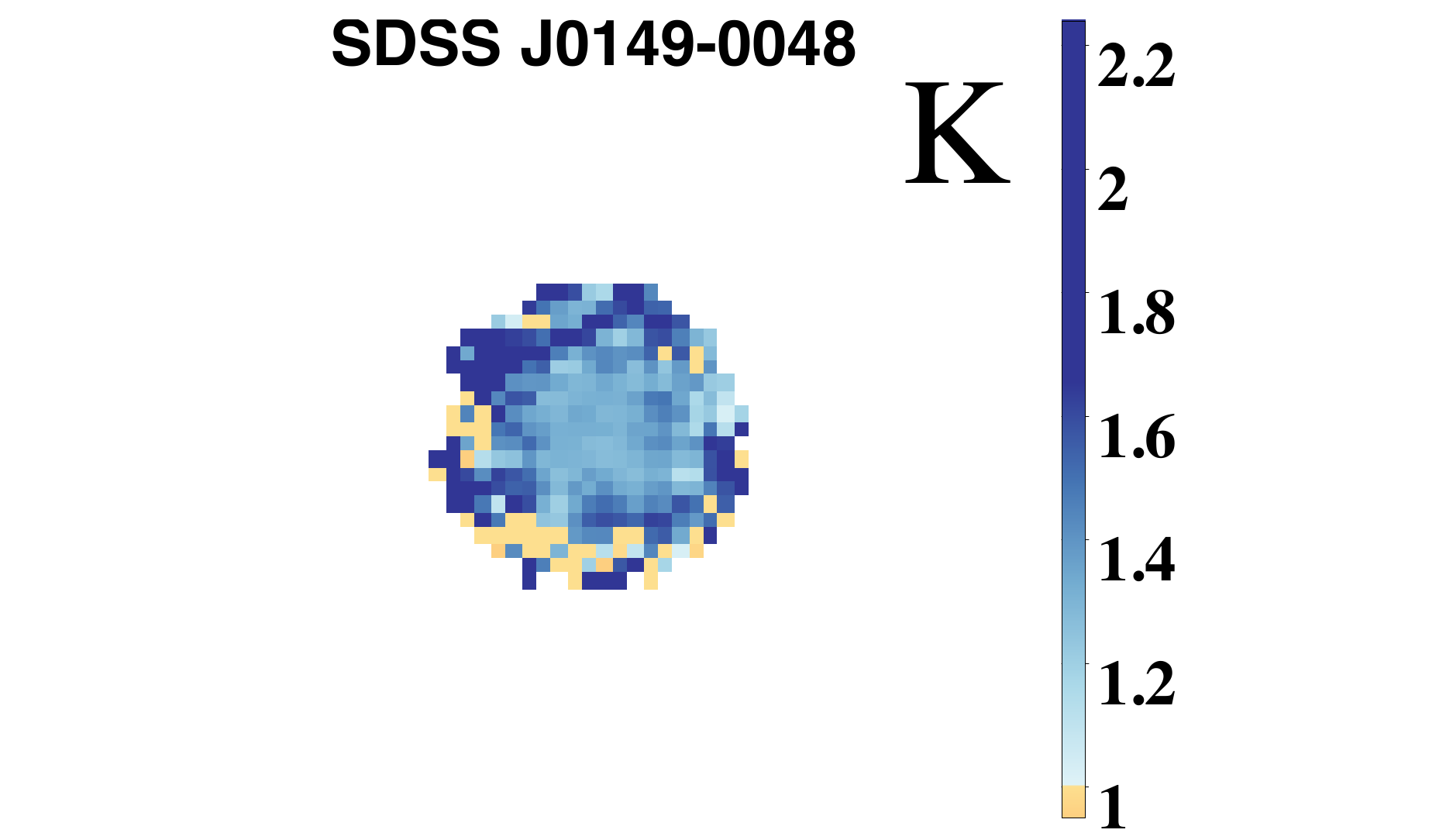}\\
\includegraphics[scale=0.28,clip=clip,trim=0mm 0mm 3.5cm 0mm]{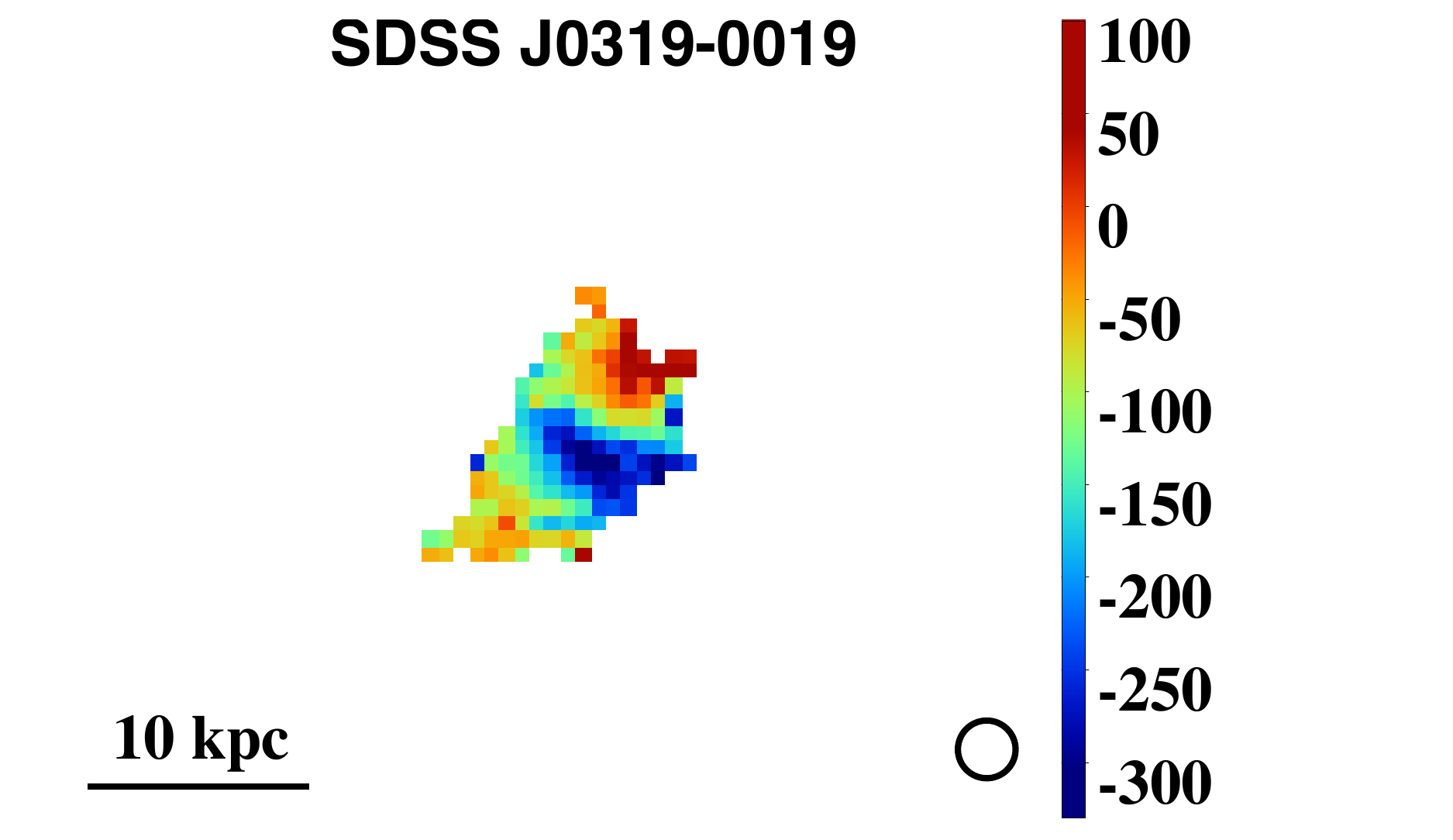}%
\includegraphics[scale=0.28,clip=clip,trim=0mm 0mm 3.5cm 0mm]{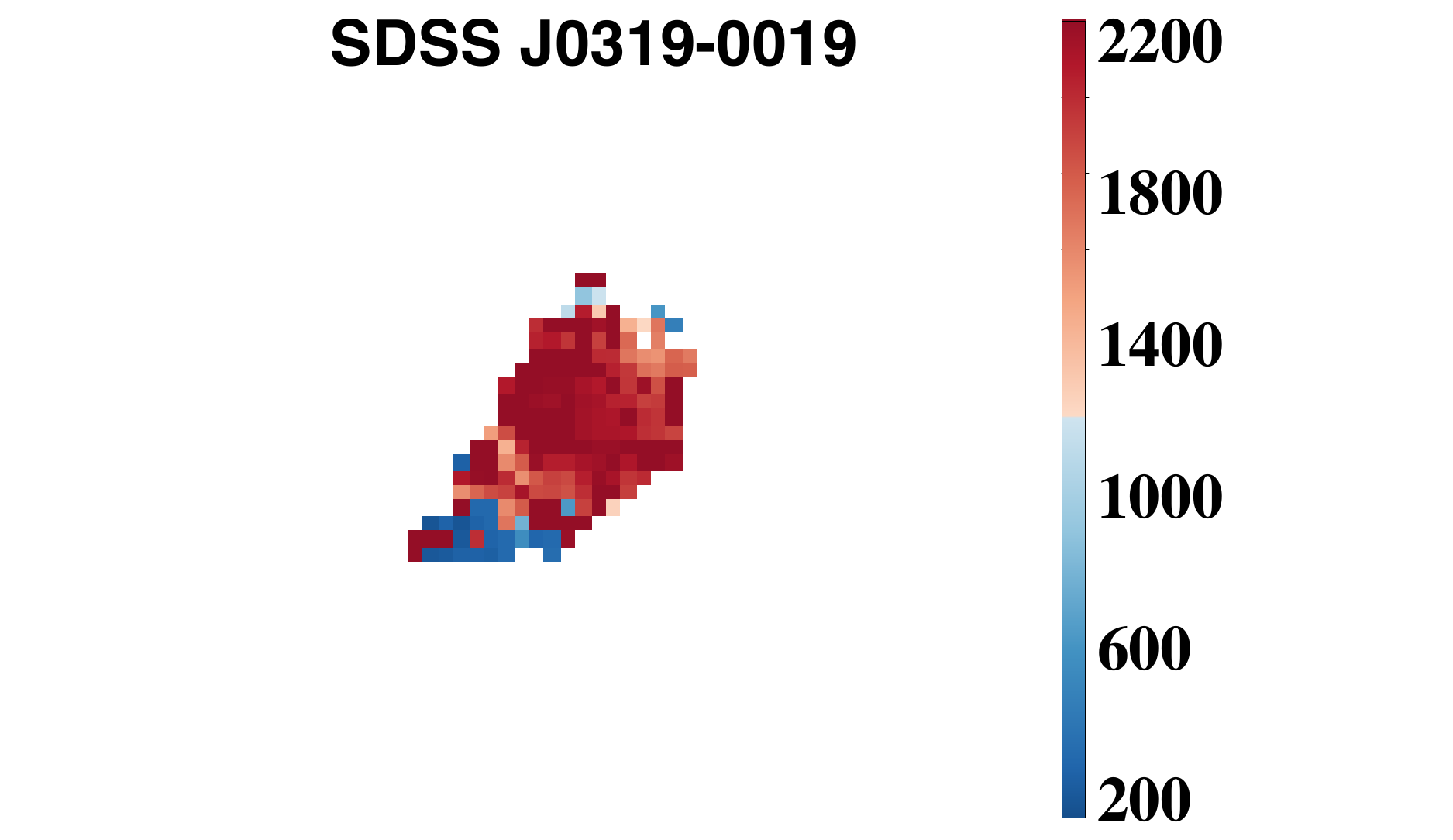}%
\includegraphics[scale=0.28,clip=clip,trim=0mm 0mm 3.5cm 0mm]{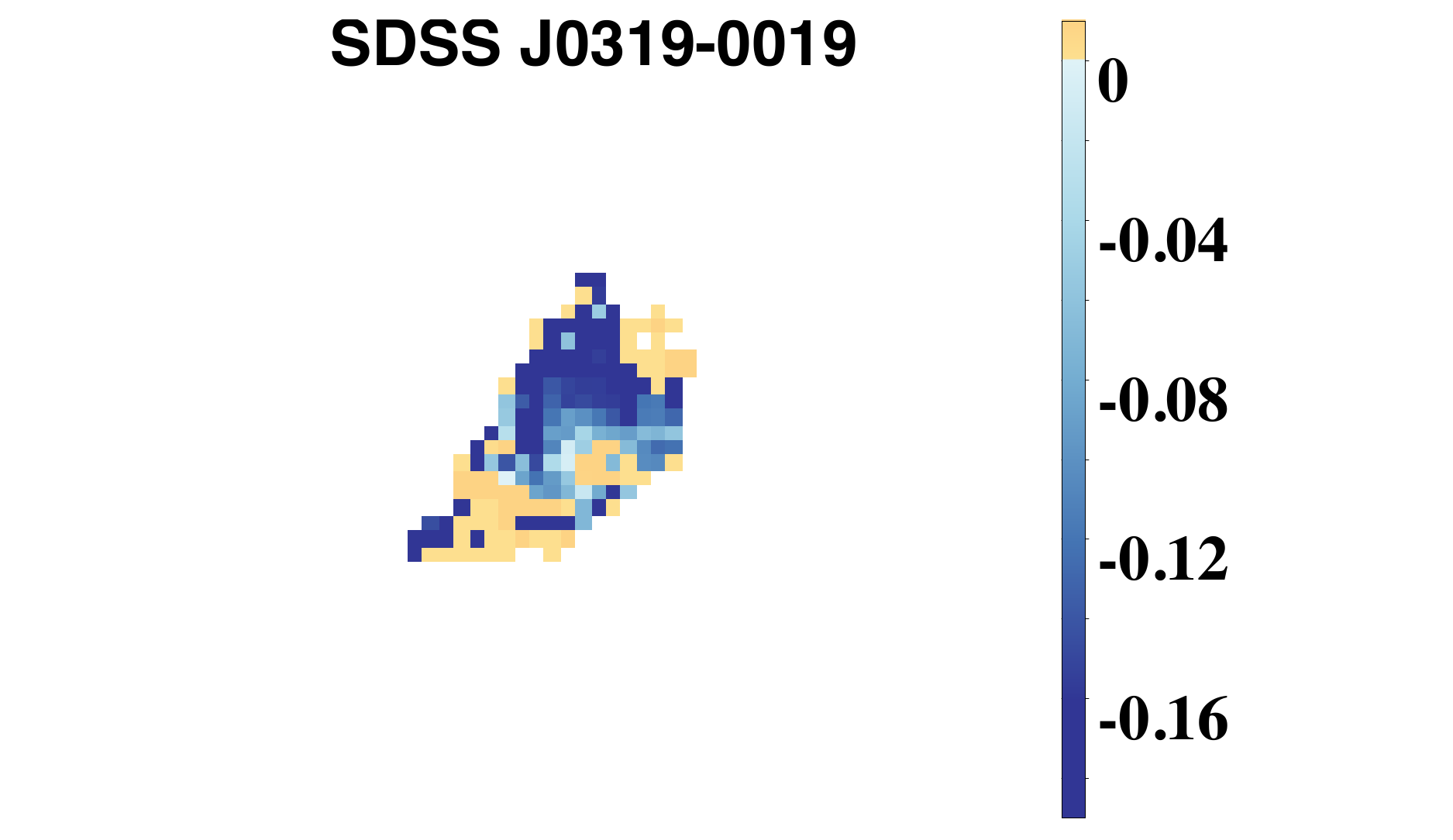}%
\includegraphics[scale=0.28,clip=clip,trim=0mm 0mm 3.5cm 0mm]{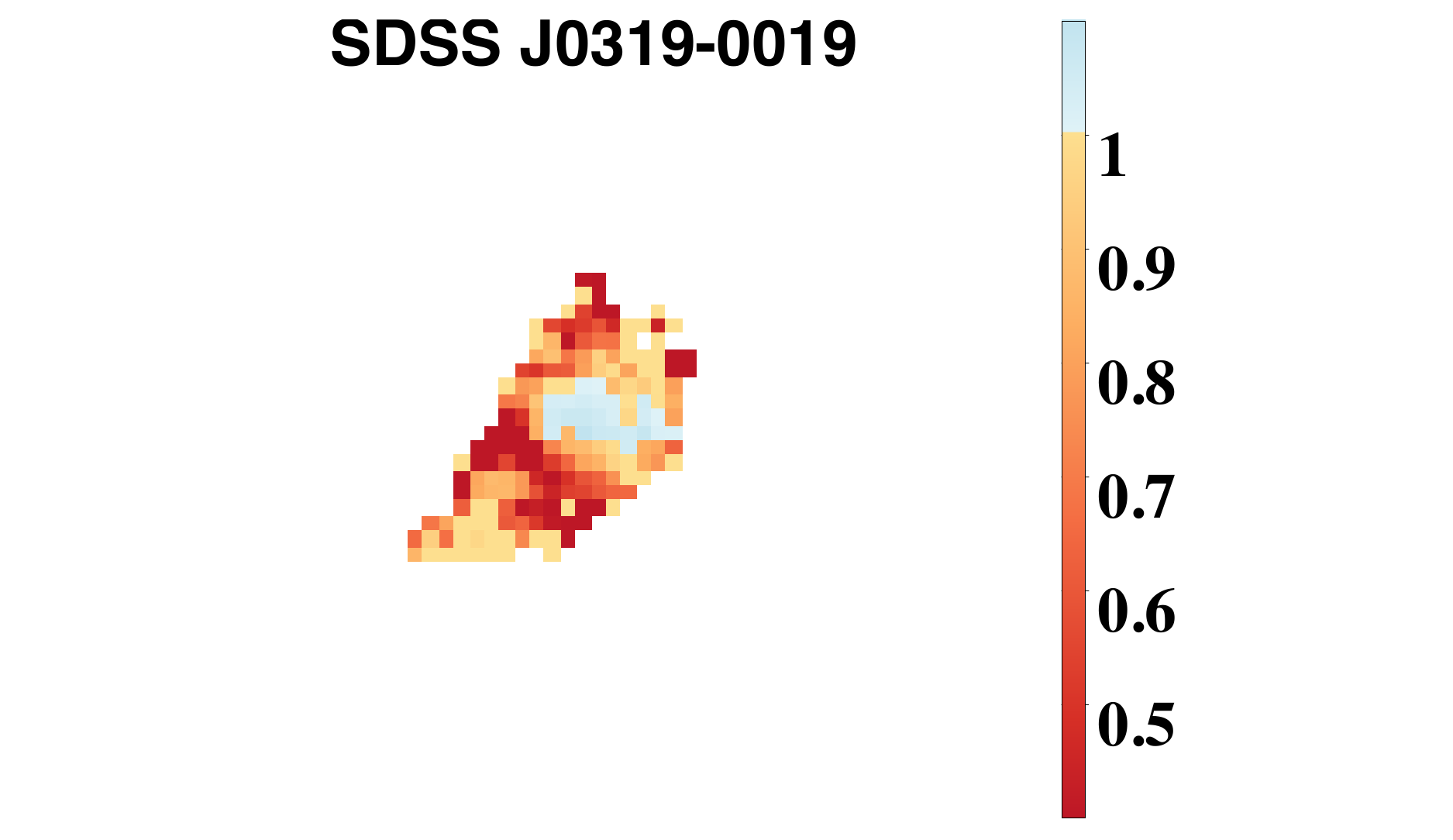}\\
\includegraphics[scale=0.28,clip=clip,trim=0mm 0mm 3.5cm 0mm]{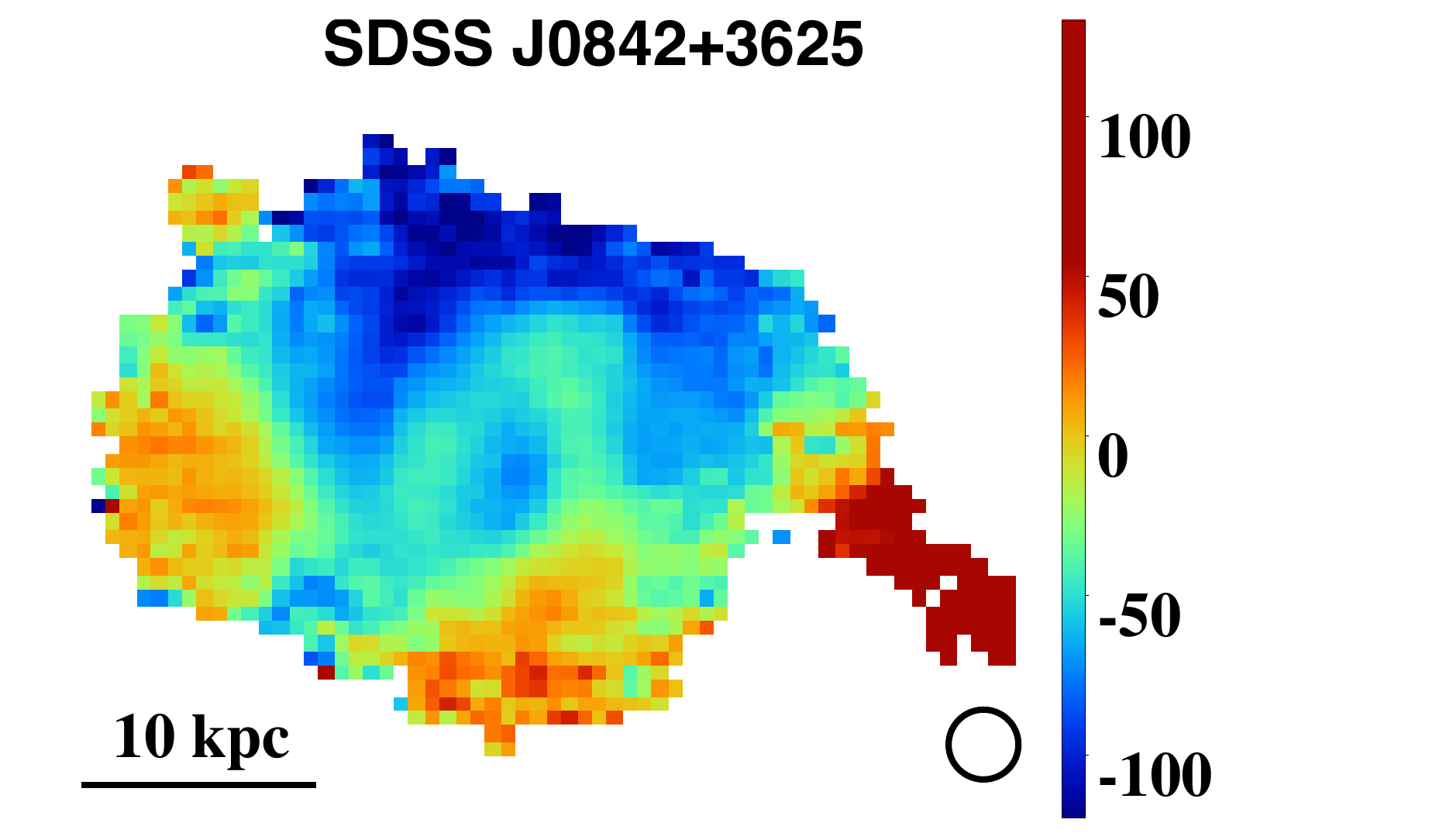}%
\includegraphics[scale=0.28,clip=clip,trim=0mm 0mm 3.5cm 0mm]{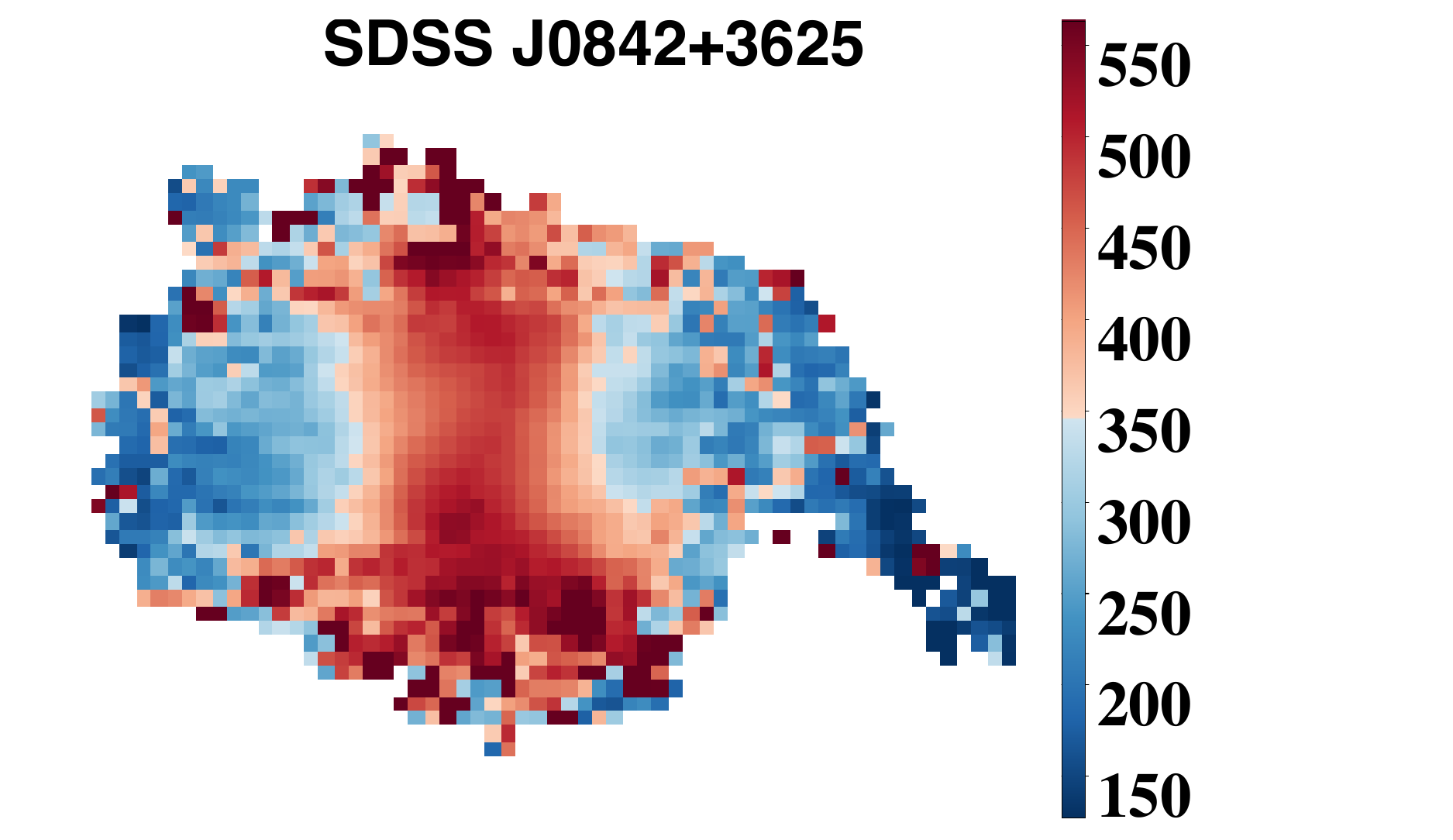}%
\includegraphics[scale=0.28,clip=clip,trim=0mm 0mm 3.5cm 0mm]{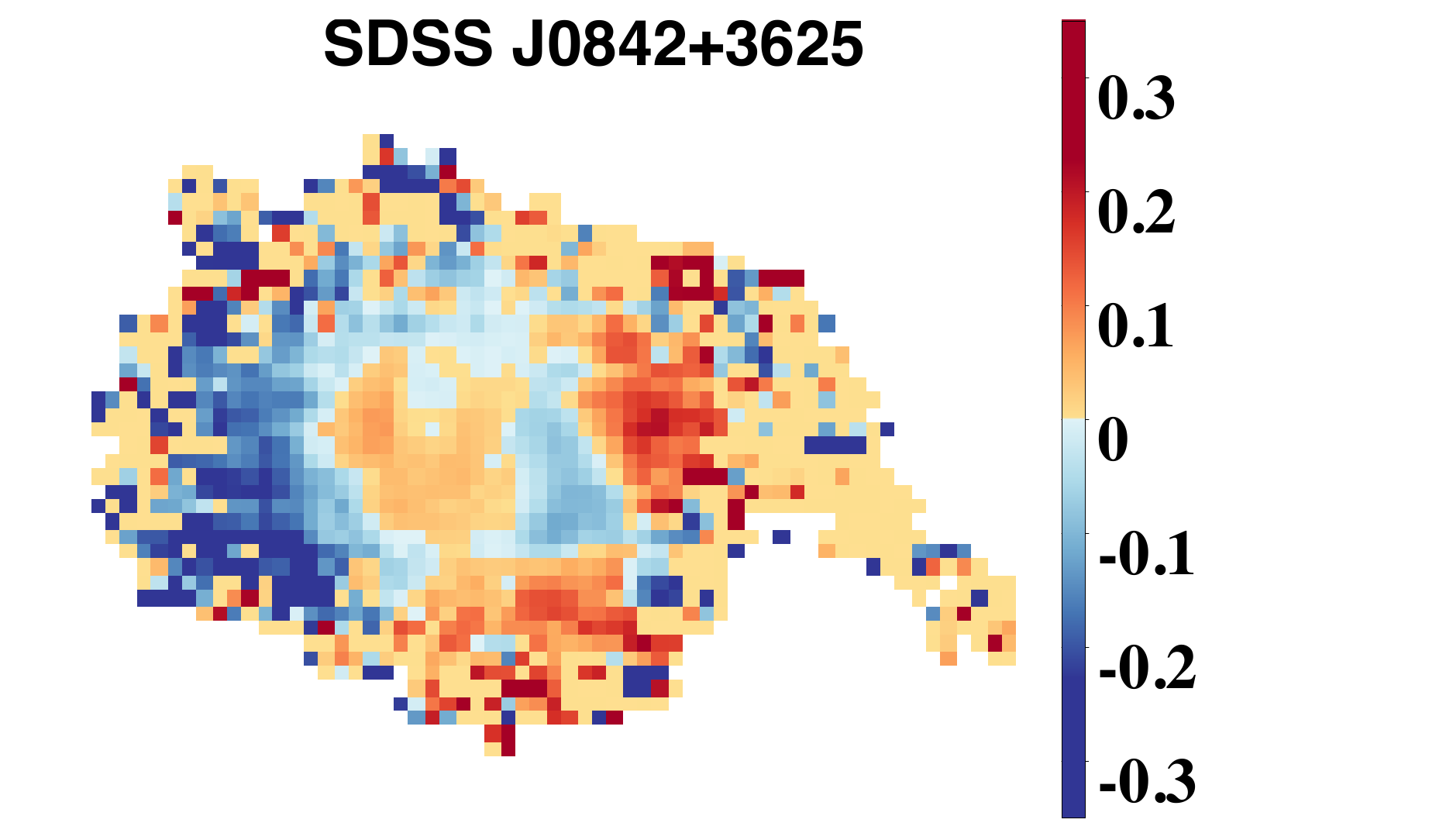}%
\includegraphics[scale=0.28,clip=clip,trim=0mm 0mm 3.5cm 0mm]{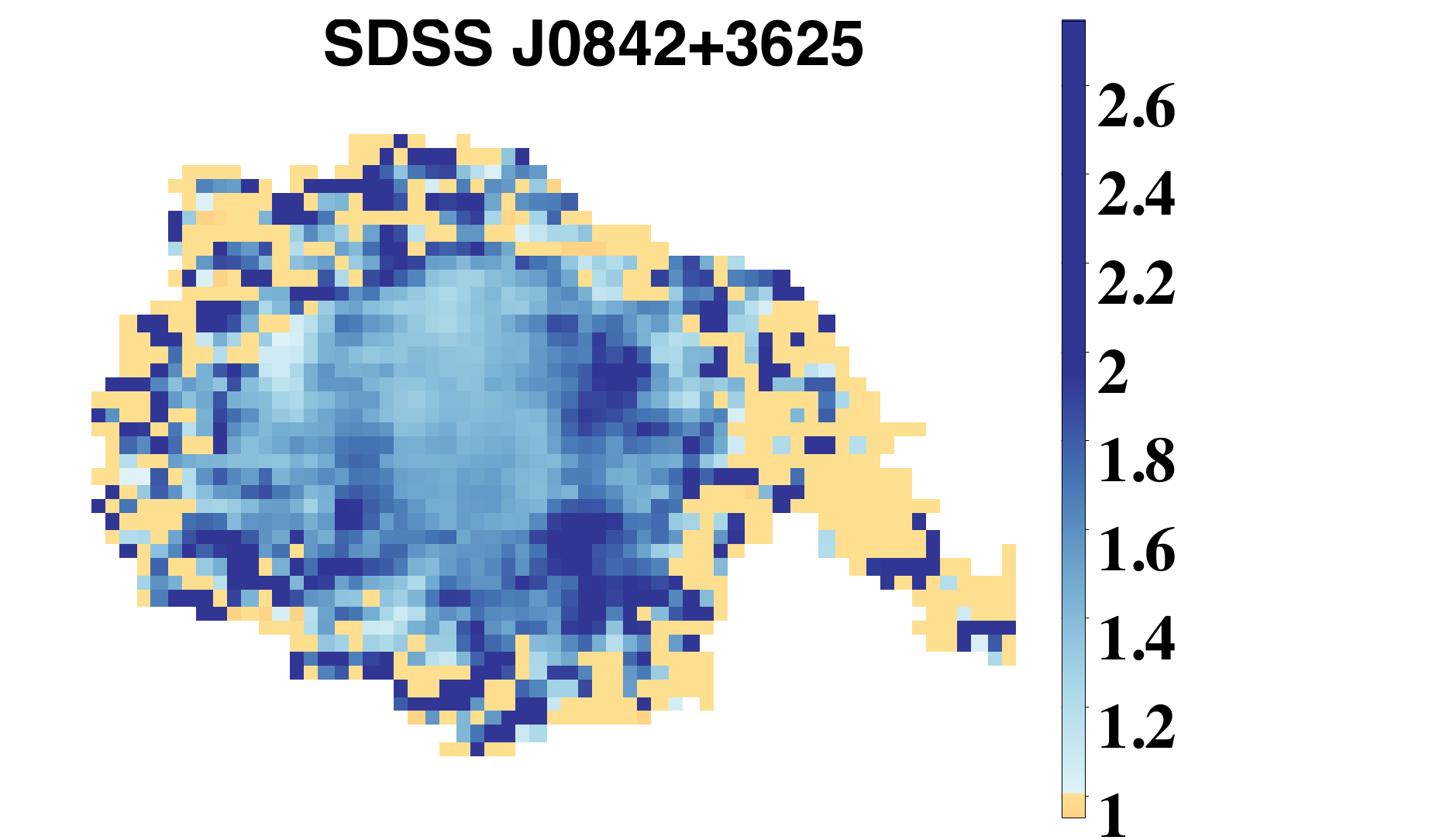}\\
\caption{Corrected Figure 3 in \citet{Liu_2013a, Liu_2013b} for the 3 quasars involved in this 
issue, though changes in these non-parameteric measurement maps are negligible. 
Left to right are: median velocity (km s$^{-1}$), line width (W80, km s$^{-1}$), asymmetry (A) and shape parameter (K) maps. The figure format is matched to Figure 3 in \citet{Liu_2013b} and more details can be found there.} 
\label{kin_new}
\end{figure*}
\begin{figure*}[H]
\centering
\includegraphics[scale=0.27,trim=0mm 0mm 22mm 0mm]{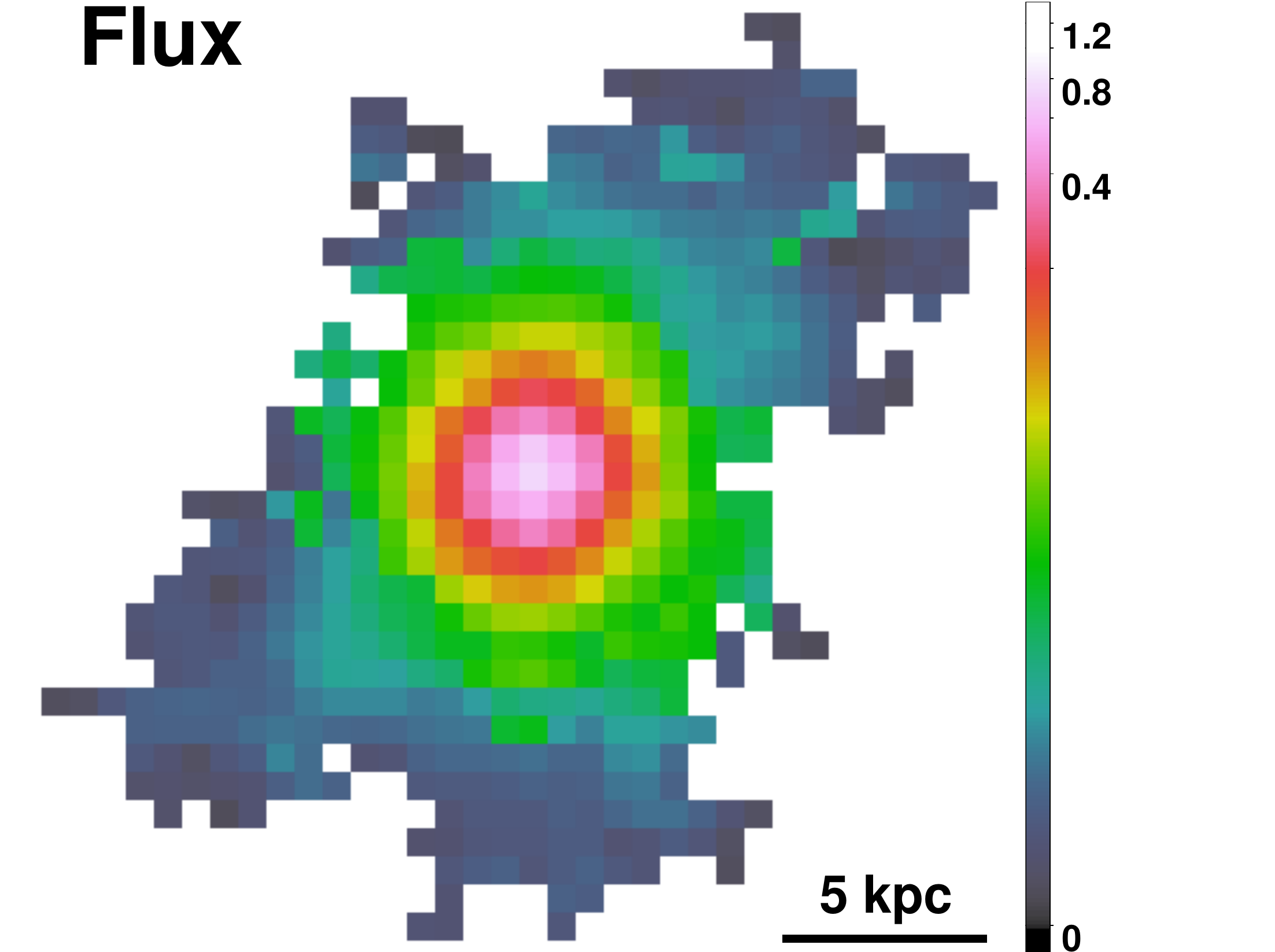}%
\includegraphics[scale=0.27,trim=0mm 0mm 22mm 0mm]{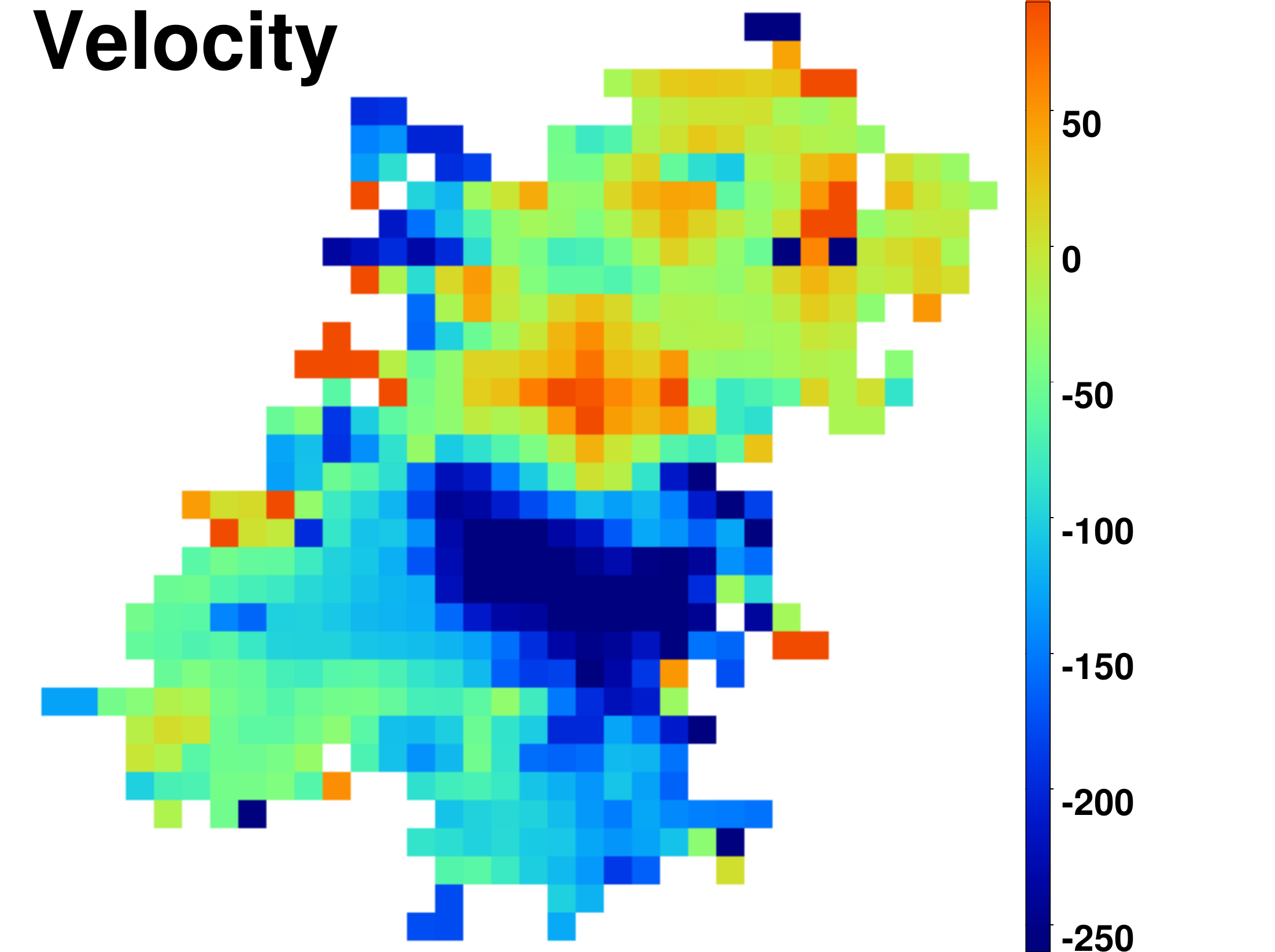}%
\includegraphics[scale=0.27,trim=0mm 0mm 22mm 0mm]{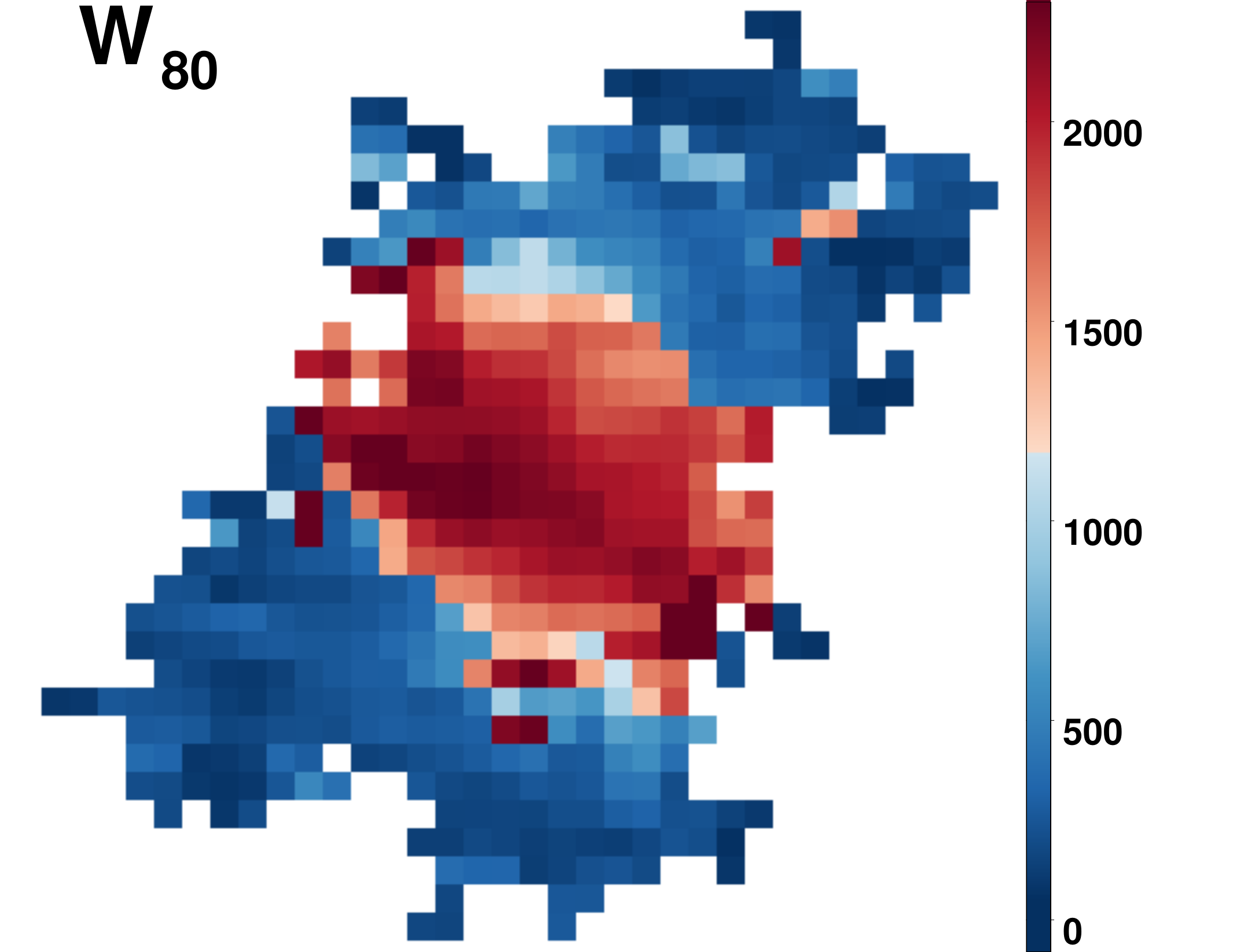}\\
\caption{Corrected Figure 8 in \citet{Liu_2013b} for the superbubble
candidate SDSS J0319$-$0019. We show the surface brightness (logarithmic scale, in units of $10^{-14}$ erg s$^{-1}$ cm$^{-2}$ arcsec$^{-2}$), radial velocity and W80 maps (both in units of km s$^{-1}$). The PSF anormaly in the [O III] surface brightness map is fixed, but the other two maps are minimally affected. The figure format is matched to Figure 8 in \citet{Liu_2013b} and more details can be found there.} 
\label{maps_new}
\end{figure*}
\begin{figure*}[H]
\centering
\includegraphics[scale=0.35,trim=0mm 0mm 3cm 0mm]{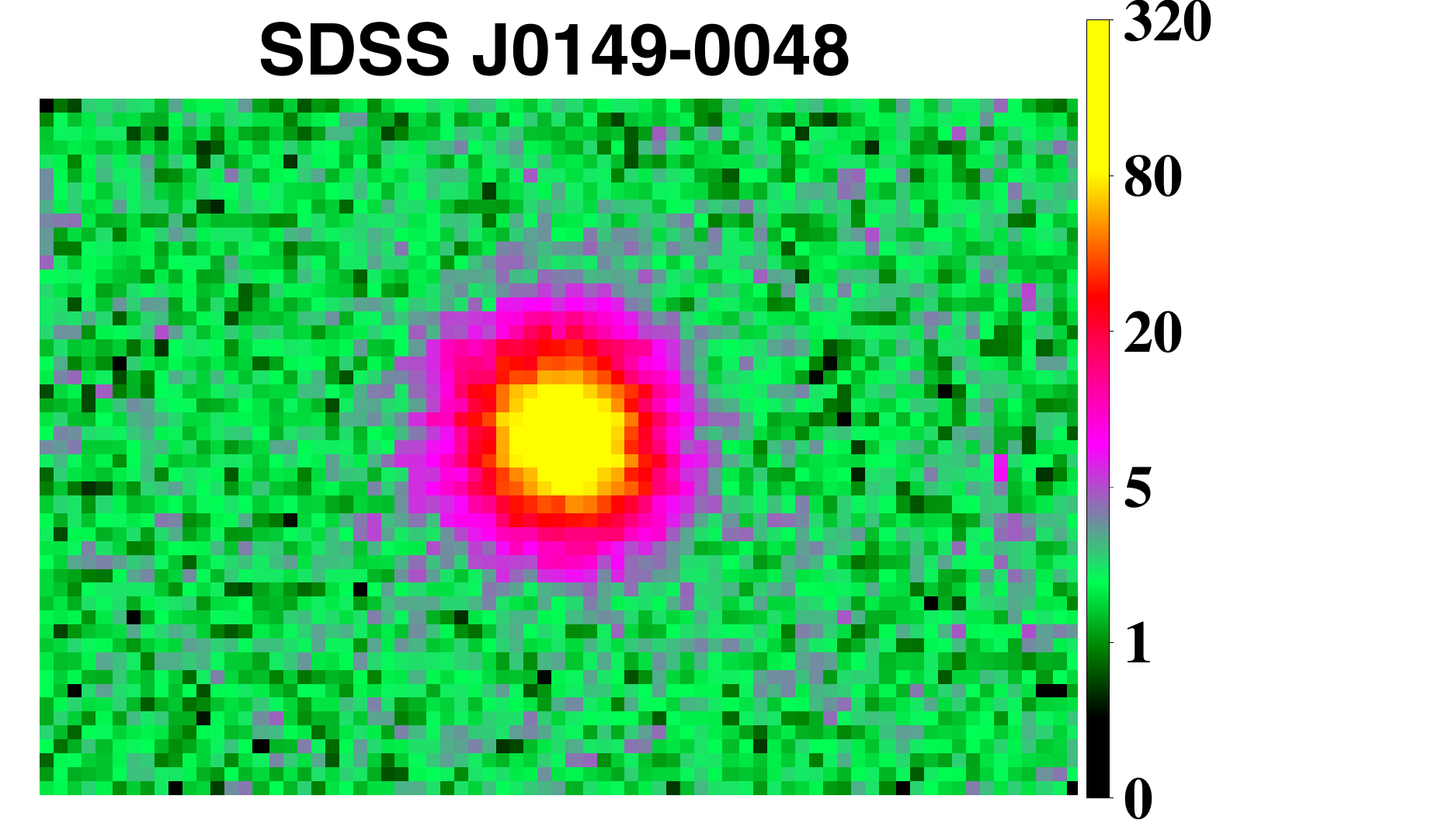}%
\includegraphics[scale=0.35,trim=0mm 0mm 3cm 0mm]{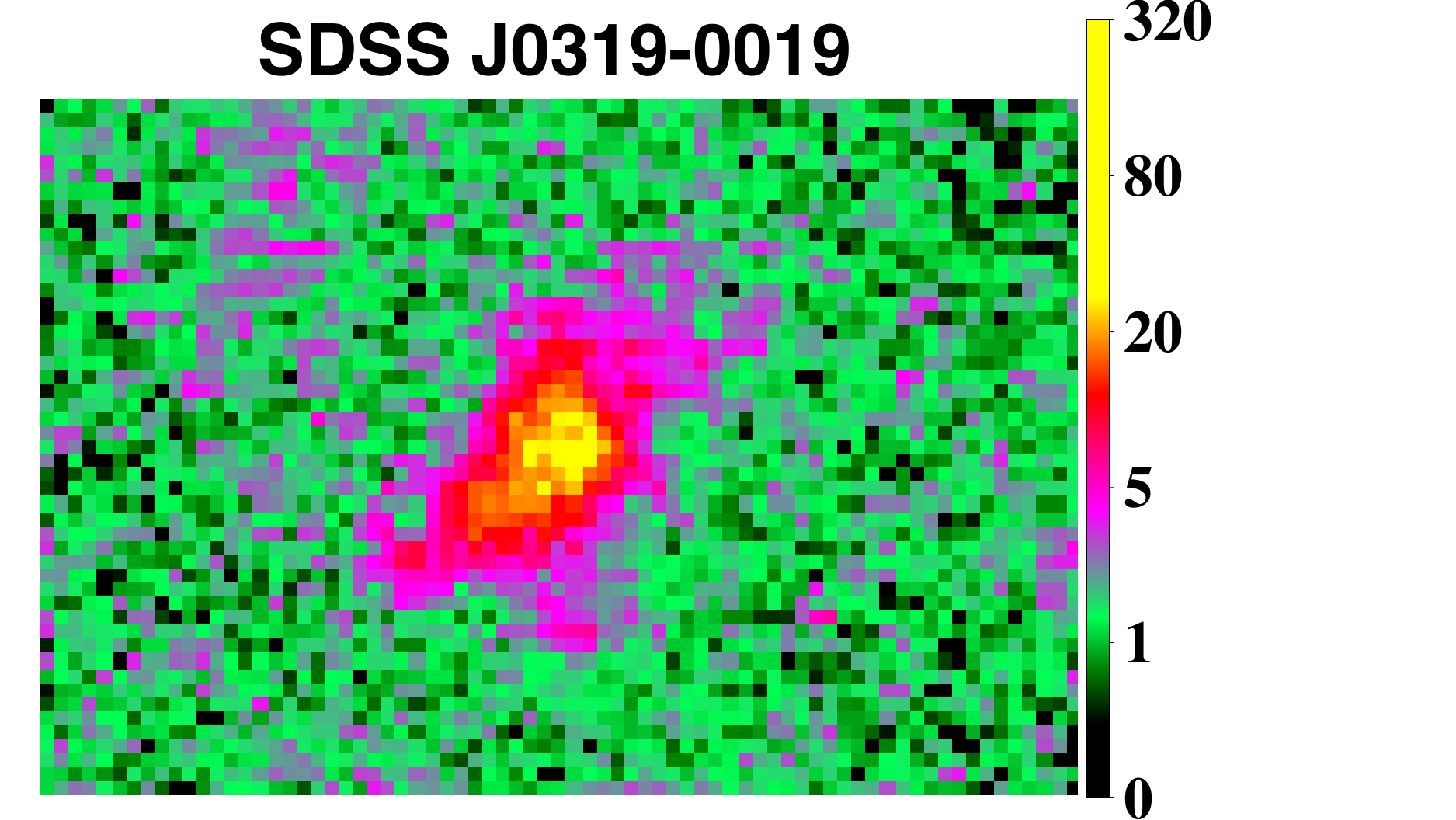}%
\includegraphics[scale=0.35,trim=0mm 0mm 3cm 0mm]{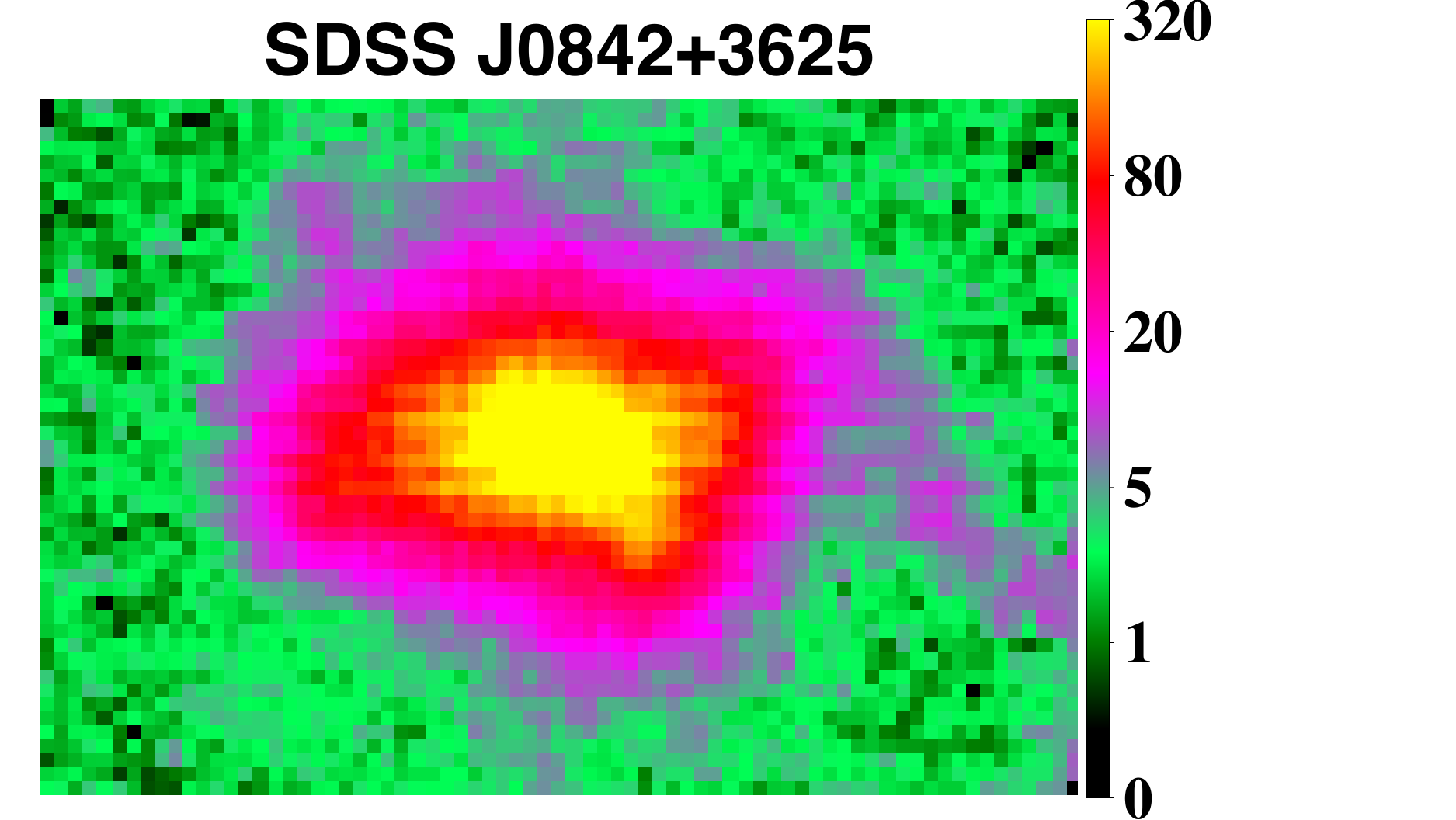}\\
\caption{Corrected Figure 9 in \citet{Liu_2013b} for the 3 relevant quasars.We show maps of S/N ratios at the peak of the [O III] emission line profile. The figure format is matched to Figure 9 in \citet{Liu_2013b} and more details can be found there.} 
\label{maps_new2}
\end{figure*}

\section{Source Flux Densities}

\begin{table*}[H]
\caption{SDSS and UKIDDS flux densities and $1\sigma$ uncertainties of Sample I and II}
\centering
{\scriptsize
\begin{tabular}{llllllllll}
\hline\hline
Object Name  & u      & g      & r      & i       & z  & Y      & J              & H              & K\\    
 & $\mu$Jy& $\mu$Jy&$\mu$Jy&$\mu$Jy&$\mu$Jy&$\mu$Jy&$\mu$Jy&$\mu$Jy& $\mu$Jy\\
\hline
SDSS J0149-0048	&			3.8	$\pm$	1.1	&	9.4	$\pm$	0.4	&	22.7	$\pm$	0.8	&	52.5	$\pm$	1.0	&	50.6	$\pm$	3.2	&	78.8	$\pm$	4.0	&	57.3	$\pm$	5.1	&	85.1	$\pm$	3.9	&	91.9	$\pm$	6.4	\\
SDSS J0210-1001	&			13.1	$\pm$	1.3	&	15.1	$\pm$	0.7	&	25.4	$\pm$	0.7	&	77.3	$\pm$	1.4	&	27.0	$\pm$	3.5	&	-		-	&	-			&	-			&	-			\\
SDSS J0319-0019	&			12.9	$\pm$	1.7	&	20.3	$\pm$	0.7	&	56.0	$\pm$	1.0	&	103.8	$\pm$	1.9	&	106.7	$\pm$	3.9	&	132.2	$\pm$	2.9	&	122.8	$\pm$	2.6	&	163.0	$\pm$	4.7	&	196.0	$\pm$	5.4	\\
SDSS J0319-0058	&			4.1	$\pm$	1.0	&	6.0	$\pm$	0.5	&	13.2	$\pm$	0.7	&	37.7	$\pm$	1.4	&	24.9	$\pm$	4.2	&	36.5	$\pm$	2.7	&	28.8	$\pm$	2.5	&	45.6	$\pm$	6.2	&	54.7	$\pm$	6.3	\\
SDSS J0321+0016	&			5.5	$\pm$	1.7	&	8.2	$\pm$	0.6	&	17.1	$\pm$	0.8	&	35.0	$\pm$	1.3	&	40.9	$\pm$	4.3	&	48.8	$\pm$	2.7	&	53.3	$\pm$	2.7	&	86.5	$\pm$	4.8	&	121.9	$\pm$	6.3	\\
SDSS J0759+1339	&			3.1	$\pm$	0.8	&	8.6	$\pm$	0.4	&	16.6	$\pm$	0.6	&	37.7	$\pm$	1.0	&	48.3	$\pm$	3.0	&	-			&	-			&	-			&	-			\\
SDSS J0841+2042	&			1.6	$\pm$	0.8	&	9.2	$\pm$	0.4	&	20.1	$\pm$	0.7	&	42.1	$\pm$	1.2	&	30.5	$\pm$	4.2	&	-			&	-			&	-			&	-			\\
SDSS J0842+3625	&			13.7	$\pm$	1.2	&	18.7	$\pm$	0.5	&	27.0	$\pm$	0.7	&	83.2	$\pm$	1.5	&	22.7	$\pm$	2.6	&	-			&	-			&	-			&	-			\\
SDSS J0858+4417	&			39.1	$\pm$	1.4	&	54.0	$\pm$	1.5	&	71.1	$\pm$	1.3	&	127.1	$\pm$	1.2	&	124.7	$\pm$	5.6	&	-			&	-			&	-			&	-			\\
SDSS J1039+4512	&			5.1	$\pm$	0.9	&	9.6	$\pm$	0.4	&	17.9	$\pm$	0.7	&	57.0	$\pm$	1.0	&	27.5	$\pm$	2.2	&	-			&	-			&	-			&	-			\\
SDSS J1040+4745	&			55.0	$\pm$	2.0	&	65.5	$\pm$	1.2	&	92.9	$\pm$	1.7	&	182.0	$\pm$	3.3	&	152.8	$\pm$	4.2	&	-			&	-			&	-			&	-			\\
	&	&																						&	&	&	&	&	&	&	\\						
SDSS J0123+0044	&			6.1	$\pm$	1.2	&	17.1	$\pm$	0.6	&	32.8	$\pm$	0.9	&	81.7	$\pm$	1.5	&	93.8	$\pm$	5.9	&	70.2	$\pm$	4.0	&	98.9	$\pm$	5.8	&	180.5	$\pm$	7.5	&	354.0	$\pm$	7.2	\\
SDSS J0920+4531	&			20.0	$\pm$	1.8	&	32.2	$\pm$	0.6	&	75.2	$\pm$	1.4	&	129.4	$\pm$	1.2	&	161.4	$\pm$	5.8	&	-			&	-			&	-			&	-			\\
SDSS J1039+6430	&			21.9	$\pm$	1.2	&	35.0	$\pm$	0.6	&	52.5	$\pm$	1.0	&	105.7	$\pm$	1.9	&	94.6	$\pm$	4.3	&	-			&	-			&	-			&	-			\\
SDSS J1106+0357	&			24.4	$\pm$	2.4	&	84.7	$\pm$	1.6	&	307.6	$\pm$	2.8	&	436.5	$\pm$	4.0	&	591.6	$\pm$	10.8	&	218.6	$\pm$	2.8	&	326.3	$\pm$	3.3	&	438.7	$\pm$	4.4	&	525.0	$\pm$	5.9	\\
SDSS J1243-0232	&			15.7	$\pm$	1.8	&	59.7	$\pm$	1.1	&	205.1	$\pm$	1.9	&	248.9	$\pm$	2.3	&	369.8	$\pm$	6.8	&	167.8	$\pm$	3.3	&	229.3	$\pm$	3.7	&	-			&	383.3	$\pm$	5.5	\\
SDSS J1301-0058	&			20.1	$\pm$	1.4	&	57.0	$\pm$	1.0	&	183.7	$\pm$	1.7	&	194.1	$\pm$	1.8	&	242.1	$\pm$	6.6	&	118.6	$\pm$	4.1	&	159.7	$\pm$	5.2	&	220.7	$\pm$	4.2	&	280.1	$\pm$	5.3	\\
SDSS J1323-0159	&			9.1	$\pm$	1.2	&	19.8	$\pm$	0.5	&	53.0	$\pm$	1.0	&	43.7	$\pm$	1.2	&	104.7	$\pm$	4.7	&	54.1	$\pm$	3.1	&	77.5	$\pm$	4.3	&	-			&	-			\\
SDSS J1413-0142	&			5.8	$\pm$	1.1	&	18.9	$\pm$	0.7	&	36.6	$\pm$	1.0	&	62.5	$\pm$	1.7	&	105.7	$\pm$	4.8	&	54.9	$\pm$	3.0	&	-			&	125.7	$\pm$	3.9	&	160.6	$\pm$	6.2	\\
SDSS J2358-0009	&			18.9	$\pm$	1.3	&	32.2	$\pm$	0.6	&	73.1	$\pm$	1.3	&	133.1	$\pm$	1.2	&	154.2	$\pm$	5.6	&	96.9	$\pm$	3.1	&	124.3	$\pm$	3.7	&	164.9	$\pm$	7.3	&	224.2	$\pm$	6.6	\\
\hline
\end{tabular}}
\label{multi_lambda}
\end{table*}

\begin{table*}[H]
\scriptsize
\caption{\textit{WISE} and \textit{Herschel} (or \textit{Spitzer}) flux densities and $1\sigma$ uncertainties of Sample I and II}
\begin{center}
\begin{tabular}{lllllll}
\hline\hline
Object Name   & WISE1             & WISE2             & WISE3              & WISE4              & PACS 70$\mu$m     & PACS 160$\mu$m      \\
 &mJy&mJy&mJy&mJy&mJy&mJy\\
\hline
SDSS J0149-0048	&	0.11	$\pm$	0.02	&	0.23	$\pm$	0.04	&	2.1	$\pm$	0.3	&	12.1	$\pm$	1.8	&	34.6	$\pm$	0.3	&	21.7	$\pm$	3.8	\\
SDSS J0210-1001	&	0.18	$\pm$	0.03	&	0.41	$\pm$	0.06	&	2.2	$\pm$	0.3	&	4.9	$\pm$	0.7	&	8.88$^{*}$	$\pm$	4.3	&	-			\\
SDSS J0319-0019	&	0.28	$\pm$	0.04	&	0.30	$\pm$	0.05	&	1.2	$\pm$	0.2	&	5.0	$\pm$	0.8	&	7.3	$\pm$	1.2	&	16.2	$\pm$	6.1	\\
SDSS J0319-0058	&	0.11	$\pm$	0.02	&	0.17	$\pm$	0.03	&	0.9	$\pm$	0.1	&	4.1	$\pm$	0.6	&	8.06$^{*}$	$\pm$	5.4	&	-			\\
SDSS J0321+0016	&	0.27	$\pm$	0.04	&	0.38	$\pm$	0.06	&	1.8	$\pm$	0.3	&	5.8	$\pm$	0.9	&	9.0	$\pm$	0.7	&	11.4	$\pm$	0.3	\\
SDSS J0759+1339	&	0.12	$\pm$	0.02	&	0.22	$\pm$	0.03	&	2.3	$\pm$	0.4	&	14.9	$\pm$	2.2	&	21.2	$\pm$	2.7	&	19.8	$\pm$	3.4	\\
SDSS J0841+2042	&	0.09	$\pm$	0.01	&	0.36	$\pm$	0.05	&	1.5	$\pm$	0.2	&	-			&	10.6	$\pm$	1.3	&	-			\\
SDSS J0842+3625	&	0.12	$\pm$	0.02	&	0.13	$\pm$	0.02	&	1.8	$\pm$	0.3	&	8.3	$\pm$	1.2	&	18.01$^{*}$	$\pm$	5.7	&	19.1	$\pm$	16.4	\\
SDSS J0858+4417	&	1.98	$\pm$	0.30	&	4.32	$\pm$	0.65	&	22.5	$\pm$	3.4	&	73.6	$\pm$	11.0	&	109.5	$\pm$	1.6	&	61.1	$\pm$	3.7	\\
SDSS J1039+4512	&	0.13	$\pm$	0.02	&	0.40	$\pm$	0.06	&	3.3	$\pm$	0.5	&	12.6	$\pm$	1.9	&	22.3	$\pm$	2.8	&	9.6	$\pm$	1.4	\\
SDSS J1040+4745	&	0.58	$\pm$	0.09	&	0.72	$\pm$	0.11	&	6.1	$\pm$	0.9	&	35.2	$\pm$	5.3	&	-			&	-			\\
	&	&&&&& \\																							
SDSS J0123+0044	&	0.97	$\pm$	0.02	&	1.73	$\pm$	0.04	&	3.8	$\pm$	0.1	&	9.1	$\pm$	0.8	&	-			&	-			\\
SDSS J0920+4531	&	1.09	$\pm$	0.03	&	1.89	$\pm$	0.04	&	5.7	$\pm$	0.2	&	21.9	$\pm$	1.3	&	-			&	-			\\
SDSS J1039+6430	&	1.34	$\pm$	0.03	&	2.95	$\pm$	0.06	&	9.1	$\pm$	0.2	&	38.5	$\pm$	1.3	&	-			&	-			\\
SDSS J1106+0357	&	0.70	$\pm$	0.02	&	0.80	$\pm$	0.02	&	3.0	$\pm$	0.2	&	15.2	$\pm$	1.1	&	-			&	31.5$^{*}$	$\pm$	6.3	\\
SDSS J1243-0232	&	0.44	$\pm$	0.01	&	0.36	$\pm$	0.02	&	1.0	$\pm$	0.1	&	8.5	$\pm$	1.2	&	-			&	-			\\
SDSS J1301-0058	&	0.41	$\pm$	0.01	&	0.57	$\pm$	0.02	&	2.0	$\pm$	0.1	&	12.7	$\pm$	0.9	&	-			&	-			\\
SDSS J1323-0159	&	0.22	$\pm$	0.01	&	0.42	$\pm$	0.02	&	2.7	$\pm$	0.2	&	13.2	$\pm$	1.1	&	-			&	58.7	$\pm$	11.7	\\
SDSS J1413-0142	&	0.18	$\pm$	0.01	&	0.18	$\pm$	0.01	&	2.5	$\pm$	0.1	&	17.0	$\pm$	0.9	&	-			&	104.8	$\pm$	21.0	\\
SDSS J2358-0009	&	0.36	$\pm$	0.01	&	0.40	$\pm$	0.02	&	1.5	$\pm$	0.2	&	5.7	$\pm$	1.0	&	-			&	-			\\
\hline 
\end{tabular}
\end{center}
\label{multi_lambda2}
\begin{tablenotes}
\item \textit{Notes:}  Flux densities marked with $^{*}$ are 70~$\mu$m and 160~$\mu$m flux densities from the \textit{Spitzer Space Telescope}.
        \end{tablenotes}
\end{table*}


\bsp	
\label{lastpage}
\end{document}